\documentclass[useAMS,usenatbib]{mn2e}
\usepackage{graphicx}
\usepackage{amsmath}
\usepackage{epstopdf}
\usepackage{grffile}
\usepackage{amssymb}
\usepackage{lscape}
\usepackage{array}
\title[SPIDERS: X-ray galaxy cluster follow-up]{SPIDERS: the spectroscopic follow-up of X-ray selected clusters of galaxies in SDSS-IV}
\author[N. Clerc et al. ]{N. Clerc$^{1}$\thanks{E-mail:
nclerc@mpe.mpg.de}, A. Merloni$^{1}$, Y.-Y. Zhang$^{2}$, A. Finoguenov$^{1,3}$, T. Dwelly$^{1}$, K. Nandra$^{1}$,
\newauthor C. Collins$^{4}$, K. Dawson$^{9}$, J.-P. Kneib$^{10,11}$, E. Rozo$^{5}$, E. Rykoff$^{6}$, T. Sadibekova$^{7,8}$,
\newauthor  J. Brownstein$^{9}$, Y.-T. Lin$^{12}$, J. Ridl$^{1}$, M. Salvato$^{1}$, A. Schwope$^{13}$,
\newauthor  M. Steinmetz$^{13}$, H.-J. Seo$^{14}$, J. Tinker$^{15}$
\\
$^{1}$ Max-Planck Institut f{\"u}r extraterrestrische Physik, Postfach 1312, 85741 Garching bei M{\"u}nchen, Germany.\\
$^{2}$ Argelander-Institut f{\"u}r Astronomie, Universit{\"a}t Bonn, Auf dem H{\"u}gel 71, 53121 Bonn, Germany \\
$^{3}$ Department of Physics, University of Helsinki, Gustaf H{\"a}llstr{\"o}min katu 2a, FI-00014 Helsinki, Finland \\
$^{4}$ Astrophys. Research Instit., Liverpool John Moores Univ., IC2, Liverpool Science Park, Brownlow Hill, Liverpool, L5 3AF, UK\\
$^{5}$ Department of Physics, University of Arizona, Tucson AZ, 85721, USA\\
$^{6}$ SLAC National Accelerator Laboratory, Menlo Park, CA 94025, USA \\
$^{7}$ Service {d'a}strophysique, IRFU, CEA Saclay, 91191 Gif-sur-Yvette, France \\
$^{8}$ Ulugh Beg Astronomical Institute of Uzbekistan Academy of Science, 33 Astronomicheskaya str., Tashkent, 100052, Uzbekistan \\
$^{9}$ Department of Physics and Astronomy, University of Utah, Salt Lake City, UT 84112, USA \\
$^{10}$ Laboratoire {d'A}strophysique, Ecole Polytechnique F{\'e}d{\'e}rale de Lausanne, Observatoire de Sauverny, 1290 Versoix, Switzerland \\
$^{11}$ Aix Marseille Universit{\'e}, CNRS, LAM (Laboratoire {d'A}strophysique de Marseille), UMR 7326, 13388, Marseille, France \\
$^{12}$ Academia Sinica Institute of Astronomy and Astrophysics, P.O. Box 23-141, Taipei 10617, Taiwan \\
$^{13}$ Leibniz-Institut f{\"u}r Astrophysik Potsdam (AIP), An der Sternwarte 16, D-14482 Potsdam, Germany \\
$^{14}$ Department of Physics and Astronomy, Ohio University, 251B Clippinger Labs, Athens, OH 45701, USA \\
$^{15}$ Center for Cosmology and Particle Physics, Dep. of Phys., New York Univ., 4 Washington Place, New York, NY 10003, USA \\
}
\begin{document}

\date{Accepted -. Received -; in original form 2016 August 31}

\pagerange{\pageref{firstpage}--\pageref{lastpage}} \pubyear{2002}

\maketitle

\label{firstpage}

\begin{abstract}
SPIDERS (The SPectroscopic IDentification of \emph{eROSITA} Sources) is a program dedicated to the homogeneous and complete spectroscopic follow-up of X-ray AGN and galaxy clusters  over a large area ($\sim$7500\,deg$^2$) of the extragalactic sky. SPIDERS is part of the SDSS-IV project, together with the Extended Baryon Oscillation Spectroscopic Survey (eBOSS) and the Time-Domain Spectroscopic Survey (TDSS).
This paper describes the largest project within SPIDERS before the launch of \emph{eROSITA}: an optical spectroscopic survey of X-ray selected, massive ($\sim 10^{14}$ to $10^{15}~M_{\odot}$) galaxy clusters discovered in \emph{ROSAT} and \emph{XMM-Newton} imaging.
The immediate aim is to determine precise ($\Delta_z \sim 0.001$) redshifts for 4,000--5,000 of these systems out to $z \sim 0.6$. The scientific goal of the program is precision cosmology, using clusters as probes of large-scale structure in the expanding Universe.
We present the cluster samples, target selection algorithms and observation strategies. We demonstrate the efficiency of selecting targets using a combination of SDSS imaging data, a robust red-sequence finder and a dedicated prioritization scheme.
We describe a set of algorithms and work-flow developed to collate spectra and assign cluster membership, and to deliver catalogues of spectroscopically confirmed clusters. We discuss the relevance of line-of-sight velocity dispersion estimators for the richer systems.
We illustrate our techniques by constructing a catalogue of 230 spectroscopically validated clusters ($0.031 < z < 0.658$), found in pilot observations. We discuss two potential science applications of the SPIDERS sample: the study of the X-ray luminosity-velocity dispersion ($L_X-\sigma$) relation and the building of stacked phase-space diagrams. 
\end{abstract}

\begin{keywords}
cosmology: observations -- catalogues -- galaxies: clusters: general -- X-rays: galaxies: clusters.
\end{keywords}


\section{Introduction}

In recent years the field of galaxy cluster surveys has been re-energised by the realisation that a well-measured cluster population places strong independent constraints on cosmological models \citep[e.g.][]{boehringer2004, vikhlinin2009, mantz2010, weinberg2013}. As the largest known bound objects, clusters can be used to simultaneously probe the cosmic expansion rate and the gravitational mechanisms responsible for the growth of structure in the Universe. The evolution of the galaxy cluster mass function across cosmological times and the distribution of clusters within the tridimensional large-scale structure are two key observables, since they are readily predicted by theoretical models and simulations.

Galaxy cluster cosmology studies start with constructing cluster samples. Fortunately for observers, the hot baryonic gas trapped in galaxy clusters emits large amounts of X-ray photons, in great part due to bremsstrahlung processes. Extended X-ray objects are thus the signpost of deep potential wells, and their X-ray luminosity directly relates to the mass of the dark matter halo in which they reside. Therefore, large surveys of the sky at the high-energy end of the electromagnetic spectrum (i.e.~X-ray) permit a complete census of clusters covering a wide range of masses and redshifts, and survey data themselves can provide estimates of the mass of these objects. This is why large surveys in the X-ray wavelengths aimed at constraining cosmology with galaxy clusters have been developed since the early years of X-ray astronomy \citep{henryarnaud1991, bahcallcen1993, jonesforman1984}, with a notable step-change brought about by the ROSAT all-sky survey \citep{ebeling2000, ikebe2002, reiprichboehringer2002, schuecker2003} and ROSAT serendipitous surveys \citep{rosati1998, romer2000,burke2003, burenin2007} and the support of Chandra and XMM-Newton \citep{pacaud2006, vikhlinin2009, mantz2010, finoguenov2010, clerc2014, pierre2016}.
The next major advance in the field will be offered by \emph{eROSITA} \cite[extended ROentgen Survey with an Imaging Telescope Array,][]{predehl2014} which will survey the entire sky in the 0.3-10~keV energy range at depths 10 to 30 times deeper than ROSAT. The combination of eROSITA's field of view, angular resolution and sensitivity will lead to the detection of $\sim 100,000$ galaxy clusters down to $[0.5-2]$~keV fluxes of $\sim 3\times 10^{-14}$~ergs\,s$^{-1}$ cm$^{-2}$ and up to redshifts of unity and beyond \cite[see][]{merloni2012,borm2014}.

However, X-ray observations alone, in general, are not sufficient to fully assess the nature of the emitting sources and, most importantly, to determine their redshifts. Therefore, optical observations play a critical role in complementing surveys of galaxy clusters in X-rays. Whilst multi-filter optical imaging proves efficient at detecting and characterizing galaxy clusters -- notably through their ubiquitous red sequence \citep[e.g.][]{gladdersyee2000, rykoff2014} --, ultimate confirmation of a galaxy cluster is achieved by optical spectroscopy. Spectroscopic observations of cluster members can be used to disentangle projection effects and substructures from real concentrations, and they also provide the precise redshift of the halo, and therefore lead to precise luminosities and masses once they are combined with X-ray measurements. Obtaining spectroscopic redshifts for galaxy cluster members is recognized as a major bottleneck in X-ray cluster surveys, because of the double need for deep imaging data to select targets, and the deep spectroscopic exposures necessary for redshift determination.

The SPIDERS (SPectroscopic IDentification of \emph{eROSITA} Sources) cluster program is specifically designed to overcome this bottleneck. It relies on the BOSS spectrograph mounted on the SDSS-2.5m telescope at Apache Point Observatory \citep{gunn2006} to follow-up galaxies detected in the large area of extragalactic sky imaged in $ugriz$ filters by the same telescope. The SDSS/BOSS instrumentation and infrastructure are used in combination with most recent techniques in finding X-ray galaxy clusters and their photometric members, in order to perform an unprecedentedly wide spectroscopic survey of X-ray galaxy clusters. Advanced techniques used in this work include the wavelet filtering of X-ray maps, and the use of a series of matched filters to look for red-sequence galaxies in a multi-variate optical parameter space (colour, position and magnitude).

This paper (one of a series of SDSS-IV technical papers), describes the targeting and analysis steps leading to the construction of a large, spectroscopically validated sample of X-ray selected galaxy clusters within SPIDERS. In the preparation phase for \emph{eROSITA}, these samples are drawn from ROSAT and XMM data. Throughout this paper we use a small pilot survey dubbed SEQUELS (Sloan Extended Quasar, ELG, and LRG Survey, a precursor to the main SPIDERS/eBOSS/TDSS program) to illustrate the efficacy of our targeting and analysis approach. We present the first scientific results from the SPIDERS cluster program.

The paper is structured as follows. Section~\ref{sect:spiders} is devoted to the general presentation of the SPIDERS cluster survey and the samples of X-ray clusters it is based upon. The selection of targets for the shallower, pre-\emph{eROSITA} phase of the survey is described in Sect.~\ref{sect:targeting}, along with forecasts regarding the outcome of the observations. In Sect.~\ref{sect:analysis} we depict the steps envisaged to transform observations into science-oriented catalogues. We emphasize that these methods are subject to improvements in the course of the survey. In Section~\ref{sect:sequels} we present some examples for the science exploitation of the SPIDERS program, using the validated sample of clusters from the SEQUELS pilot survey, that we compare to existing cluster catalogues in Sect.~\ref{sect:mcxc}. We conclude in Sect.~\ref{sect:conclusions}.

Unless otherwise stated, we assume a flat $\Lambda$CDM cosmological model with $\Omega_m=0.3$, $\Omega_{\Lambda}=0.7$ and $H_0 = 100 h$~km\,s$^{-1}$\,Mpc$^{-1}$ with $h=0.7$. We define $L_C \equiv L_X/E(z)$, with $E(z)=H(z)/H_0$ and $L_X$ the [0.1-2.4]~keV luminosity of a cluster.


\section[]{The SPIDERS cluster program}
	\label{sect:spiders}

	\subsection[]{General description}

SPIDERS is an observational program, part of the SDSS-IV project \citep{blanton2017}. The primary goal of SPIDERS is to obtain homogeneous and complete spectroscopic follow-up of extragalactic sources, both point-like and extended, using data from X-ray satellites and over the SDSS imaging footprint. Given the nature of these sources, SPIDERS naturally splits into two main components, i.e.~an AGN program and a cluster program. The SPIDERS AGN targeting strategy is described in Dwelly et al. (in prep.), and will collect $\sim$50,000 spectra of ROSAT, XMM and \emph{eROSITA} X-ray AGN. A pilot study for the SPIDERS AGN survey, based around the BOSS follow-up of X-ray selected AGN in the XMM-Newton XMM-XXL field, is presented in \citet[][]{menzel2016}.

This paper describes the targeting of X-ray extended sources identified as galaxy cluster candidates. The driving goals of the program are the confirmation of those candidates and the assignment of a precise redshift. This in turn leads to the determination of precise absolute cluster parameters (including X-ray luminosity and mass). A number of important secondary goals include the estimation of cluster dynamical masses (via line-of-sight velocity dispersion measurements), the study of the physical interplay between massive dark matter halos, the hot baryonic gas they host, and the galaxies that live therein.
From a technical point of view, the SPIDERS cluster program represents a novel approach to galaxy cluster spectroscopic follow-up: the large data volume involved (several thousands of galaxy clusters) demands innovative targeting and analysis strategies. Considering that the ultimate goal envisaged by the program is precision cosmology using galaxy clusters as tracers of the large-scale structure, we require that all the procedures involved must undergo careful control and validation.

\begin{figure*}
	\includegraphics[width=\linewidth]{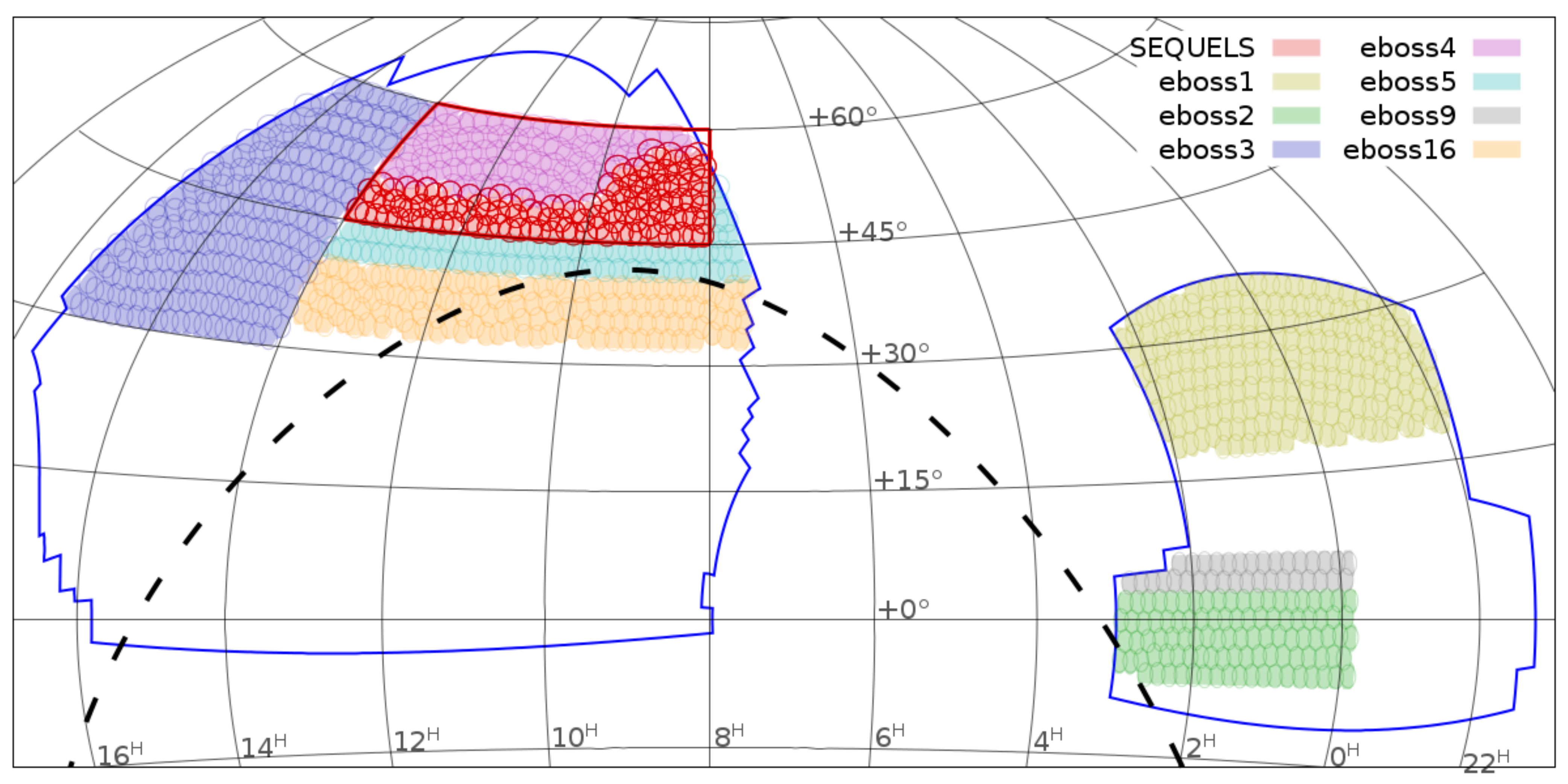}
		\caption{Location of the survey in equatorial coordinates. The blue line is the perimeter of the BOSS optical imaging area within which the eBOSS survey lies. Various distinct regions of sky (chunks) are tiled separately: in this figure chunks {\tt eboss} 1 to 5, 9 and 16 are displayed, approximating the area expected to be covered after 2 years of eBOSS/SPIDERS/TDSS survey operations. Each spectroscopic plate is represented by a circle of diameter $\sim 3 \deg$. The black dashed line indicates the boundary between the eastern and western Galactic hemispheres and delimits the German and Russian halves of the \emph{eROSITA} sky.}
	\label{fig:layout} 
\end{figure*}

SPIDERS will follow-up X-ray extended sources detected in \emph{eROSITA} data in the final years of SDSS-IV. Prior to \emph{eROSITA}'s launch, galaxy clusters identified in the shallower RASS and sparser XMM-Newton data will constitute the bulk of the SPIDERS program. There will be an incremental increase in X-ray sensitivity brought about by each of \emph{eROSITA}'s sky surveys, which will start to be be accumulated after the start of SDSS-IV. Therefore, SPIDERS is planned in three tiers. Tier~0, the shallowest tier, relies mainly on ROSAT data. When successive \emph{eROSITA} catalogues become available (following the cadence of one deeper X-ray catalogue every six months, going from eRASS:1 to eRASS:8), sources in the North Galactic Cap within the German \emph{eROSITA} sky ($180 < l < 360$)  will be added to the pool of targets. The planned launch date of \emph{eROSITA} means that Tier~1 and Tier~2 will most likely correspond to eRASS depths 2 and 4 respectively.
Fig.~\ref{fig:layout}~shows the layout of the surveys in Equatorial coordinates. The area covered in Tier~0 corresponds to the entire eBOSS footprint not covered in Tiers~1 and 2, i.e.~$>5,200~\deg^2$, while Tier~1 and 2 will lie to the South of the eROSITA-DE boundary (the black dashed curve in this figure), with exact footprints dependent on the launch date.

We detail in the following sub-sections the samples of galaxy clusters followed-up by SPIDERS in Tier~0, along with expectations regarding the deeper tiers, when \emph{eROSITA} is available.

\begin{figure}
	\includegraphics[width=84mm]{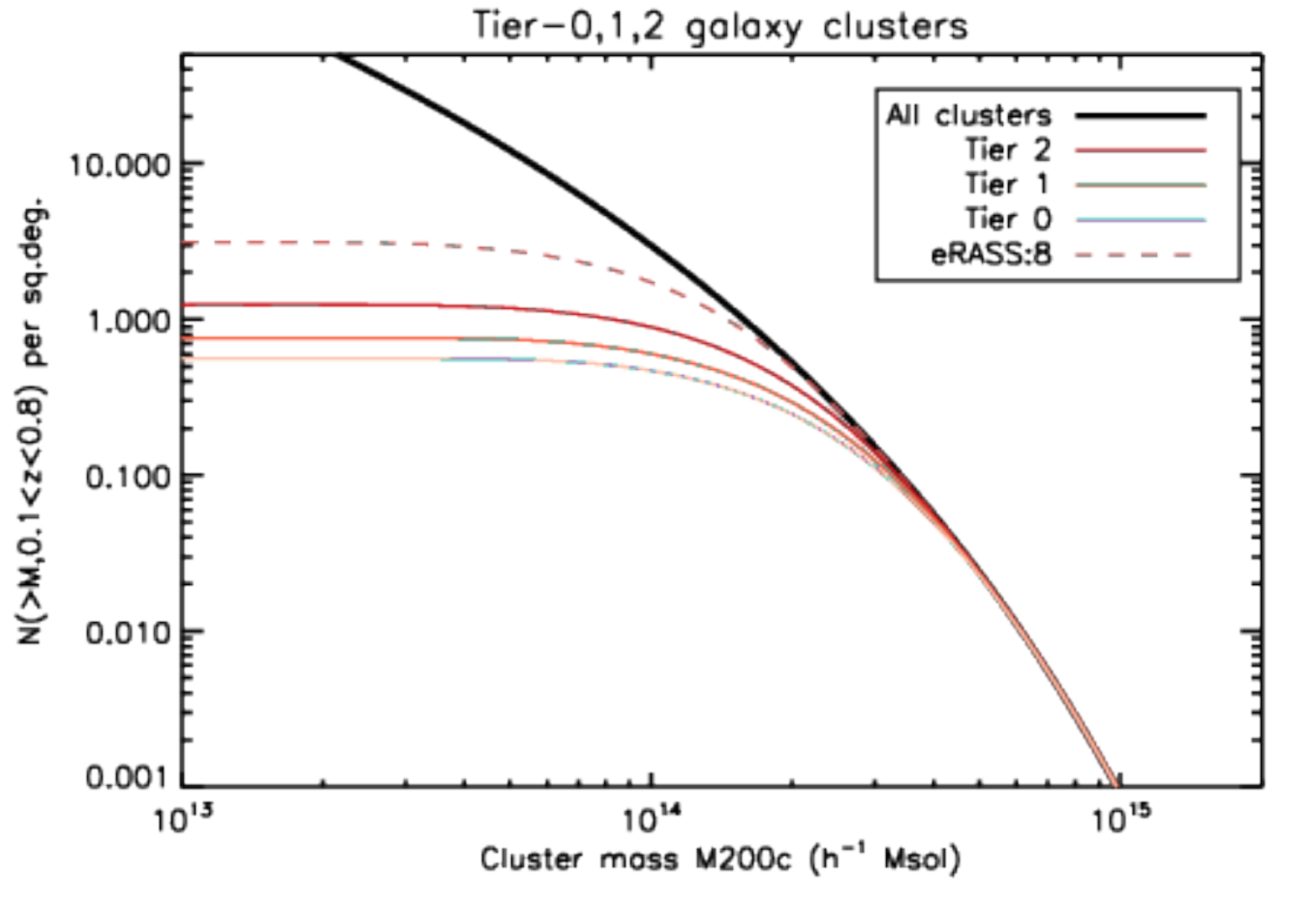}
		\caption{Cumulative mass distribution of galaxy clusters (per unit area) in the SPIDERS survey, split into three different tiers. Tier~0 stands for CODEX clusters only. Also shown is the cumulative mass function of all haloes with mass above $10^{12.5} h^{-1}$~M$_{\odot}$ and the mass distribution of \emph{eROSITA} clusters at the end of the 4-year survey (eRASS:8, not part of SPIDERS). The flux limits assumed are taken from \citet{merloni2012}.}
	\label{fig:dndm_tiers} 
\end{figure}	
	
	\subsection[]{SPIDERS Tier 0: CODEX and XCLASS}
	
Prior to the delivery of first cluster catalogues from \emph{eROSITA}, SPIDERS ensures follow-up of galaxy clusters discovered in the RASS (\emph{ROSAT} All-Sky Survey) and in XMM archival data. The two relevant cluster samples are the CODEX (Finoguenov et al., in prep.) and the XCLASS-RedMapper catalogue \citep{clerc2012b, sadibekova2014} respectively. Both are based on X-ray detections of galaxy clusters, yet they differ in their characteristics and their construction. Since they conveniently encompass the range of X-ray properties expected from \emph{eROSITA} clusters (see below), they show a particular interest in view of preparing the \emph{eROSITA} survey.

We provide details on their construction in the following paragraphs, and Table~\ref{table:codex_vs_xclass} summarizes the main characteristics of both samples. Note in particular that the same red-sequence finder was run for both samples.

\begin{figure}
	\includegraphics[width=84mm]{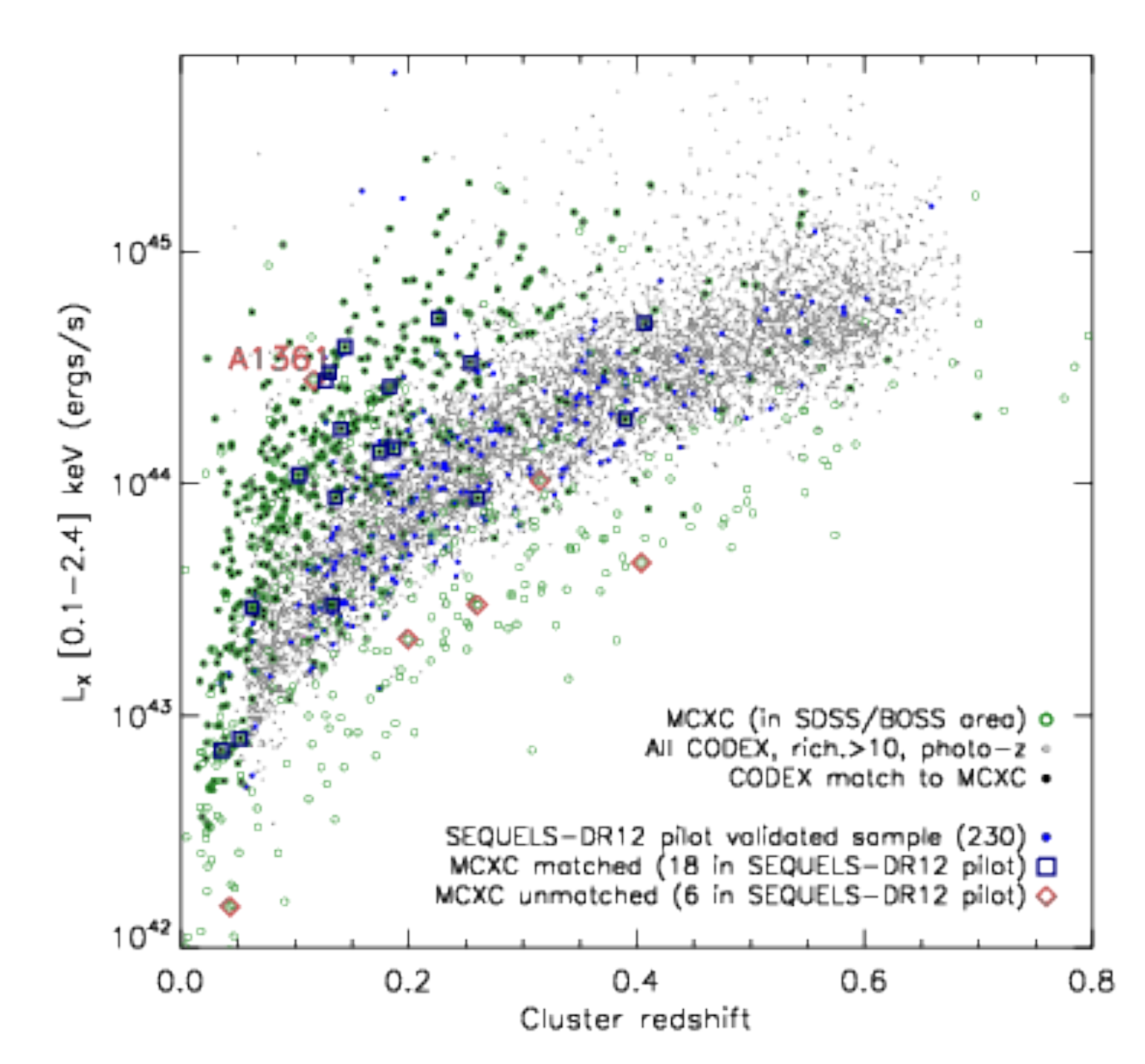}
		\caption{The distribution of SPIDERS cluster candidates in the redshift-X-ray luminosity plane. Grey points represent CODEX clusters with a richness above 10, i.e.~the main pool of targets in SPIDERS. When matched in position to a MCXC cluster, their location is taken from the MCXC meta-catalogue \citep{piffaretti2011} and marked by a darker point. Otherwise, their redshift corresponds to $z_{\lambda}$, as estimated from SDSS photometry, with typical uncertainty $\Delta_z/(1+z) \sim 0.01 - 0.02$; their rest-frame [0.1-2.4]~keV luminosity derives from the ROSAT flux using $z_{\lambda}$.
		Blue points are the 230 CODEX clusters confirmed in the SEQUELS-DR12 demonstration sample (Sect.~\ref{sect:description_sequels}) with a spectroscopic redshift (typical $\Delta_z/(1+z) \sim 0.001$). 24 MCXC clusters lie within the SPIDERS pilot footprint (Fig.~\ref{fig:pilotareasamples}). ABELL~1361 is within a masked area of the CODEX survey, hence the absence of a match despite its remarkable X-ray brightness.}
	\label{fig:lxzmcxc} 
\end{figure}

\begin{figure}
	\includegraphics[width=84mm]{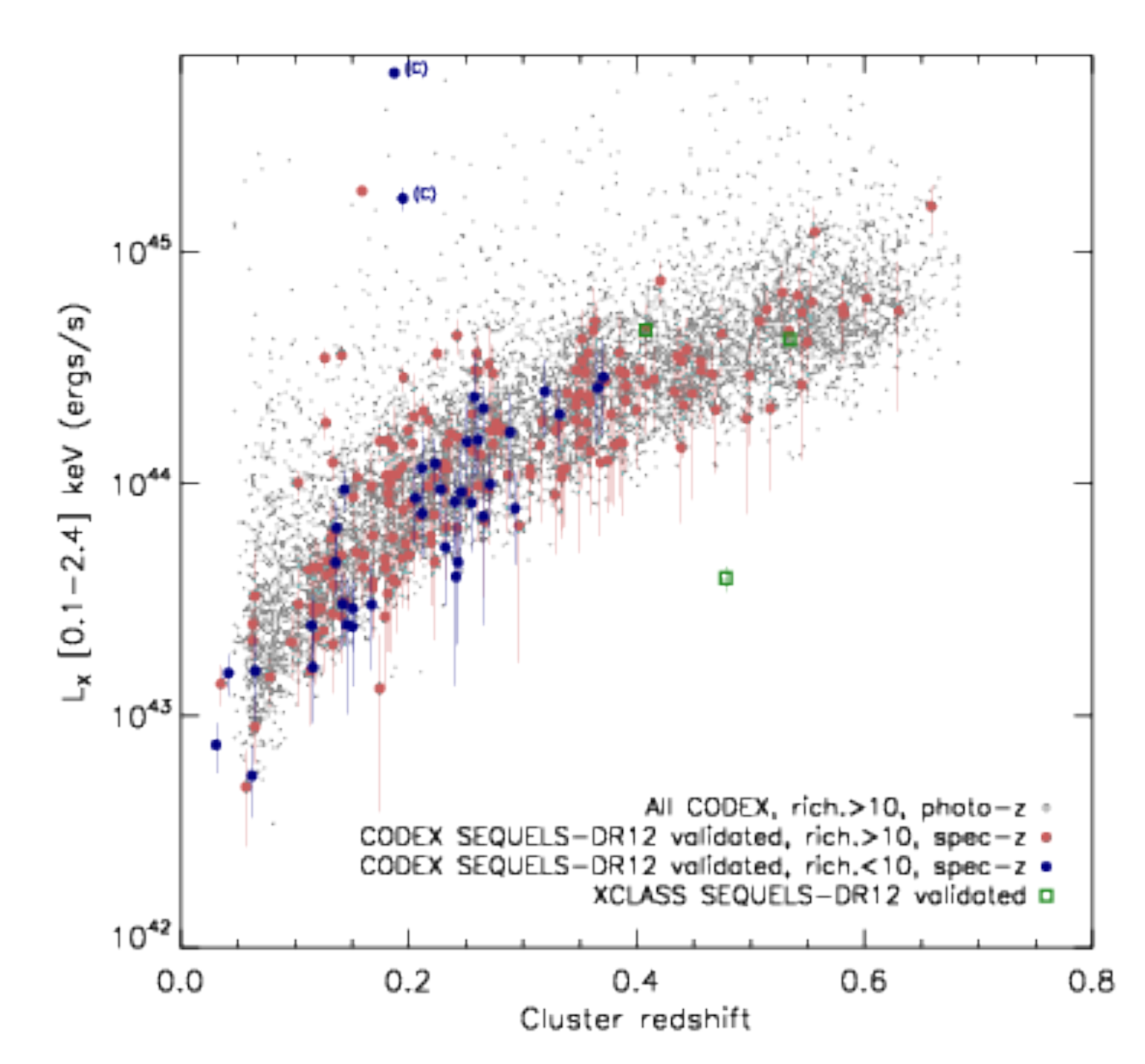}
		\caption{The distribution of SPIDERS confirmed clusters in the redshift-X-ray luminosity plane. Similarly as in Fig.~\ref{fig:lxzmcxc}, grey dots represent the main pool of targets in SPIDERS.
		Red and blue points are the 230 CODEX clusters confirmed in the SEQUELS-DR12 demonstration sample.
		The three XCLASS clusters validated as part of the demonstration sample are displayed as green squares.
		Two low-richness CODEX clusters labelled '(C)' are suffering from contamination by a point-source in RASS data, not necessarily linked to the system, and artificially boosting the X-ray luminosity measurement.}
	\label{fig:lxzdist} 
\end{figure}

	\subsubsection[]{The CODEX subsample}
CODEX (COnstrain Dark Energy with X-ray clusters, Finoguenov et al., in prep.) is an extensive search for galaxy clusters in ROSAT data, based on the association of RASS photon overdensities to red-sequence galaxies identified in SDSS. It covers the entire SPIDERS/eBOSS footprint and these detections are expected to show as the brightest, best-characterized, cluster sources in future \emph{eROSITA} data.
The current study provides the only spectroscopically complete CODEX catalogue down to low richness values. As such, this paper is the first in a series of CODEX catalogue papers. The sample construction is fully detailed in Finoguenov et al. (in prep.)~; we briefly summarize here the steps leading to the list of cluster candidates.

\begin{figure}
	\includegraphics[width=84mm]{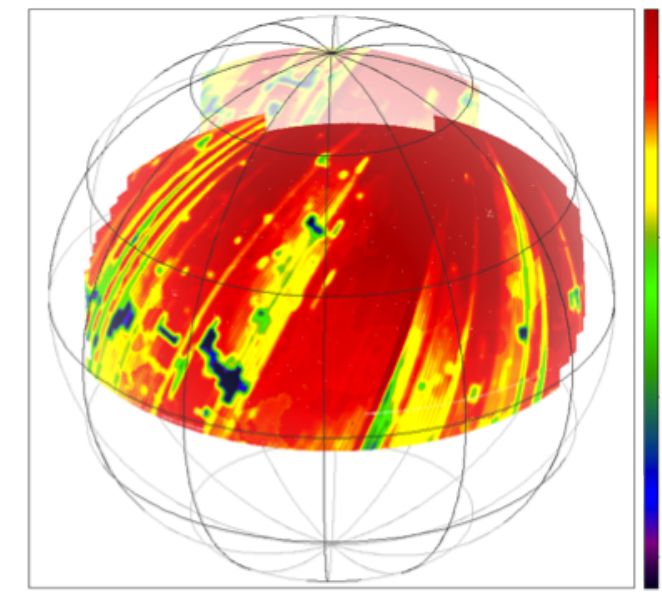}
		\caption{Rosat All-Sky Survey (RASS) sensitivity in the CODEX footprint (North galactic cap). Colour bar indicates the limiting flux in the [0.5-2]~keV band, from red ($10^{-13}$~ergs\,s\,cm$^{-2}$) to black ($8\times10^{-13}$~ergs\,s\,cm$^{-2}$).}
	\label{fig:rass_sensitivity} 
\end{figure}

As a first step, RASS data is searched for faint sources using a wavelet-based detection algorithm. The detection threshold is set to 4-$\sigma$. Sensitivity maps (as in Fig.~\ref{fig:rass_sensitivity}) are created as by-products and help in assessing the completeness of the sample. On average, the 90\% completeness level is achieved for a source delivering 8 X-ray counts, while the 10\% completeness level is reached for sources delivering 4 counts. Given these sensitivity estimates, the number of spurious sources is estimated between 500 and 1000 across the entire CODEX area and the number of X-ray AGN amounts to around 20\,000.
The RedMapper algorithm \citep{rykoff2014} looks in SDSS imaging data (Data Release 8) for galaxies with similar colours around each faint RASS source, i.e.~for a red-sequence formed by passive galaxies at the same redshift. This provides in turn an estimate for the photometric redshift of the cluster (based on the colours of the galaxies) and an optimized richness estimator. The counterpart having the highest richness is listed for each RASS X-ray source.
Given the uncertain position of RASS detections the red-sequence algorithm is then run to optimally find the cluster center. The constraint on the centre position is relaxed, to be within $3\arcmin$ from the X-ray position\footnote{The mean and 95th percentile of the RASS faint point-source 1-$\sigma$ positional uncertainty are $\sim 20$ and $\sim 35$ arcsec respectively.}. The newly found red-sequence is used to provide a fresh estimate for the cluster photometric redshift and richness (optical or "OPT" quantities: $z_{\lambda, {\rm OPT}}, \lambda_{\rm OPT}$, etc.)
In the final step, X-ray properties based on the RASS count-rate and the RedMapper redshift are calculated in optimized apertures (imposing a minimal signal-to-noise threshold of 1.6), assuming a model for the X-ray spectral emissivity. Among them stand the aperture-corrected cluster flux $f_X$ and $[0.1-2.4]$~keV luminosities $L_X$.

The average number density of CODEX sources over the BOSS imaging footprint is 0.8~deg$^{-2}$ (for candidates with richness\footnote{By 'richness' we will refer to the RedMapper richness estimator \citep{rykoff2014}. It correlates with the total cluster mass and equals the sum of the membership probabilities $p_{\rm mem}$ of galaxies within a given system.} $\lambda_{OPT} \geq 10$). Adding those with lower richness brings this number up to 1.0~deg$^{-2}$. However, due to spatial fluctuations in the RASS depth, these numbers vary as a function of sky position.

	\subsubsection[]{The XCLASS-RedMapper subsample}
XCLASS \citep[XMM CLuster Archive Super Survey,][]{clerc2012b} is a search for galaxy clusters detected in the XMM-Newton archive, based on a robust cluster detection algorithm, developed in the context of the XMM-LSS \citep[e.g.][]{pacaud2006,clerc2014} and XMM-XXL \citep{pierre2016} surveys.  Extensive simulations of XMM observations, including realistic instrumental effects and astrophysical source populations, support the construction of a pure sample of extended objects in $[0.5-2]$~keV XMM images (the "C1" selection, \citealt{pacaud2006}). Visual screening removes nearby galaxies and detector artefacts, leading to the final catalogue of XCLASS galaxy cluster candidates. The L4SDB\footnote{{http://xmm-lss.in2p3.fr:8080/l4sdb/}} database stores validated detections, along with other useful information related to the X-ray sources (redshifts, flux measurements, etc.) The XCLASS surveyed area amounts to $\sim 90$~deg$^2$ but due to its very nature, it is scattered across the extragalactic sky ($|b_{\rm galactic}| > 20 \deg$). All analyzed XMM observations were deliberately shrunk to 10~ks depths (exactly) so as to provide a survey as uniform as possible in sensitivity.

\citet{sadibekova2014} performed the correlation of XCLASS C1 sources with the RedMapper optical cluster catalogue in the regions where the two surveys overlap. A major difference with the CODEX sample consists in very reliable cluster X-ray positions (the positional uncertainty amounts to a few arcsec rms), and the secure extended nature of the X-ray detections.
Similarly to CODEX, the RedMapper algorithm provides an estimate for the photometric redshift and the optical richness ($\lambda_{XC}$) of the clusters. The SPIDERS sample contains 238 XCLASS clusters securely matched to a RedMapper candidate: i.e. $\lambda_{XC}>20$ and a correlation radius $r_{\rm corr} \leq 3 \arcmin$, or $(5<\lambda_{XC}<20)$ and $r_{\rm corr} \leq 1 \arcmin$.
We further added a group of 40 less securely matched sources, having $(5<\lambda_{XC}<20)$ and $1\arcmin \leq r_{\rm corr} \leq 3 \arcmin$.

The total number of XCLASS-RedMapper sources across the full SDSS imaging footprint amounts to 278, 84 of them are in common with the CODEX subsample described earlier. Since they are irregularly distributed on sky, their sky density is quoted over the common overlap area between XMM observations and the imaging footprint and amounts to 3-4 deg$^{-2}$.

\begin{table*}
	\centering
\caption{\label{table:codex_vs_xclass} Characteristics of the two samples of X-ray clusters followed-up by SPIDERS prior to the launch of eROSITA. Notes: $^{(a)}$: calculated over the area overlapping XMM-Newton observations analyzed in X-CLASS. $^{(b)}$ only for SEQUELS pilot area, see Sect.~\ref{sect:sequels}.}
		\begin{tabular}{@{}lcc@{}}
\hline
				&	CODEX		&	XCLASS-RedMapper			\\
\hline
Number of clusters in SDSS DR8 footprint & 10 415 & 278 \\
Sky distribution & Full SDSS area & Spatially scattered \\
Average candidate density ($\deg^{-2}$)	& 0.8 & 3-4$^{(a)}$  \\
Maximal redshift & $\sim 0.6$ & $\sim 0.6$ \\
Minimal richness & 10 (3$^{(b)}$) & 5 \\
\hline
X-ray data origin & RASS faint sources & XMM-Newton archival data \\
X-ray selection & 4-$\sigma$ above background & C1 selection (extended sources) \\
Limiting flux in X-rays (0.5--2~keV, units ergs/s/cm$^2$) & $\sim 10^{-13}$ & $\sim 10^{-14}$ \\
X-ray positional accuracy & $\sim 3\arcmin$ & $\lesssim 10\arcsec$ \\
X-ray spatial resolution & $\sim 100 \arcsec$ & $\sim 10-20 \arcsec$ \\
X-ray energy resolution ($\Delta E$ @ 1 keV) & $\sim 450$~eV & $\sim 100$~eV \\
\hline
Red-sequence finder & redMaPPer v.5.2 & redMaPPer v.5.2 \\
Optical search & Around each X-ray source & Independent from X-ray sources \\
Optical/X-ray association & Richness cut vs. chance identification & Angular distance criterion + visual checks \\
\hline
		\end{tabular}
\end{table*}

\begin{figure*}
	\begin{tabular}{cc}
	\includegraphics[width=84mm]{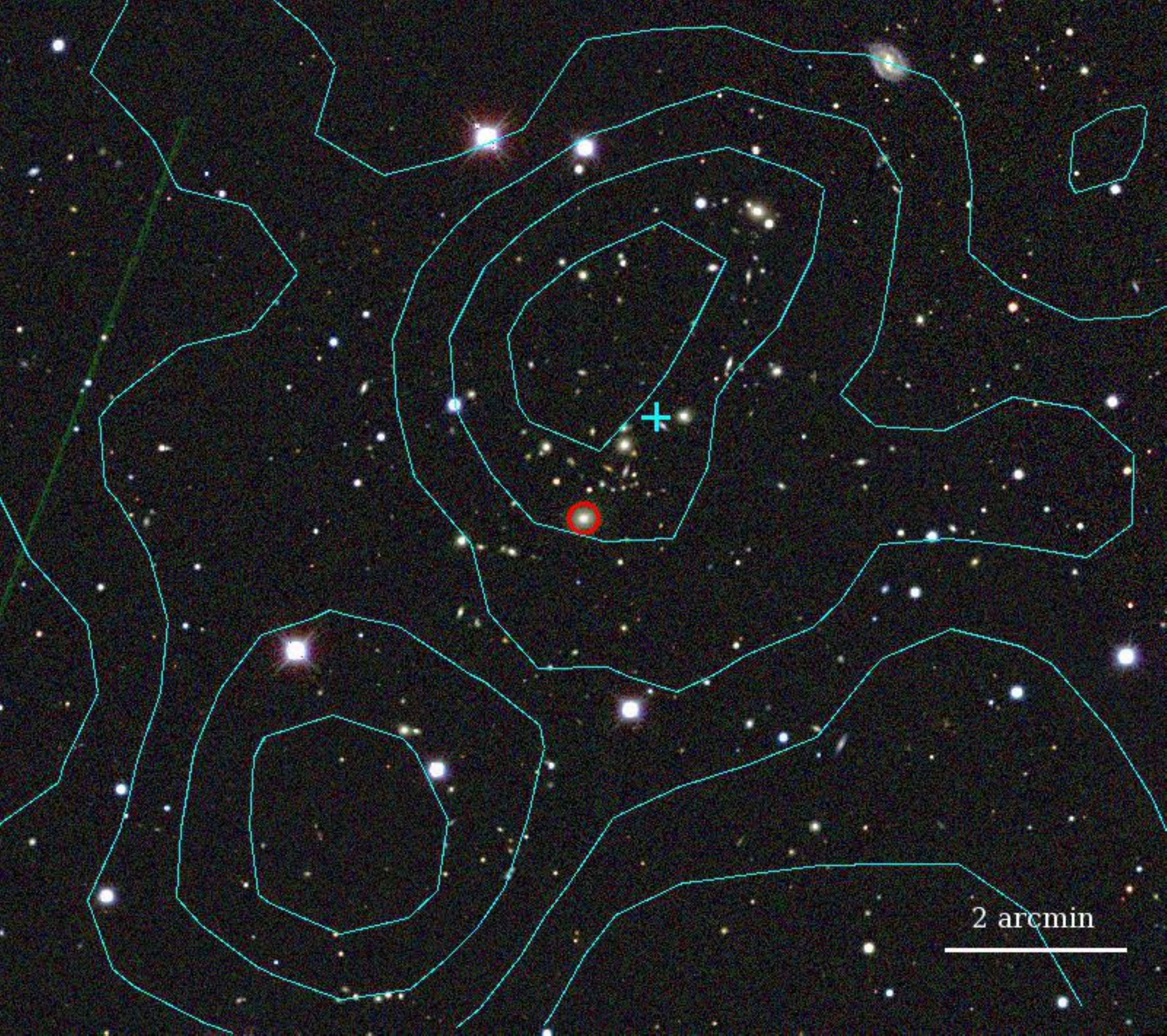} & \includegraphics[width=84mm]{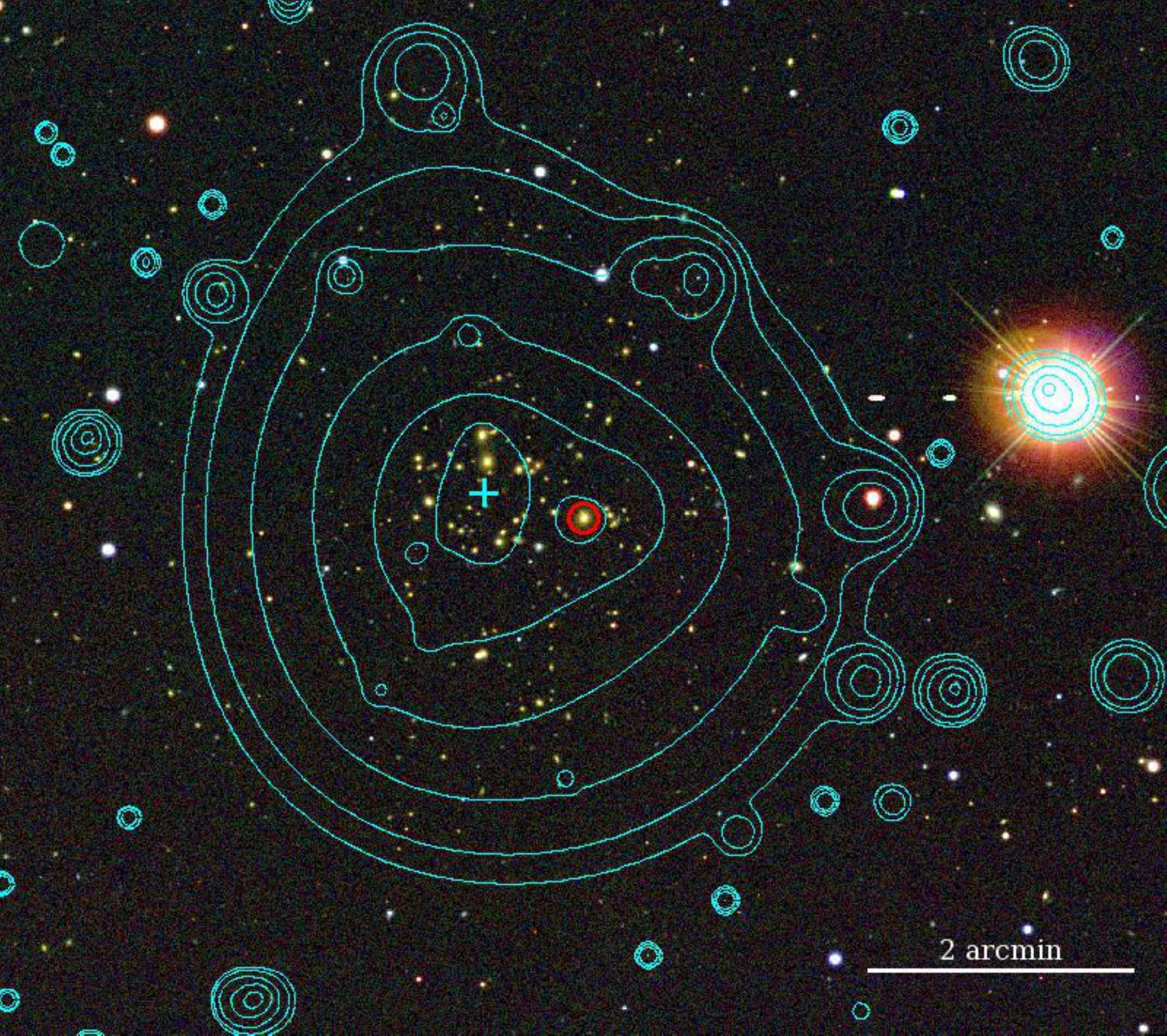} \\
	\end{tabular}
	\caption{SDSS-$gri$ composite images of two SPIDERS clusters with X-ray contours overlaid. {\it Left:} CODEX cluster (Id: 2\_2338, known as ABELL 661) at R.A.\,=\,8h27m15.5s, Dec\,=\,+$53\degr$8m53s and $z=0.121$, the contours correspond to ROSAT-All-Sky Survey [0.1-2.4]~keV smoothed image. \emph{Right:} bright XCLASS cluster (Id: XC~0062 ; RX\,J0256.5+0006 in \citealt{romer2000}) at R.A.\,=\,2h56m30.8s, Dec\,=\,+$0\degr$6m3s and $z_{\rm phot}=0.37$ (the SPIDERS-determined redshift of XC~0062 will be published at completion of the survey). The contours correpond to XMM-Newton [0.5-2]~keV smoothed image. In both panels, North is up and East is left, the cyan cross indicates the position of the original X-ray detection, the red circle the position of the optical centre. Point-like sources are easily distinguishable in XMM data (small-scale overdensities in the right panel). }
	\label{fig:example_codex_xclass} 
\end{figure*}

	\subsection[]{\emph{eROSITA} survey: eRASS samples}

\begin{figure}
	\includegraphics[width=84mm]{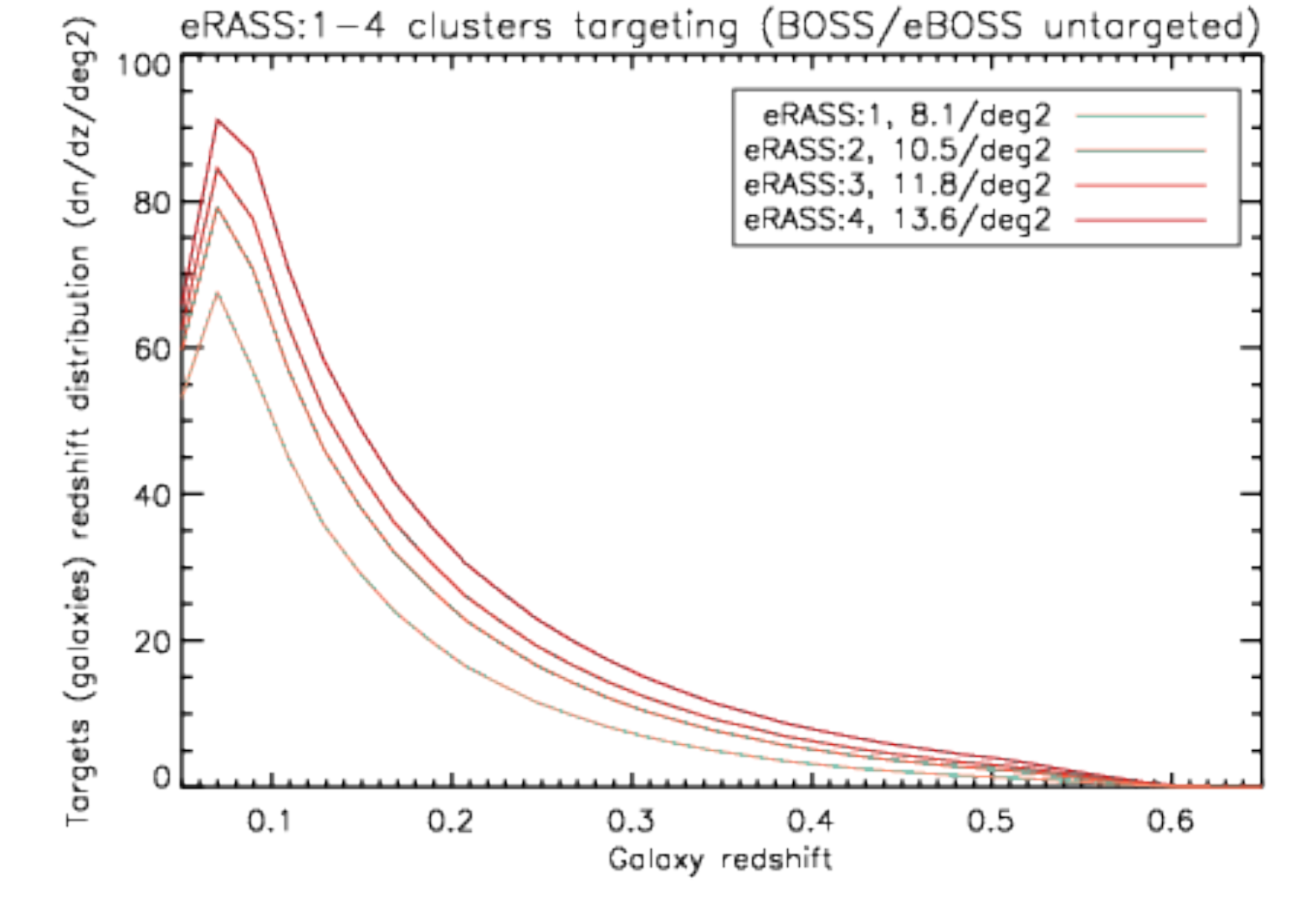}
		\caption{Redshift distribution of all targetable galaxies in the \emph{eROSITA} era of the SPIDERS survey. It includes a model for the cluster number density combined to galaxy luminosity functions and roughly accounts for the lost of flux in fibers and their maximal spacing. These numbers are indicative and need refinement in the course of the Tier-0 phase of the survey.}
	\label{fig:dndz_galaxies_erass} 
\end{figure}

Although this paper focuses mainly on the targeting of Tier~0 samples, namely CODEX and XCLASS, we forecast our target budget for the future Tiers 1 \& 2 in SPIDERS. These forecasts are based on pre-launch assumptions as for the amount and nature of \emph{eROSITA} clusters \citep{merloni2012}. A simple model, subject to the current uncertainty concerning the in-flight performances of the instrument and the actual physics of the population it will uncover, helps in deriving rough numbers and adjusting the targeting strategy.
We modeled the galaxy cluster mass distribution using \citet{tinker2008} halo mass function and converted masses ($M_{200c}$) to X-ray temperatures and luminosities using  scaling relations. The \emph{eROSITA} selection is modeled with a lower cut in soft-band flux, representative of the selection relevant to each tier ($f_{\rm lim} = 1.2$ and $0.8 \times 10^{-13}$~ergs\,s$^{-1}$\,cm$^{-2}$ for Tier 1 and 2 respectively). Integrating the resulting filtered mass distribution provided the curves shown in Fig.~\ref{fig:dndm_tiers}.
The galaxy population within clusters was simulated by means of galaxy luminosity functions parametrized as a function of cluster mass and redshift \citep{popesso2005,hansen2009}. We folded a spectral energy distribution template representative of passive galaxies \citep{maraston2009} into the SDSS filter set. Flux losses due to the finite $2\arcsec$ fiber aperture were accounted for by assuming a size-magnitude relation \citep{bernardi2007} and a typical $1.4\arcsec$ seeing (see details in \citealt{zhang2016}). We then applied a photometric selection $17 < i(2 \arcsec) < 21.2$ representative of the SPIDERS target selection (see Sect.~\ref{sect:targeting}) and excluded galaxies whose photometric properties correspond to BOSS galaxy targets \citep[the LOWZ and CMASS selections~;][]{bolton2012}. Finally, a cluster radius-dependent sampling factor was set to account for fiber collisions. The resulting redshift distribution of targetable galaxies not already targeted in BOSS is shown in Fig.~\ref{fig:dndz_galaxies_erass}, along with the densities of targets for each layer of the \emph{eROSITA} survey. These numbers are indicative and are refined within the Tier 0 phase of SPIDERS.

Based on those calculations, SPIDERS (in its Tier 1 and 2 phases) will confirm 90\% of $z \lesssim 0.6$ \emph{eROSITA} clusters by obtaining (at least) 3 spectroscopic redshifts per system (including those known from previous SDSS observations.) SPIDERS will also raise the number of spectroscopic members per cluster virial radius to 10 for 50\% of the $z<0.5$ clusters and to 20 for 20\% of the $z<0.5$ clusters.

	\subsection[]{The SEQUELS pilot program}
	\label{sect:description_sequels}

SEQUELS \citep[The Sloan Extended Quasar, ELG, and LRG Survey,][their App.~A.3]{alam2015} is an ancillary program part of BOSS (SDSS Data Release 12) and served as a pilot survey for the eBOSS, SPIDERS and TDSS programs in SDSS-IV \citep{dawson2016}.
Its initial footprint consists in the rectangle $(120 \leq RA \leq 210)$ and $(45 \leq DEC \leq 60)$. Only 300~deg$^2$ of this area were observed (corresponding to 66 plates) as part of Data Release 12 \citep[DR12,][and Fig.~\ref{fig:layout}]{alam2015}.
As a preparation for the SDSS-IV SPIDERS cluster follow-up program, SEQUELS contains a number of targets assigned to SPIDERS clusters. While the parent cluster samples are the same as for SPIDERS Tier~0 (i.e.~CODEX and XCLASS), the target selection slightly differs and it is in general broader in SEQUELS (see Sect.~\ref{sect:targeting}).
Throughout this work, we illustrate our envisaged analysis procedures with results extracted from the pilot SPIDERS program in SEQUELS DR12.
Note that 51 SEQUELS plates (about $166 \deg^2$) are observed in the course of the eBOSS survey (post-DR12) and therefore the targeting strategy for those slightly differs from the main SPIDERS survey. These objects are not considered in the following 'pilot sample'.


\section[]{Targeting strategy}
	\label{sect:targeting}

This section details the steps followed in preparing the target lists in the first phase of SPIDERS (Tier 0). The main difference with respect to conventional multi-object spectroscopic observations of galaxy clusters, consists in an ensemble treatment of the entire pool of targets. Because the exact set of targets is only known after the eBOSS tiling algorithm has run \citep{dawson2016} and accommodated for the various target classes within eBOSS, we worked out a scheme for assigning priorities to potential targets, aimed at optimizing the primary science goal, namely the number of spectroscopically confirmed clusters.

	\subsection[]{Target selection and prioritization\label{sect:targetsel}}
	
		\subsubsection[]{The CODEX and XCLASS red-sequences}

To each galaxy cluster candidate we attach a list of potential member galaxies detected over the SDSS imaging data, which form the likely red sequence of a cluster in the SDSS passbands.
Specifically, the redMaPPer algorithm assigns to each galaxy near a cluster a probability $p_{\rm mem} \in [0,1]$ \citep{rykoff2014} that it actually is a cluster member, based on its magnitude, colours and position relative to the cluster centre. This allows to rank galaxies by membership probability within each cluster, down to $p_{\rm mem}=0.05$, a ranking that we convert in terms of targeting priority, as described in the following for CODEX and XCLASS targets.

		\subsubsection[]{CODEX clusters}

The entire CODEX red-sequence member catalogue comprises 312564 objects over the entire BOSS footprint. Among them, 3797 formally belong to two or more parent clusters: however, since the membership catalogue includes objects down to membership probability $p_{\rm mem}=5\%$, this amount is not necessarily indicative of projection effects: it actually includes galaxies with low probability in one of their parent clusters, as well as galaxies that belong to several clusters on valid physical grounds (mergers).

In order to maximize the redshift determination efficiency of the targets  \citep[e.g.][]{bolton2012}, only red-sequence candidates with $17.0 < {\tt FIBER2MAG\_I} < 21.2$ are considered for targeting. The algorithm starts with the richest cluster in the sample (as defined by $\lambda_{OPT}$) and iteratively proceeds by decreasing richness. It assigns to each member an integer {\tt TARGETSELECTED} indicating its rank in the red-sequence. Members with a spectroscopic redshift determined from past SDSS/BOSS observations, with values satisfying {\tt SPECPRIMARY == 1} and {\tt ZWARNING == 0} were identified and removed from the initial list. Targets already assigned a rank within a higher-richness cluster keep the {\tt TARGETSELECTED} flag they were assigned previously. 

\begin{figure}
	\includegraphics[width=84mm]{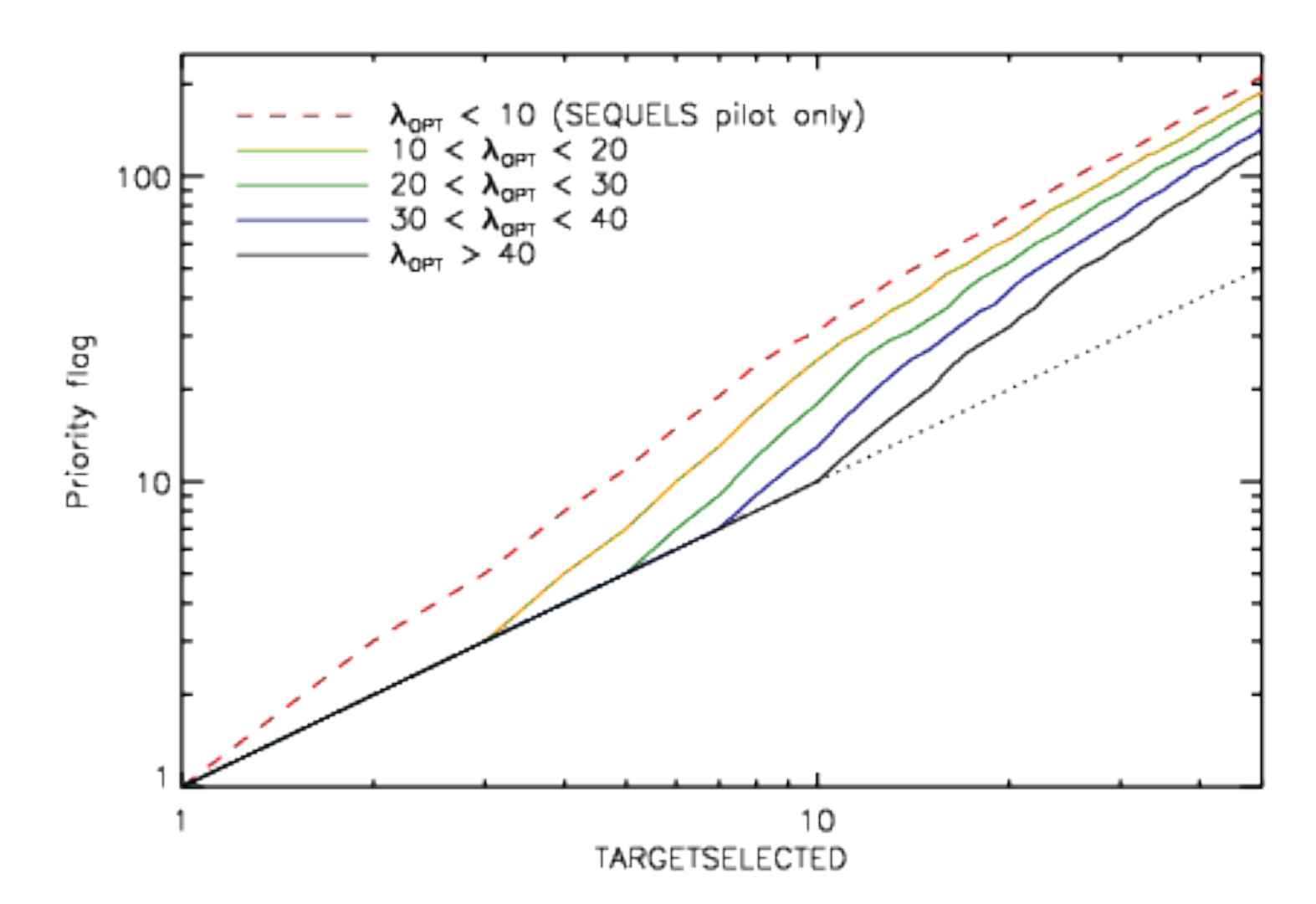}
		\caption{Scheme applied to assign a priority flag to red-sequence members in a cluster depending on its richness $\lambda_{OPT}$ (dotted line is the one-to-one relation). Members in rich clusters (lower curve, plain black) are assigned lower priority flags, ensuring a highest fraction of targeted objects. The red-dashed line stands for very poor clusters ($\lambda_{OPT} < 10$). These candidates are not included in SPIDERS but considered in the SEQUELS Selection (Sect.~\ref{sect:targeting}).}
	\label{fig:prio_rank} 
\end{figure}

The priority flag for each target is computed based on a combination of {\tt TARGETSELECTED} and the cluster richness $\lambda_{OPT}$. Fig.~\ref{fig:prio_rank} displays the relation between the priority flag and the galaxy rank in the red-sequence, as a function of cluster richness. A low priority flag indicates high targeting priority. We ensured the 3 highest-probability objects in the red-sequence are prioritized regardless of the cluster richness, in order to maximize the number of confirmed clusters in the sample.
This simple scheme ensures higher prioritization of rich clusters and galaxies relevant to cluster confirmation. A last step consists in applying a hard cut to the priority flag at the value of 80, except in the {\tt eboss3} chunk where this threshold is set at 33 (see App.~\ref{app:eboss3}). This change is motivated by the highest density of RASS-faint sources in this area, which lies close to the deep polar region of the X-ray survey.

		\subsubsection[]{XCLASS-RedMapper clusters}

XCLASS-RedMapper clusters are targeted in a way very similar to CODEX clusters, with two exceptions: no cluster richness-based selection is applied; and the conversion from {\tt TARGETSELECTED} to the actual priority flag is computed regardless of the richness and follows the lowest (plain black) curve in Fig.~\ref{fig:prio_rank}, i.e.~all XCLASS clusters are treated equivalently to rich CODEX clusters.
These two differences stem from the secure galaxy cluster nature of these objects (bona-fide extended sources in X-ray, compare both panels of Fig.~\ref{fig:example_codex_xclass}) and ensure higher internal prioritization of the overall less numerous XCLASS targets in the SPIDERS survey.

	\subsection[]{Tiling forecasts\label{sect:tiling}}

The pool of targets along with the priority flag is submitted to the eBOSS tiling algorithm. Given their relative sparsity (less than $10 \deg^{-2}$), and because the high-level requirement for SPIDERS is a high completeness level in spectroscopic confirmation of clusters, SPIDERS cluster targets are assigned first among other eBOSS targets.

\begin{figure}
	\includegraphics[width=84mm]{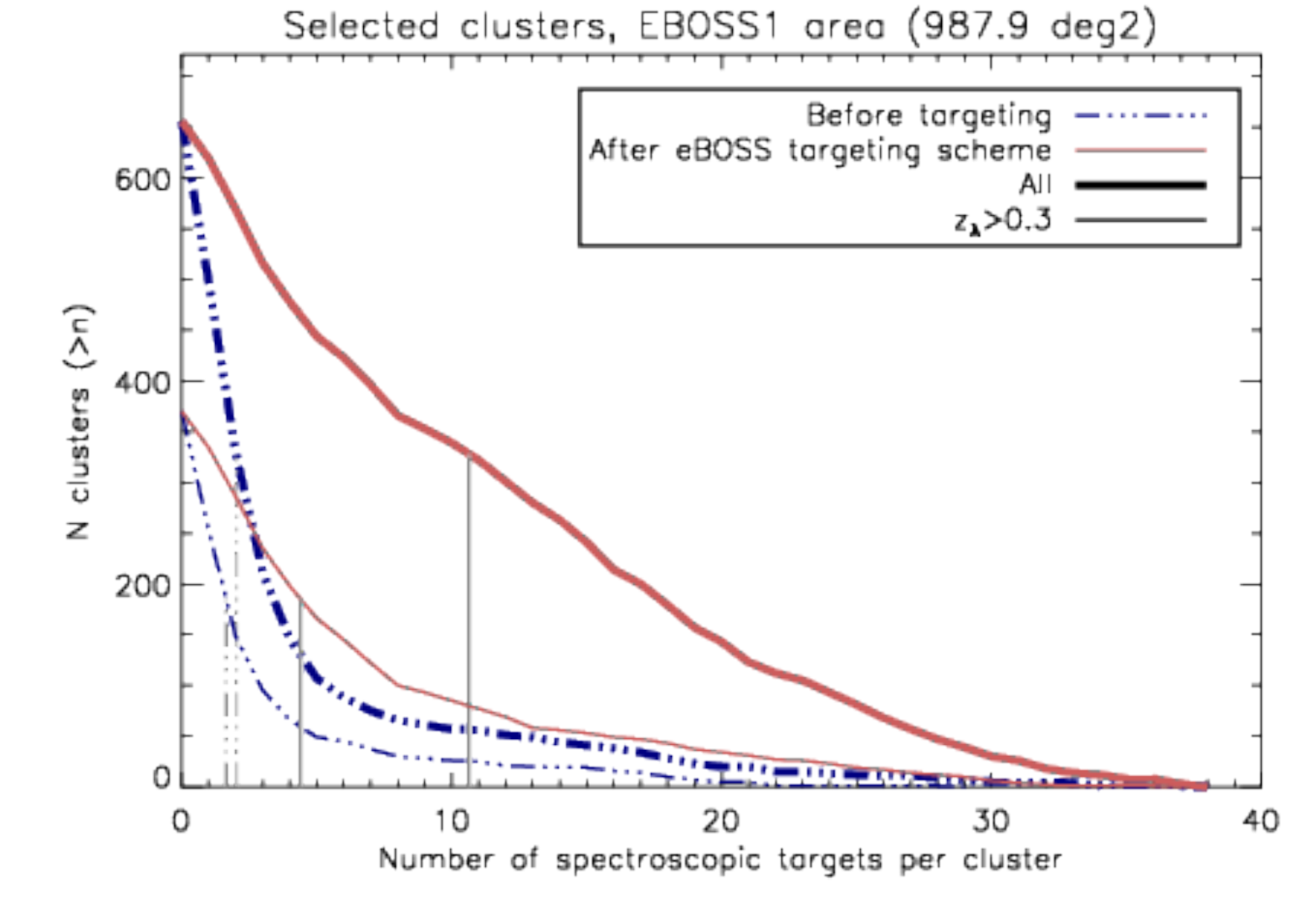}
		\caption{Number of clusters that will have more than $n$ spectroscopic members in their red-sequence after full observation of the {\tt eboss1} chunk. The dot-dashed curve shows the situation prior to the start of SPIDERS (with a median of 2 spectroscopic redshifts per cluster), while the solid red curve is a forecast based on eBOSS tiling results (median of 10 spectroscopic redshifts per cluster, including possible interlopers). These median numbers change from 2 to 4 when considering only high-redshift candidates (thin lines, as compared to thick lines representing objects at all redshifts).}
	\label{fig:eboss1_tilingforecasts} 
\end{figure}

\begin{figure}
	\includegraphics[width=84mm]{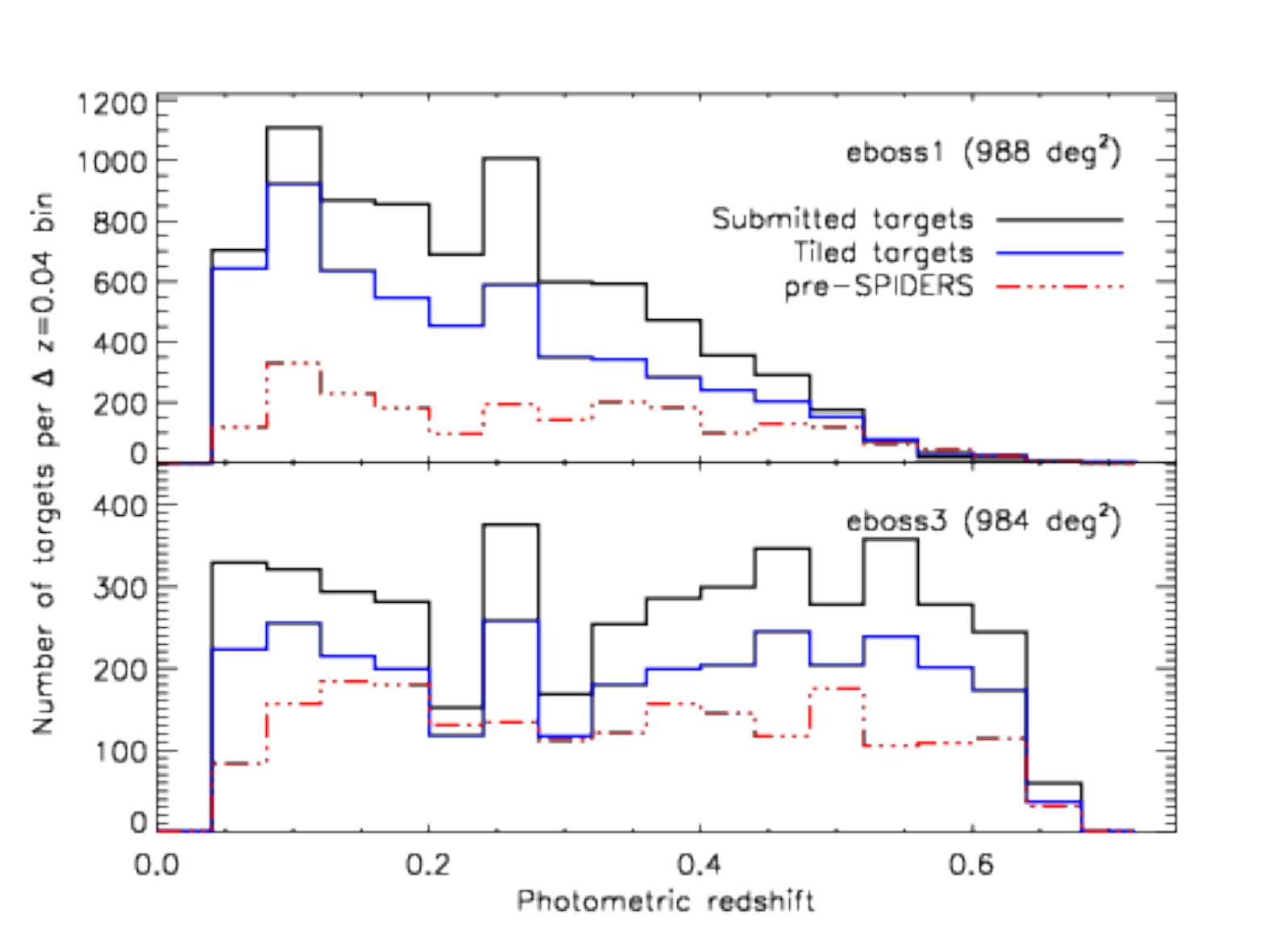}
		\caption{Redshift distribution of SPIDERS\_RASS\_CLUS target in all clusters in chunk {\tt eboss1} (656 systems) and chunk {\tt eboss3} (943 systems). Targets submitted to the fiber allocation algorithm are shown with a black line, tiled targets with a blue line. The redshift distribution of red-sequence members with a redshift prior to SPIDERS observations is shown as a red dashed line.}
	\label{fig:eboss13_targetdistributionredshift} 
\end{figure}

Fig.~\ref{fig:eboss1_tilingforecasts} shows the expected number of spectroscopic redshifts of red-sequence galaxies per CODEX cluster in the {\tt eboss1} chunk, based on the plate tiling. Fig.~\ref{fig:eboss3_tilingforecasts} shows the equivalent for the {\tt eboss3} chunk, which is obtained with a lower priority threshold due to the increased depth of RASS in this area of sky (App.~\ref{app:eboss3}).
Within the {\tt eboss1} area, the number of spectroscopic redshifts in the red-sequence increases from 2 (median) prior to observations to about 10 (median) after observation, while in the {\tt eboss3} area this number amounts to 8.
Fig.~\ref{fig:eboss13_targetdistributionredshift} shows the photometric redshift distribution of SPIDERS\_RASS\_CLUS targets in both chunks, i.e. the photometric redshift of the galaxy cluster they are attached to. Noticeably, SPIDERS will increase the fraction of cluster member with redshift up to $z\sim 0.6$. The deeper X-ray data in chunk {\tt eboss3} enables the use of spectroscopic targets for targeting more distant clusters.

\begin{figure}
	\includegraphics[width=84mm]{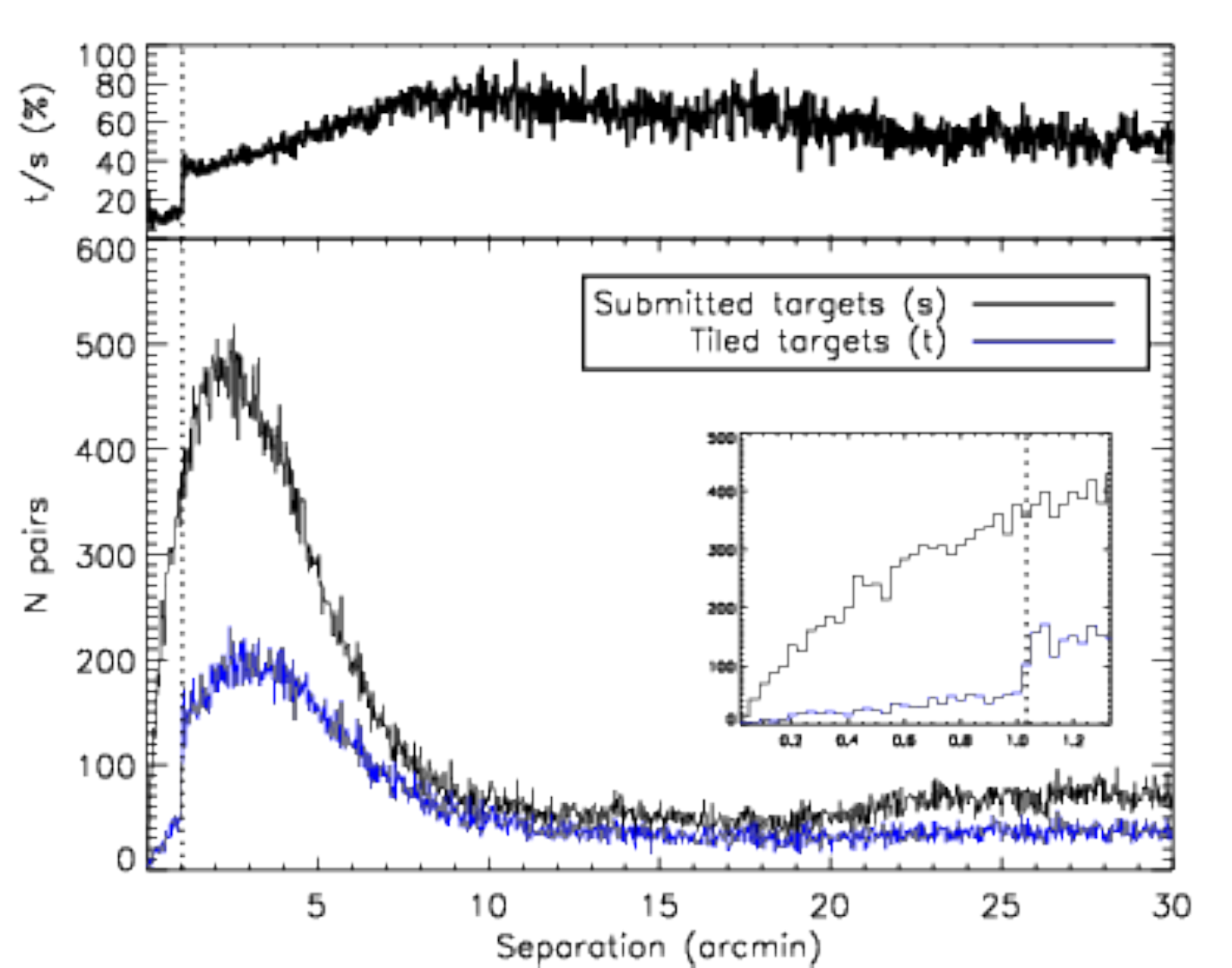}
		\caption{Histograms of pairwise separations ($2\arcsec$ bins) between SPIDERS submitted targets (black curve) and between tiled targets in SPIDERS clusters (lower curve), the top panel shows the target allocation completeness. The inset shows a zoom over the first $\sim 80\arcsec$ and the dashed vertical line indicates the eBOSS fiber collision radius ($62\arcsec$).}
	\label{fig:eboss1_targetdistance} 
\end{figure}

\begin{figure*}
\begin{tabular}{c}
	\includegraphics[width=\linewidth]{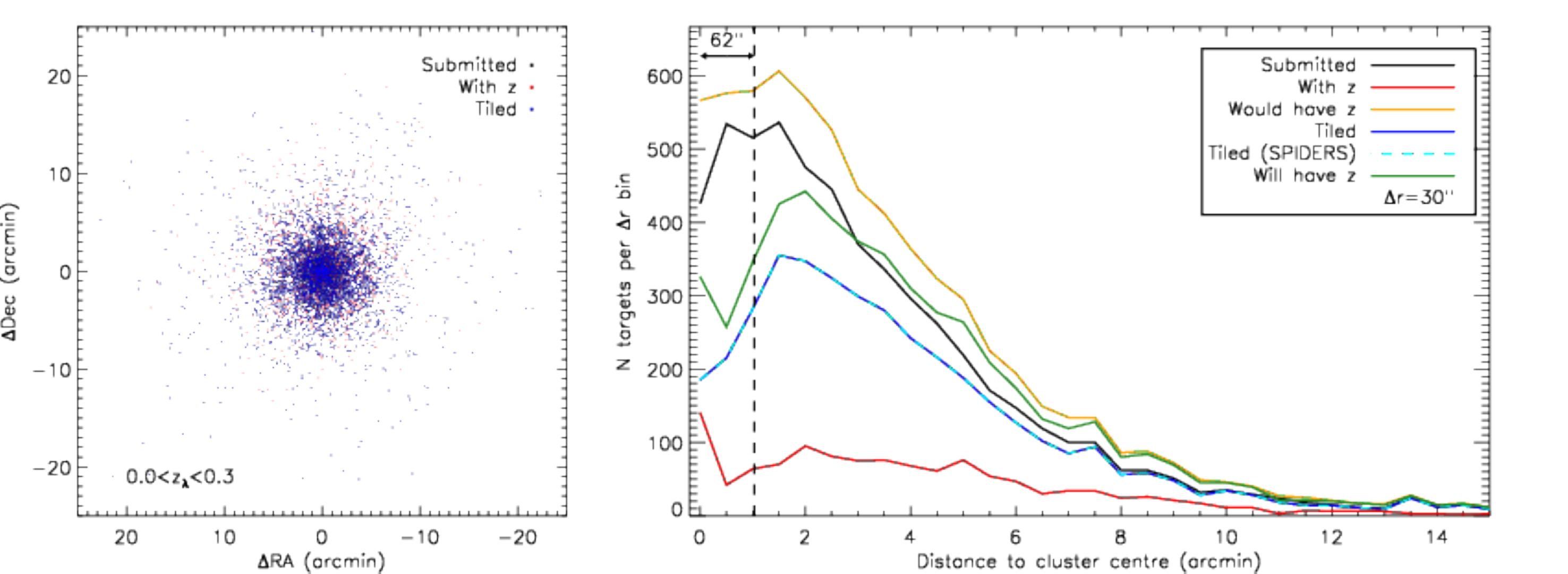} \\
	\includegraphics[width=\linewidth]{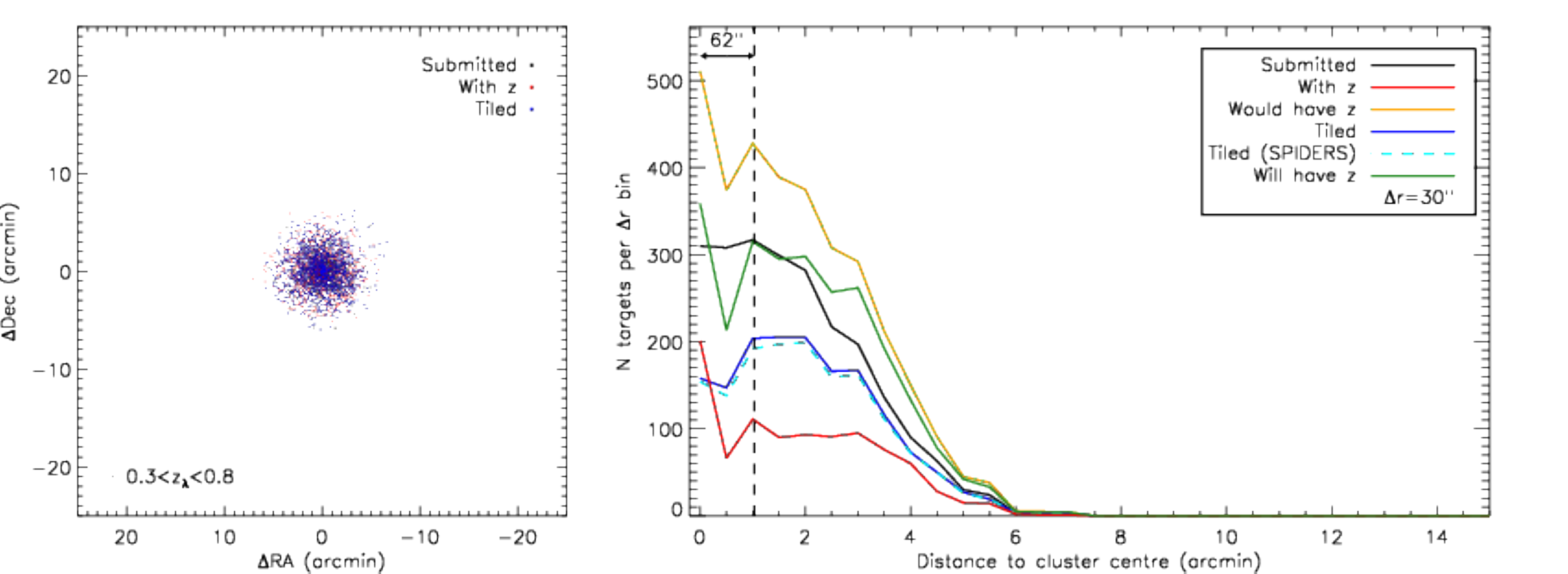}	
	\end{tabular}
		\caption{Sky distribution of SPIDERS\_RASS\_CLUS targets in all clusters in chunk {\tt eboss1} (656 systems), displayed relative to the cluster "optical" centre (RA\_OPT, DEC\_OPT) and split in two photometric redshift bins (upper and lower row). The left panel shows the distribution of submitted (black), tiled (blue) targets and those with a redshift known prior SPIDERS (red). The right panel shows those distributions as a function of distance to the cluster centre. \emph{"Would have z"} corresponds to the sum of submitted and known-redshift targets. \emph{"Will have z"} corresponds to the sum of tiled and known-redshift targets.}
	\label{fig:eboss1_targetsky} 
\end{figure*}

Figure~\ref{fig:eboss1_targetdistance} shows (for the {\tt eboss1} chunk) the pairwise separation between targets submitted to the eBOSS tiling algorithm, and the separation between cluster targets that will be assigned a fibre. Targets closer than the fiber collision radius are too close to each other (collided targets) and cannot be observed on a single plate. This explains the jump in the histogram of tiled targets at $62\arcsec$ separation. Multiple plate overlaps resolve a fraction of those collisions and enable access to smaller separations, hence a non-zero completeness for the collided set of SPIDERS targets.
Fig.~\ref{fig:eboss1_targetsky} shows the distribution of targets relative to their parent cluster centre. A substantial fraction of central galaxies already have a redshift determined from previous observations, and SPIDERS will observe almost all remaining cluster central galaxies. Because of the fiber collisions, a noticeable dip in completeness is expected at cluster distances $\lesssim 1\arcmin$, and the completeness increases with increasing radial distance.

	\subsection[]{Illustration: the SEQUELS-DR12 pilot}
	\label{sect:sequels-pres}

A total of 918 CODEX and 28 XCLASS-RedMapper clusters lie within the rectangle $(120 \leq RA \leq 210)$ and $(45 \leq DEC \leq 60)$, being the target selection area for SEQUELS (Sect.~\ref{sect:sequels-pres}).
The selection of galaxy cluster candidates in the SEQUELS pilot survey slightly differs from the main SPIDERS selection. All CODEX clusters have been considered regardless of their richness, and a richness cut $\lambda_{XC} \geq 20$ has been applied to select XCLASS-RedMapper clusters.
The galaxy targeting strategy in SEQUELS is similar to, but not identical to, the final SPIDERS targeting algorithm. The two differences are:
\begin{itemize}
\item a cut in {\tt FIBER2MAG\_I} set at 21.0 instead of 21.2
\item a target list being trimmed at a priority $\leq 50$ instead of $\leq 80$ ($\leq 33$ in the case of eboss3 chunk)
\end{itemize}

\begin{figure}
	\includegraphics[width=84mm]{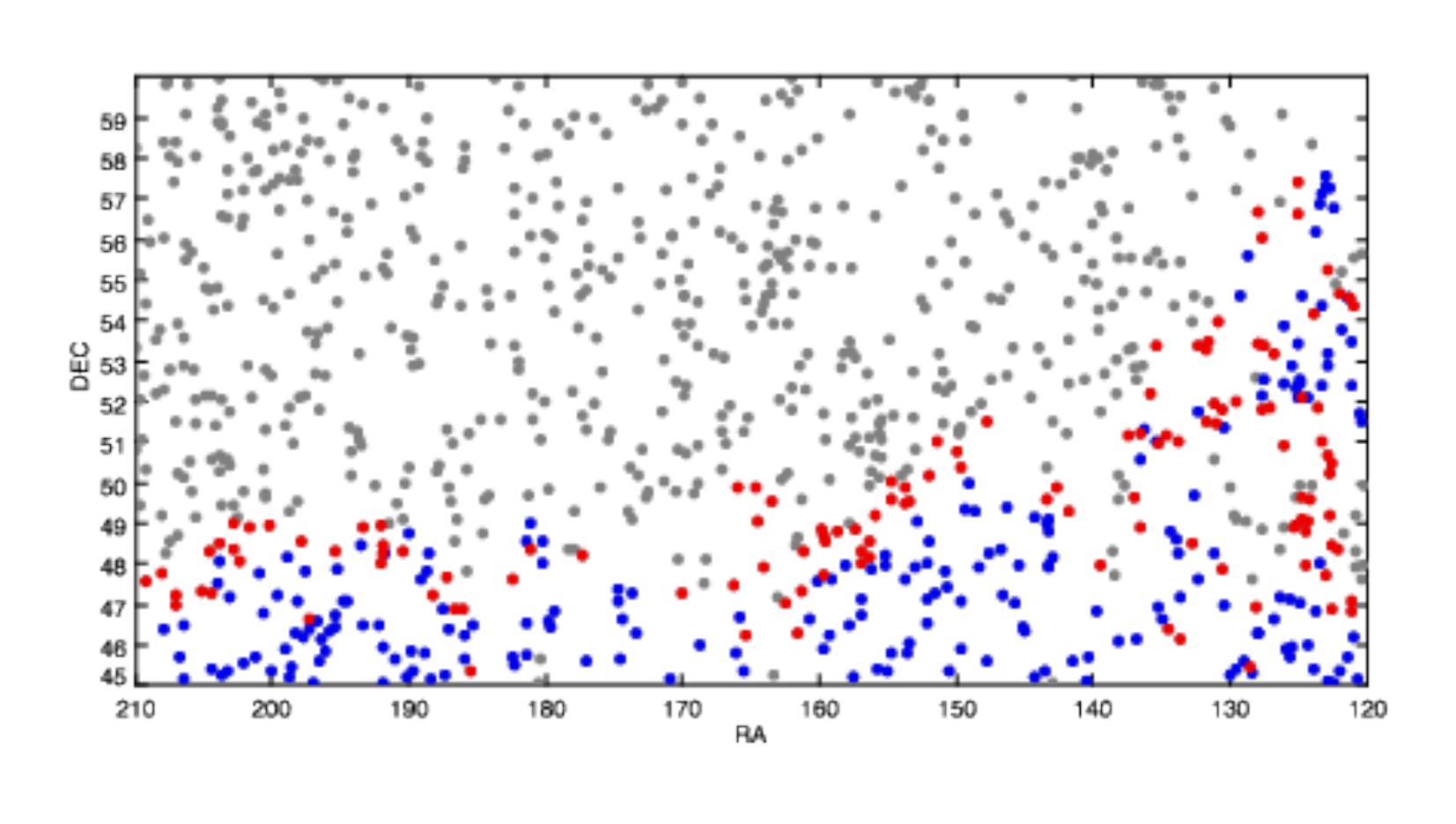}
		\caption{Sky distribution of SPIDERS/CODEX clusters in the SEQUELS pilot area. Each point is one CODEX cluster (918 objects, no cut on richness). Coloured points correspond to clusters within (at least) one SEQUELS-DR12 plate. Blue are "fully observed" clusters (230/918): the number of obtained SEQUELS-DR12 spectra equals the number of tiled targets. Red are "partially observed" (121/918) and missing redshifts will be obtained in the course of the SPIDERS survey.}
	\label{fig:sequels_clusters} 
\end{figure}

We show in Fig.~\ref{fig:sequels_clusters} the sky distribution of CODEX clusters in the SEQUELS area. Those having at least one new redshift from SEQUELS observations are shown with colours. Within SDSS Data Release 2012, 230 of them are completely observed (i.e.~all tiled targets have been acquired) and 121 are pending completion. Throughout this paper we will illustrate the procedure envisaged to build the SPIDERS cluster catalogue using this sample of 351 galaxy clusters in the frame of Data Release 12. As shown in Fig.~\ref{fig:lxzdist}, this illustration sample is representative of the complete SPIDERS sample.

\section[]{Analysis steps: from SPIDERS spectra to cluster properties}
	\label{sect:analysis}

SPIDERS observations deliver spectra of sources identified as red-sequence members in CODEX and XCLASS X-ray clusters. This section describes the steps needed to reach the primary goals of SPIDERS, namely confirmation and redshift determination of X-ray selected galaxy clusters. This procedure is illustrated throughout this section with results from the SEQUELS-DR12 sample of CODEX candidate clusters, made of 351 objects in total. The algorithms presented here are prototypical and adapted to this illustration sample. They will benefit from developments in the course of the SPIDERS survey.

		\subsection[]{Data reduction}

SEQUELS-DR12 data are processed identically as in BOSS, namely using the {\tt idlspec2d} routines \citep{dawson2016}. Redshifts and classifications of sources are obtained after fitting a set of templates to the reduced spectra \citep{bolton2012}. Fits excluding quasar templates ("{\tt \_NOQSO}" values) provide reliable redshifts for targets known to be galaxies. The final SPIDERS data reduction and spectral classification will rely on the eBOSS improved pipeline developments and be backward compatible with previous BOSS data.

	\subsection[]{Spectra and redshift collection}

Conversely to other target classes in BOSS and eBOSS, each SPIDERS cluster is a collection of spectroscopic targets (potential cluster members), instead of a proper target itself. Redshifts are collected in the vicinity of a (candidate) cluster, and listed while keeping track of relevant associated information (magnitude, photometric and spectroscopic flags, etc.)

\begin{figure*}
	\includegraphics[width=\linewidth]{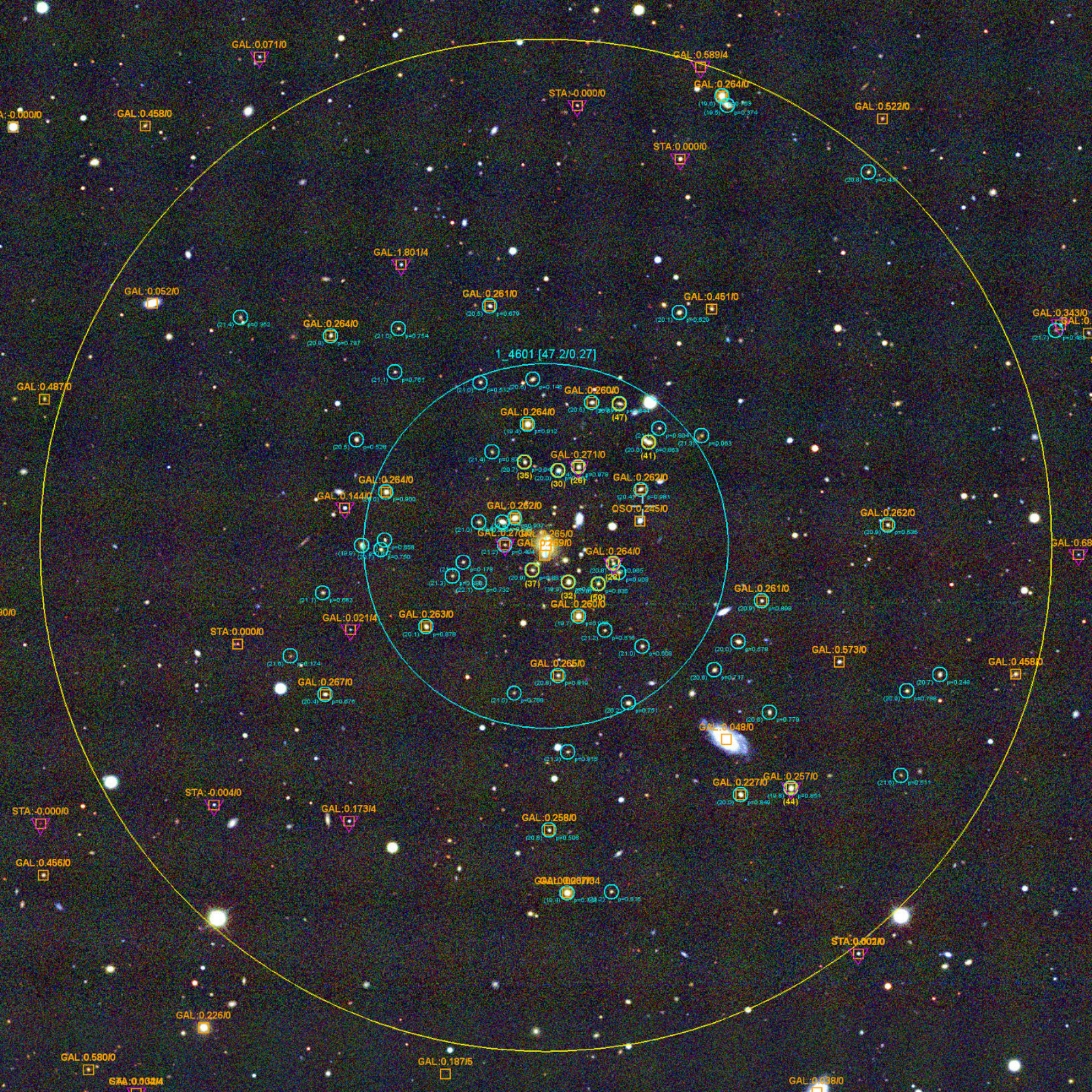}
		\caption{$12\arcmin \times 12\arcmin$ $(g,r,i)$ composite image of CODEX 1\_4601 at R.A.\,=\,12h22m05.3s, Dec\,=\,+$45\degr$18m36s, $z_{\lambda, OPT}=0.27$, $\lambda_{OPT}=47.2$, as observed and confirmed by SPIDERS at $z_{spec}=0.2630 \pm 0.0009$ (see Fig.~\ref{fig:example_diagnostic}). The 2~arcmin large cyan circle is centred on the cluster "optical" centre. The original X-ray position is materialized by the thick blue cross ($\sim 1$ arcmin to the north-west).
		Cyan circles indicate red-sequence members, numbers below correspond to their {\tt fiber2mag\_i} magnitudes and membership probability $p_{\rm mem}$.
		Gold circles are SEQUELS submitted targets, the number in parenthesis is the $priority$ (1 stands for SEQUELS AGN, SPIDERS\_RASS\_AGN).
		Orange boxes are SDSS spectroscopic redshifts. Text above indicates the best-fit type (GAL, QSO, STA for galaxy, quasar and star templates) with associated {\tt Z}/{\tt ZWARNING}. The "NOQSO" fits are preferred when available (i.e.~for all SDSS-BOSS data).
		Green boxes are targets with the {\tt EBOSS\_TARGET0} bit set to 21, i.e.~SPIDERS\_RASS\_CLUS targets.
		Tiled targets in SEQUELS appear as magenta triangles.
		The large yellow circle shows the $R_{200c} = 1.3$~Mpc radius of the cluster as derived from X-ray data.
		}
	\label{fig:example_image_cluster} 
\end{figure*}

\begin{figure}
	\includegraphics[width=84mm]{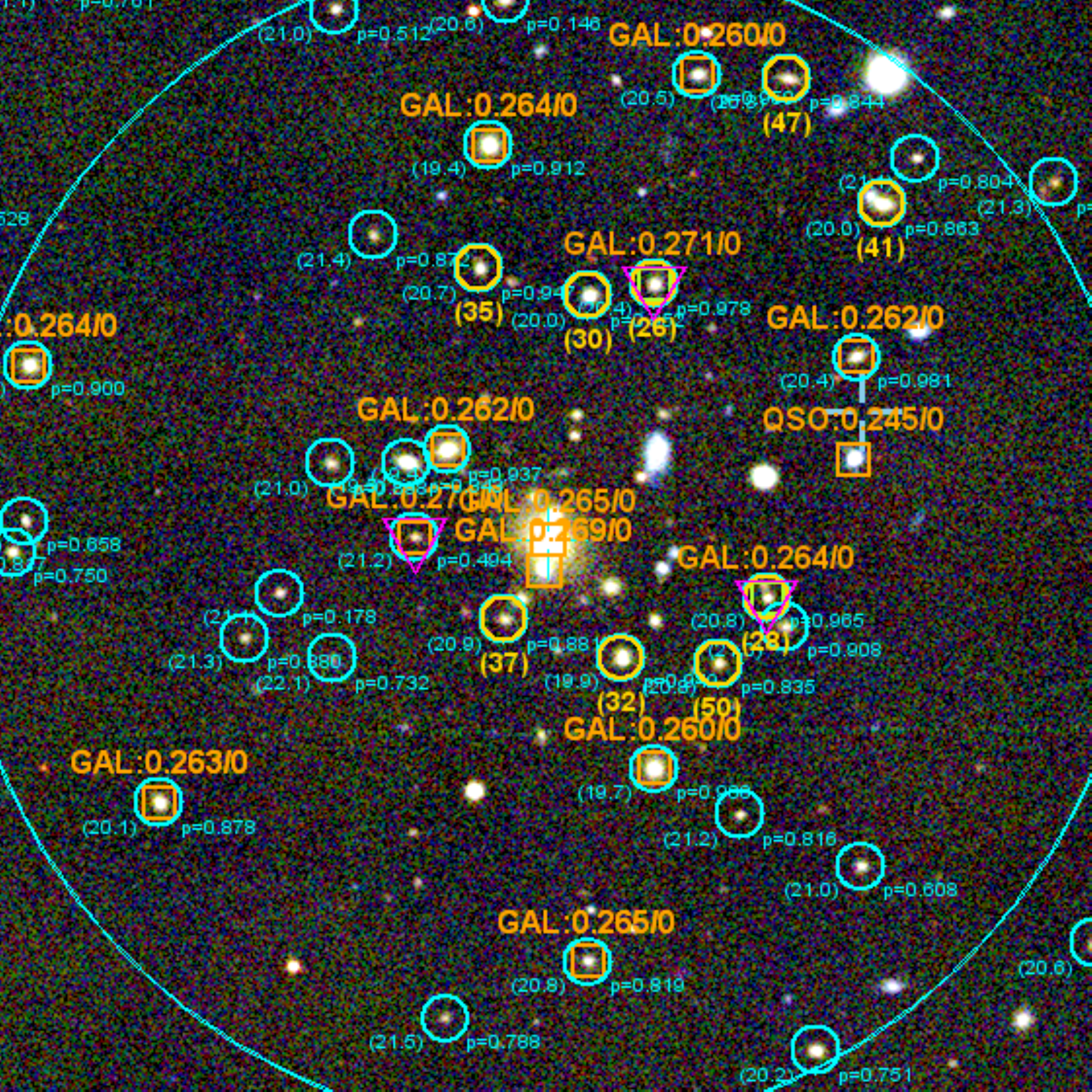}
		\caption{Zoom over the central part of Fig.~\ref{fig:example_image_cluster}, centred on the cluster "optical" position. The large cyan circle has a radius 2~arcmin (see legend of Fig.~\ref{fig:example_image_cluster} for explanations on symbols.)}
	\label{fig:example_image_cluster_inset} 
\end{figure}

Fig.~\ref{fig:example_image_cluster} and Fig.~\ref{fig:example_image_cluster_inset} display 3-colour $gri$ images of a CODEX cluster observed and confirmed in SEQUELS-DR12. The overlays correspond to: the CODEX \emph{cluster} catalogue (large cyan circle), the CODEX \emph{cluster photometric member} catalogue (red-sequence members, small cyan circles), the list of \emph{submitted targets} (gold circles), the list of \emph{tiled targets} (magenta triangles) and \emph{spectroscopic redshifts} from SDSS (red and orange squares, the latter correspond to SDSS data up to DR11).
In case a target was observed multiple times over the course of the SDSS programs, the eBOSS {\tt SPECPRIMARY} flag is considered and higher priority is given to higher signal-to-noise spectra.
For the SEQUELS-DR12 sample that is used as an illustrative example in this paper, only galaxies identified as members of the cluster red-sequence ($p_{\rm mem} \geq 5\%$) are taken into account. Therefore "{\tt NOQSO}" values are considered (i.e.~{\tt Z\_NOQSO, CLASS\_NOQSO, ZWARNING\_NOQSO}), except in the case of redshifts whose origin is SDSS-I/II for which we use the standard {\tt Z, CLASS, ZWARNING} values.

The final SPIDERS data collection procedure will improve on specific points, in particular by investigating the benefits of including galaxies with a spectroscopic redshift excluded from the RedMapper red-sequence (because of the colour selection, or cluster-centric distance cuts, etc.) While this may prove advantageous in confirming the cluster redshift, their selection is more heterogeneous and more difficult to track back.

	\subsection[]{Automatic membership}

Most surveys of X-ray galaxy clusters rely on an ultimate validation by one or several trained astronomers based on spectroscopic redshifts of individual galaxies \citep[e.g.][for recent applications]{guzzo2009,adami2011}. This is needed in order to disentangle dubious cases, carefully inspect interlopers and members and classify the reliability of the cluster redshift. The limited manpower imposes limits on the large amount of galaxy clusters involved in SPIDERS that can be visually screened; however this can be alleviated by running an automatic procedure in first place. This algorithm must be able to separate the secure and easy cases, only requiring quick eyeballing, from the more difficult ones demanding deeper inspection. In the former situations, the automatic procedure must be able to address the membership of red-sequence galaxies. In order to account for the variety of cluster masses, physical states, richnesses and redshifts in the sample, we decided to adopt a broad approach, preparative of the visual inspection of every individual cluster. In the following, we demonstrate its main features and its applicability with the SEQUELS-DR12 sample.

Our procedure runs on each galaxy cluster individually, based on the list of red-sequence members associated to a spectroscopic redshift (see above).
The bi-weight average \citep[][]{beers1990} of those $N_{zpsec,0}$ redshifts provides the starting point (first guess) of an iterative clipping procedure. It performs an initial rejection of members with velocities offsets greater than $5000$~km/s (relative to this first guess mean redshift). The bi-weight average of the resulting $N_{zspec,1}$ potential members is computed. An estimate of the velocity dispersion \citep[][]{beers1990} is also computed and results from the bi-weight variance (if $N_{zspec,1} \geq 15$) or the gapper estimator (if $N_{zspec,1} < 15$). Objects lying further away than 3 times the velocity dispersion from the average velocity are rejected ("3-$\sigma$ clipping"). This procedure is iterated until convergence or stops after 10 steps. The remaining objects are called members ($N_{zspec,k} = N_{\rm mem}$, $k \leq 10$).
In the course of the iterative procedure described above, several cases may arise:
\begin{itemize}
\item $N_{zspec,0}<3$: the cluster is left for visual inspection
\item $N_{zspec,1}=0$, i.e.~the initial 5000 km/s clipping rejected all members: the procedure stops, a flag is issued.
This may correspond to the case in which groups of galaxies are too far from each other in velocity space, for instance in case of several distinct structures along the line of sight.
\item $0<N_{zpsec,i}<3$, i.e.~only 1 or 2 members are left after $i$ steps: the iteration process stops and returns the member list without estimating the mean nor the velocity dispersion.
\item $N_{zspec,i}=0$, i.e.~no member is left after $i$ steps: a flag is issued indicating failure of the $\sigma$-clipping method.
\item $N_{zspec,k} \geq 3$: the process succesfully converges, a cluster redshift is estimated from the biweight-average of the $N_{\rm mem}$ galaxies and the biweight-variance (or gapper estimator if $N_{\rm mem}<15$) serves as an estimate for the velocity dispersion.
\end{itemize}

\begin{figure*}
	\includegraphics[width=\linewidth]{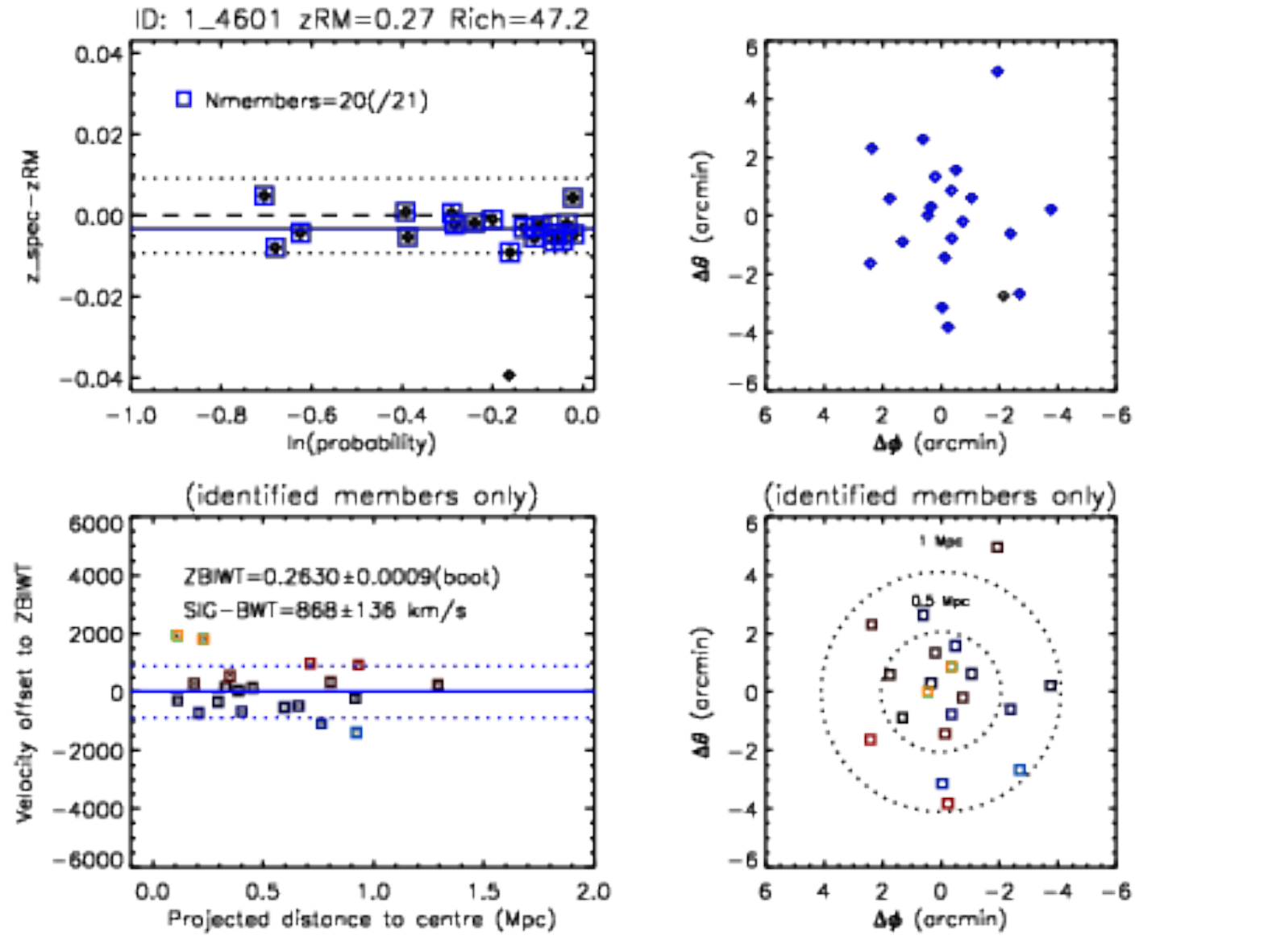}
		\caption{Example diagnostic plots outcome of the automated membership procedure, for one particular cluster in the SEQUELS-DR12 sample (1\_4601, $z_{\lambda, OPT}=0.27$, $\lambda_{OPT}=47.2$, see Fig.~\ref{fig:example_image_cluster} and~\ref{fig:example_image_cluster_inset}).
		\emph{{\bf Top left} -- "redshift/probability plot":} offset of all red-sequence members with spectroscopic redshifts (21 objects, black diamonds) relative to the RedMapper photometric redshift ($z_{\lambda,OPT}$, applicable to all members). Error bars (often too small to be visible) are displayed for each data point. The x-axis shows $\ln(p_{\rm mem})$ the (logarithmic) probability of each red-sequence member, as computed by RedMapper. The photo-z uncertainty is represented by the two horizontal dotted lines. Blue squares are selected spectroscopic members (20 objects). A grey bar indicates the initial 5000 km/s selected range before iterative clipping. The blue horizontal bar represents the spectroscopic redshift value (bi-weight average) calculated with the spectro-members.
		\emph{{\bf Top right} -- "sky location of members":} All spectroscopic red-sequence members (21 objects) are displayed in a $12 \times 12$~arcmin projected map. Blue symbols are identified members. The centre of the map corresponds to the optical cluster centre.
		\emph{{\bf Bottom left} -- "velocity-distance plot":} this plot considers only identified members (20 objects). The cluster redshift (bi-weight average) is taken as a reference and velocity offsets are indicated on the y-axis. Error bars on the individual velocities are represented as vertical lines (insivble in this figure). The x-axis displays the projected distance to the cluster centre. Blue horizontal line shows the 0 offset, dashed lines show the velocity dispersion value ($\pm 1 \sigma$). The colour-code indicates blue-/redshifted objects. The bi-weight average redshift ZBIWT and the bootstrap uncertainty on this value are indicated in the panel. The velocity dispersion estimate (Bi-weight variance if $N_{spec} \geq 15$, Gapper estimate if $N_{spec}<15$) is given along with its uncertainty (see text).
		\emph{{\bf Bottom right} -- "sky projection of spectroscopic members":} similar as the above panel, but only for selected spectroscopic members and reproducing the colour code (blue-/redshifted objects). Circles indicate projected physical distances to the cluster centre.}
	\label{fig:example_diagnostic} 
\end{figure*}

Fig.~\ref{fig:example_diagnostic} provides a detailed illustration of the results output of the procedure in a successful case, extracted from the SEQUELS-DR12 sample. For this cluster, one of the 21 members of the red-sequence with a spectroscopic redshift was flagged as an interloper, it has $p_{\rm mem}=0.85$. The cluster spectroscopic and photometric redshifts are compatible within their $1 \sigma$ uncertainty. We note the presence of a Seyfert 1 galaxy located $\sim 1\arcmin$ to the West of the cluster core, possibly contaminating the X-ray emission of the galaxy cluster (as hinted by the ROSAT soft X-ray contours). The rest-frame velocity of this object relative to the cluster redshift is above 4000~km/s, hence consistent with it not being included in the dynamical analysis of the cluster. We discuss and model X-ray AGN contamination later in this study (see App.~\ref{app:selbias}).

The automatic procedure delivers a redshift for 219 out of the 351 candidates with $N_{\rm mem} \geq 3$.  We note that our choice for an initial 5000~km/s rejection criterion is more inclusive than other studies relying instead on a lower threshold, usually 3000~km/s \citep[e.g.][]{}. We checked that changing to this value provides similar results, except in cases requiring human decision. For 194 systems, the final cluster redshifts agree within $10^{-3}$ relative difference. The other systems are complex or poor systems, either discarded or refined while performing the visual confirmation (as described in the next section).

	\subsection[]{Manual steps and refinements}

Validation of the galaxy cluster and final assessment of its redshift are achieved through visual screening of the outcome of the automatic procedure. This process should allow a number of refinements inaccessible to algorithms. In particular, the inspection of individual galaxy spectra may refine or discard the result of the eBOSS fitting algorithm, based on e.g.~the knowledge of the cluster photometric redshift and the probability $p_{\rm mem}$ that the object belongs to the cluster. The object can therefore be added or removed from the list upon which the cluster validation is performed. Inclusion or removal of members as well as particular weights given to members (e.g.~depending on their $p_{\rm mem}$ value, or in the case of a BCG) help in deciding the validation status and mean redshift of the cluster. Line-of-sight projection effects not disentangled by the photometric membership algorithm can also be identified and split into several components. Finally, a comment can be set by the inspector.  We anticipate such inspection to be collaborative, final decisions should be taken based on the judgement of independent inspectors.

\begin{figure}
	\includegraphics[width=84mm]{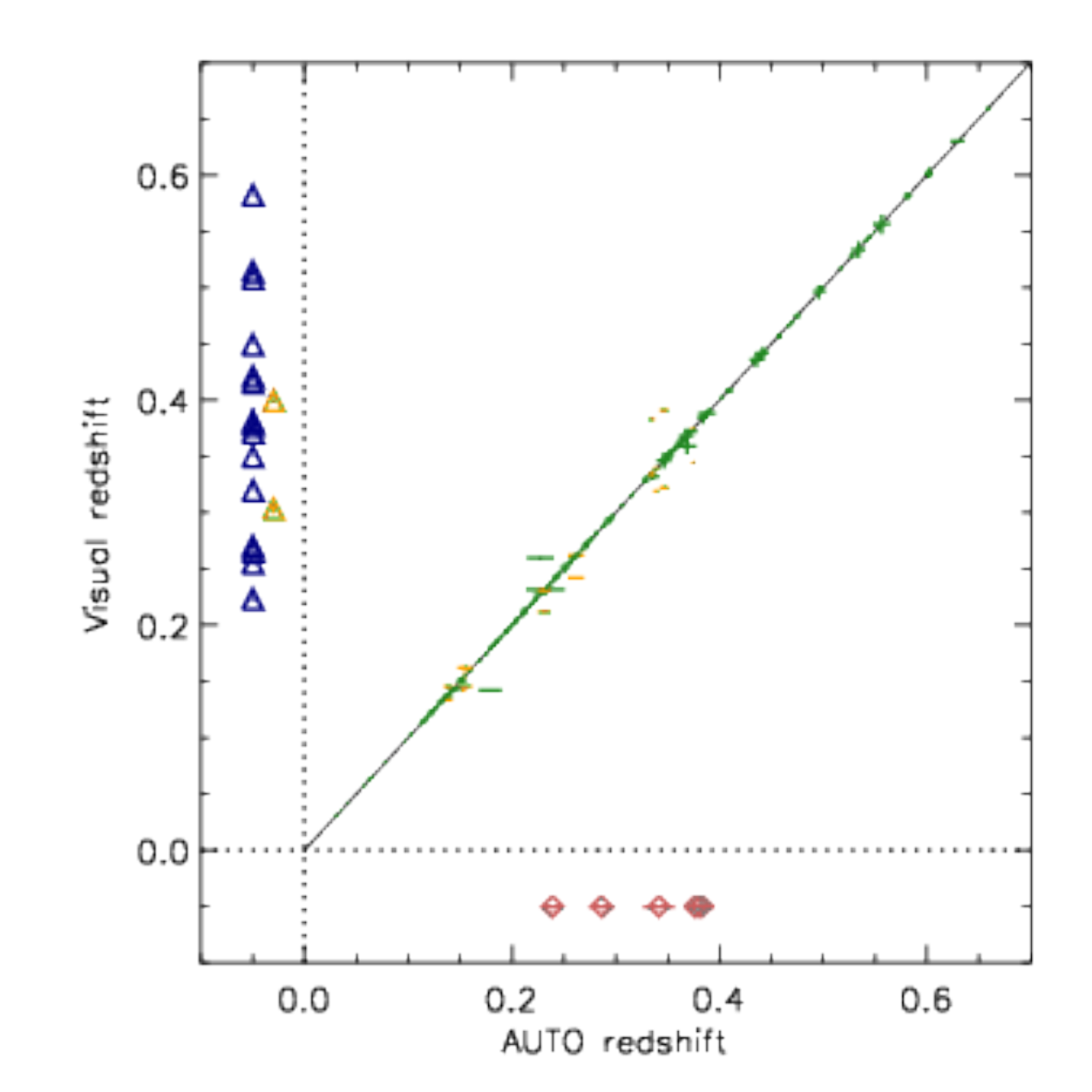}
		\caption{Result of the visual inspection (see Table~\ref{tab:visual_result}) as a function of the cluster spectroscopic redshift, shown with outputs from the automatic procedure (x-axis) and of the visual procedure (y-axis). Green points (and error bars) stand for clusters initially validated by the algorithm and confirmed by visual inspection (205). Blue triangles are cluster recovered by visual inspection (15) and orange triangles are systems split in 2 components (9 initially validated as single-component and 1 initially discarded). Red diamonds are candidates initially validated and discarded by visual inspection (5).}
	\label{fig:visual_zzplot} 
\end{figure}

We illustrate the validation process with the SEQUELS-DR12 sample of 351 clusters. Since these clusters will be re-inspected within the complete SPIDERS Tier-0 survey with more redshifts, only one inspector participated in this exercise. Table~\ref{tab:visual_result} shows that a large fraction of the algorithm decisions are confirmed by visual screening, while 10 candidates were split into multiple distinct components, 5 were discarded and 15 promoted. In 30 cases no spectroscopic redshifts were found in the red-sequence, leaving the cluster status as non-validated (these clusters are mostly high-redshift $z \gtrsim 0.6$ candidates whose members are too faint to be spectroscopically observed). Fig.~\ref{fig:visual_zzplot} displays the same result as a function of the automatic and final cluster spectroscopic redshift.

\begin{table}
	\centering
\caption{\label{tab:visual_result}Result of the visual inspection of 351 CODEX cluster candidates in SEQUELS-DR12, split according to the outcome of the automatic algorithm. Note that only 230/351 candidates are completely observed within SEQUELS-DR12, i.e.~all of their tiled targets have been acquired within the program, their statistics are indicated in parentheses. "Non validated" does not necessarily means non-existence of a cluster and may result from too low a number of spectra in the red-sequence.}
		\begin{tabular}{@{}lcc@{}}
\hline
	{\bf Auto-validation status:}&	{\bf Validated}	&	{\bf Pending}	\\
\hline
\hline
Visual inspection status:	&		&		\\
 - Single-component, validated	&	205 (119)	&	15 (11)	\\
 - 2-component split			&	9 (7)	&	1 (1)	\\
 - Non validated				&	5 (2)	&	86 (65)	\\
 - No spec-z (non validated) 	&	-	&	30 (26)	\\
\hline
		\end{tabular}
\end{table}

	\subsection[]{Redshift and velocity dispersion estimates}
		\label{sect:zveldisp}

		\subsubsection[]{Cluster redshift estimates}

\begin{figure}
	\includegraphics[width=84mm]{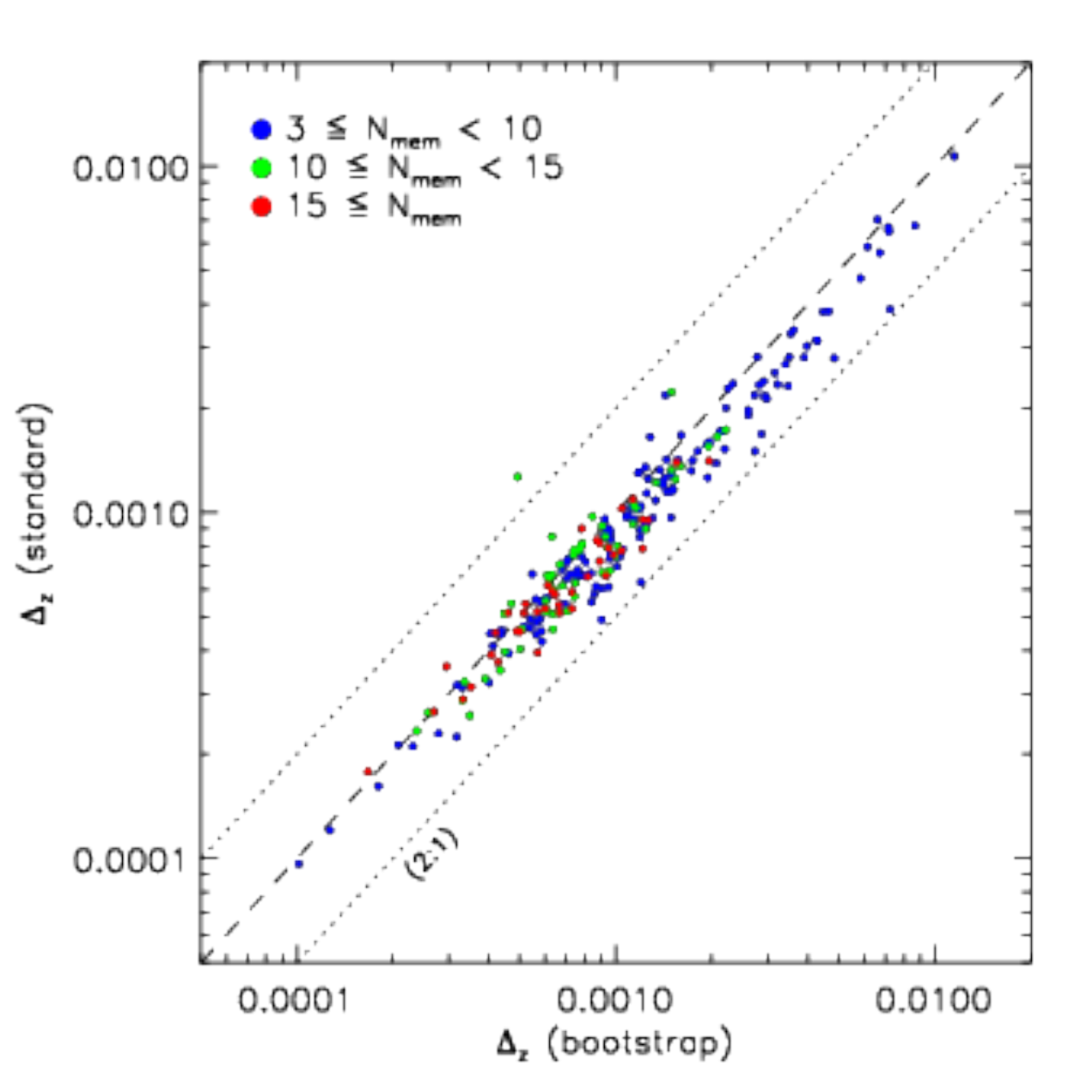}
		\caption{Comparison of statistical uncertainties on the cluster spectroscopic redshift. The x-axis shows the bootstrap error (our baseline value throughout this work) while the y-axis shows the result from Eq.~\ref{eq:ruelzerr}, involving an estimate of the standard deviation of velocities. The dashed line shows equality, dotted lines represent a factor 2 between plotted quantities.}
	\label{fig:compa_deltazestimates} 
\end{figure}

Final cluster redshift estimate (hereafter $z_{\rm BIWT}$, or simply $z$) is based on the bi-weight average \citep[][]{beers1990} of all red-sequence galaxies selected as cluster members, in the cases where $N_{\rm mem} = 3$ or more members are identified. Cases with 1 or 2 members only correspond to a redshift set manually, typically equal to that of the BCG.
When $N_{\rm mem} \geq 3$, the statistical uncertainty $\Delta_z$ of the cluster redshift is computed by bootstrap resampling of the $N_{\rm mem}$ velocities. Fig.~\ref{fig:compa_deltazestimates} compares these uncertainties with a more common estimator \citep[see e.g.][their Eq.~4]{ruel2014}, involving the standard deviation of velocities ($\sigma$):
\begin{equation}
\label{eq:ruelzerr}
\Delta_{z} {\rm (standard)} = \frac{1}{c} \frac{\sigma \cdot (1+z)}{\sqrt{N_{\rm mem}}}
\end{equation}
with $\sigma$ given by the bi-weight variance estimator if $N_{\rm mem} \geq 15$ and by the gapper estimator otherwise (see Sect.~\ref{sect:zveldisp}).

The two estimates are in good agreement with each other. In almost all cases, the bootstrap technique provides slightly more conservative uncertainty estimates than the standard one and we consider the former as our baseline redshift error.

\begin{figure}
	\includegraphics[width=84mm]{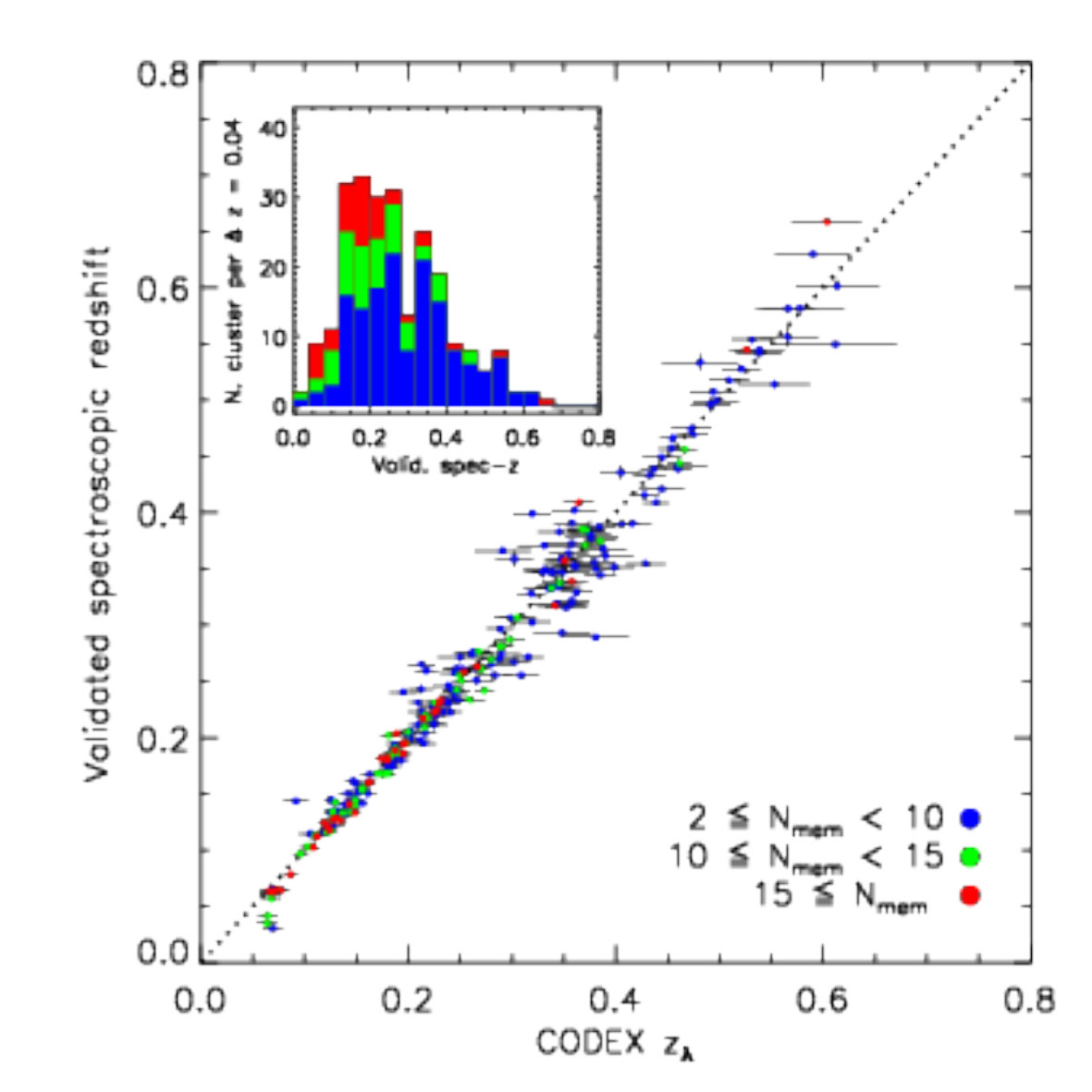}
		\caption{Comparison between the photometric redshift $z_{\lambda}$ output of RedMapper and the spectroscopic redshift of the 230 CODEX clusters validated by visual inspection in SEQUELS-DR12. The spectroscopic redshift error is the bootstrap uncertainty. The colour code reflects the number of spectroscopic members retained in validating the cluster. The histogram in the inset shows the distribution of spectroscopic redshifts, with the same colour coding.}
	\label{fig:zrm_vs_zspec} 
\end{figure}

The typical cluster redshift statistical uncertainty is $\Delta_z/(1+z) \lesssim 10^{-3}$, a factor 10 lower than the typical cluster photometric redshift error, with a median number of 10 members. Figure~\ref{fig:zrm_vs_zspec} compares the photometric and spectroscopic redshift estimates for each of the validated clusters: the very good agreement between them is not surprising \citep[e.g.][]{rykoff2014}, although this comparison emphasizes a noticeable improvement brought by spectroscopic redshifts at $z \geq 0.2 - 0.3$, both in terms of accuracy and precision.
The theoretical quantity of interest for the SPIDERS clusters is the redshift of the halo in which the X-ray gas and the galaxies are hosted. An uncertainty of $\Delta_z = 10^{-3}$ on the redshift of an object at $z=0.3$ corresponds to a velocity offset of 230~km/s, hence a few times smaller than the typical velocity dispersion of a galaxy cluster. It also corresponds to $\sim 4$~Mpc comoving radial distance, hence slightly larger than the typical size of a galaxy cluster.

\begin{figure*}
	\includegraphics[width=\linewidth]{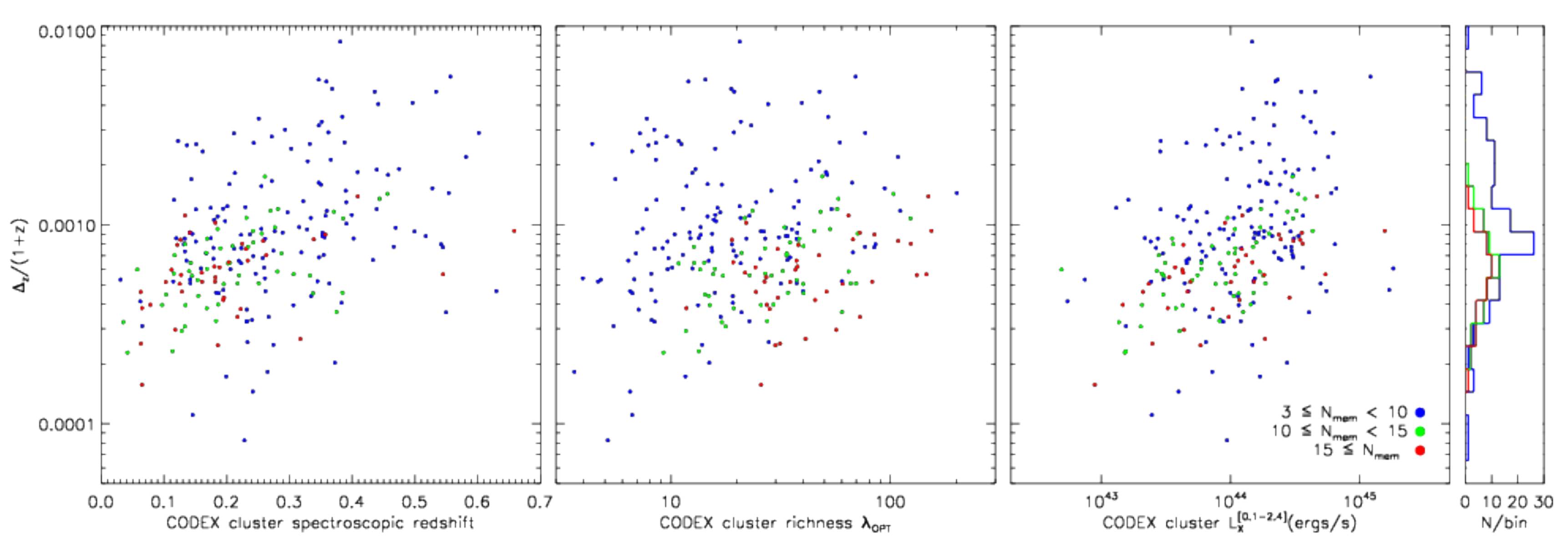}
		\caption{Distribution of statistical errors $\Delta_z/(1+z)$ in the cluster spectroscopic redshift among the SEQUELS-DR12 sample of validated clusters. These uncertainties are estimated through bootstrap resampling of the $N_{\rm mem}$ redshifts identified as cluster members. In particular they do not include additional sources of uncertainties due to potential systematic effects, e.g.~presence of substructures, inhomogeneous sampling, etc. The right panel is an histogram of sources binned by values of the redshift uncertainty.}
	\label{fig:zerr_vs_x} 
\end{figure*}

The statistical uncertainties on cluster redshifts are shown in Fig.~\ref{fig:zerr_vs_x}, as a function of redshift, cluster richness and cluster X-ray luminosity, and colour-coded by the number of members $N_{\rm mem}$ entering their computation. As expected, higher $N_{\rm mem}$ lead to lower uncertainties on the redshift estimates, which favors low-redshift clusters. Because of the various selection effects involved in detecting clusters (flux-limit in X-rays, red-sequence in optical), trends in the redshift uncertainty versus richness or luminosity do not appear clearly.

		\subsubsection[]{Radial velocity dispersions}
				
Once cluster members are identified, one estimates their line-of-sight velocities $v_i$, defined as \citep{danese1980}:
\begin{equation}
\frac{v_i}{c} = \frac{z_i - z_{\rm BIWT}}{1+z_{\rm BIWT}}
\end{equation}

\begin{figure}
	\includegraphics[width=84mm]{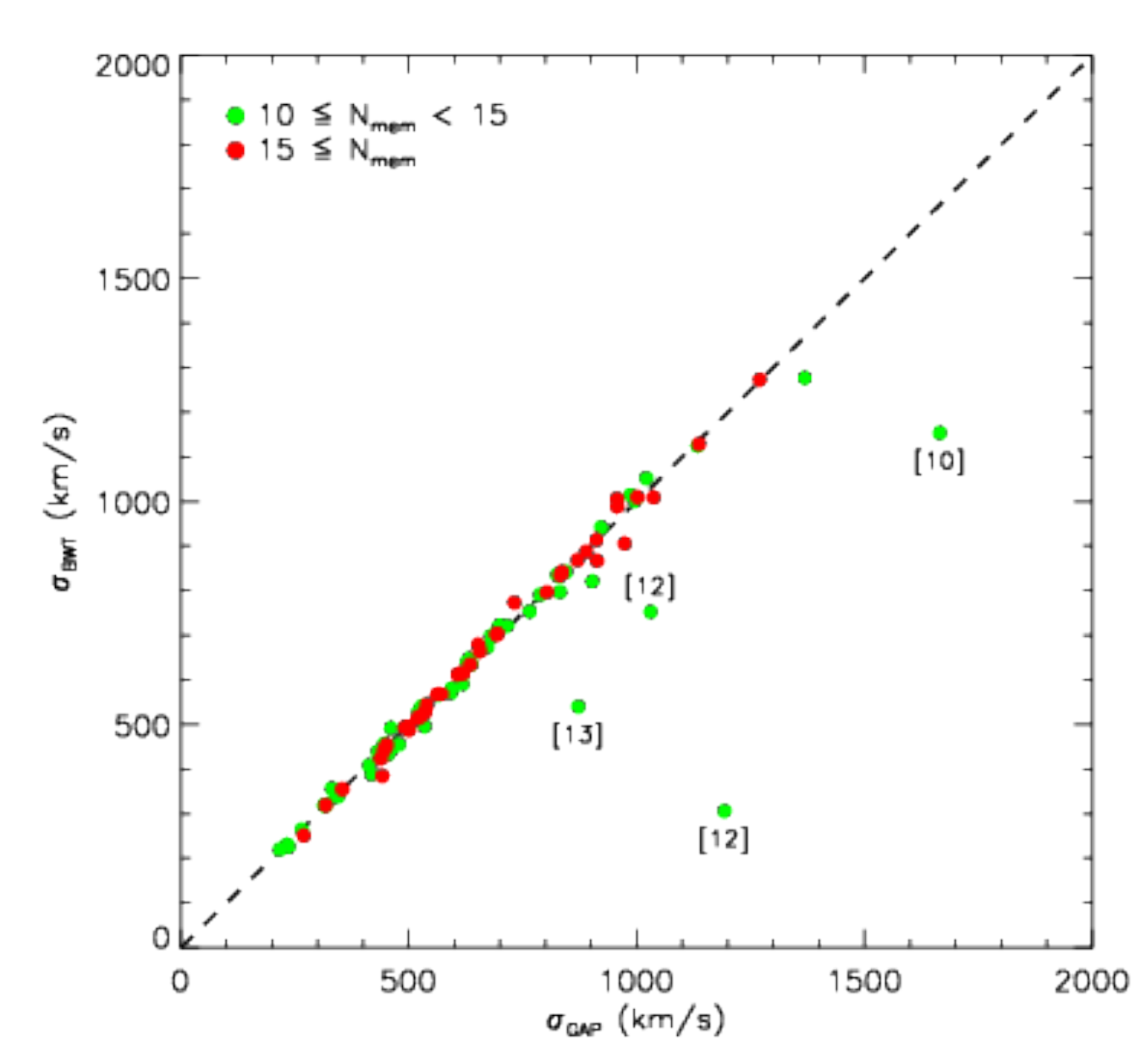}
		\caption{Comparison of velocity dispersion estimates obtained from the Gapper method ($\sigma_{\rm GAP}$) and the biweight sample variance ($\sigma_{\rm BWT}$) for individual CODEX clusters in the SEQUELS-DR12 sample. Only systems validated with more than 10 and 15 spectroscopic members are displayed. There is good agreement between the two estimators, although outliers are present, indicative of badly determined velocity dispersions due to, e.g.~substructure or presence of interlopers (the number of members is indicated in brackets).}
	\label{fig:gapper_vs_biwt} 
\end{figure}

We use two of the most common estimators for the dispersion of velocities, namely the "gapper" ($\sigma_{\rm GAP}$) and the "bi-weight variance" ($\sigma_{\rm BWT}^2$).
We refer to \citet{beers1990} for details in their computation and the algorithm\footnote{We used a Fortan version of ROSTAT adapted to our purposes.} used for these calculations. We refer the reader to \citet[][]{ruel2014} for a discussion of measurements of velocity dispersions in the regime of low number of spectroscopic members, in the context of galaxy clusters selected by Sunyaev-Zeldovich effect in the South Pole Telescope data.
Both estimators are computed for each cluster, although it is clear that a high enough number of members must enter the derivation to ensure robust measurements.
Fig.~\ref{fig:gapper_vs_biwt} demonstrates the good agreement between the two measurements provided that $N_{\rm mem} \geq 15$. While the majority of $10 \leq N_{\rm mem} < 15$ clusters also lie on the one-to-one line in this figure, a number of them stand as outliers, possibly impacted by the presence of interlopers or substructures in their list of spectroscopic members.

Evaluating the uncertainties and biases linked to cluster velocity dispersion measurements performed with a small number of spectroscopic members is a rather complex task. It requires in particular an understanding of the selection and sampling processes leading to the list of members entering the catalogue. Ideally, one would want to design end-to-end simulations reproducing all of the steps described above, from the cluster selection in X-rays down to the calculation of velocity dispersions.
Such procedures are feasible, for instance by combining N-body simulations and semi-analytical models \citep[e.g.][]{biviano2006,saro2013}.
An alternative, simpler, approach consists in resampling dense observations of clusters with high numbers of spectroscopic members and well-determined velocity dispersions $(\sigma^{true}_{\rm BWT})$, ensuring the target sampling reproduces that of SPIDERS. We follow this approach in the present work, bearing in mind the opportunities for further, more detailed developments.
We resample the observations of the HIFLUGCS sample of clusters \citep{zhang2011}, imposing a limiting magnitude and a minimal fiber distance corresponding to SPIDERS observations and accounting for the mass and redshift distribution of clusters in SPIDERS. The corresponding procedure is fully described in \citet{zhang2016}. It leads for each HIFLUGCS cluster to 500 resampled realizations, each realization leads in turn to an estimate of an observed $\sigma^{obs}_{\rm BWT}$. Grouping results by the number of members $N_{\rm mem}$ remaining after resampling, we derive the average value and the spread in $\sigma^{obs}/\sigma^{true}$.
This calculation provides therefore a baseline for the bias-correction and 1-$\sigma$ uncertainty on individual velocity dispersion measurements. We note that the catalog of the galaxy redshifts of the HIFLUGCS is a rather clean member galaxy input catalogue, likely almost free from interlopers. Uncertainties derived from the scatter in the down-sampling thus do not account for the effect of interlopers.

	\subsection[]{Catalogue production}
	\label{sect:catalogue_prod}

The updated, accurate, cluster spectroscopic redshifts enter as input of a new computation of X-ray cluster properties.

For the CODEX subsample, this procedure follows the same route as when starting from photometric redshifts ($z_{\lambda}$). Details on the procedure can be found in \citet{mirkazemi2015}: assuming a cosmological model, ROSAT fluxes are converted into rest-frame $[0.1-2.4]$~keV luminosities and scaling relations allow an estimate of the cluster mass and typical radius $R_{500}$ and $R_{200}$. The typical uncertainty on the luminosities of CODEX clusters amounts to $\sim 35$\%, as computed from the Poissonian fluctuation of number counts in ROSAT data. As an illustration, Fig.~\ref{fig:lxzdist} highlights the position of the SEQUELS-DR12 confirmed clusters in the luminosity-redshift plot, along with the corresponding error bars.

The XCLASS galaxy clusters benefit from high-quality X-ray data, thanks to the exquisite spatial and spectral resolution of XMM: the angular point-spread function FWHM is around $10-20 \arcsec$, depending on the off-axis angle of the cluster, and the spectral line spread function FWHM is around 100~eV at 1~keV energy. X-ray surface-brightness profiles and spectra are therefore the primary observables from which cluster physical properties are derived. In addition to $\lesssim 10\%$ accurate bolometric luminosities, surface brightness-averaged temperatures of C1 clusters can be measured with relatively good accuracy ($\sim 15$\%), depending on the actual cluster temperature, the number of counts collected by the instruments and the uncertainties in background subtraction \citep[e.g.][]{clerc2014}.

The \emph{eROSITA} data will be similar to the XMM data, although with a spatial resolution $\sim 1.5-2$ times lower. The methodology to compute X-ray cluster properties by combining SPIDERS spectroscopic redshifts and \emph{eROSITA} data is expected to lie between that of XCLASS clusters and CODEX clusters.


\section[]{Results from SEQUELS-DR12 sample}
	\label{sect:sequels}

Throughout this paper we illustrated the SPIDERS targeting strategy and plans for data analysis by means of the SEQUELS-DR12 pilot sample. We now elaborate on the use of such a sample of spectroscopically confirmed clusters, and present possible science applications with the perspective of the much larger, upcoming, SPIDERS sample.

	\subsection[]{Catalogue presentation}

		\subsubsection[]{The SPIDERS-CODEX clusters}

\begin{figure}
	\includegraphics[width=84mm]{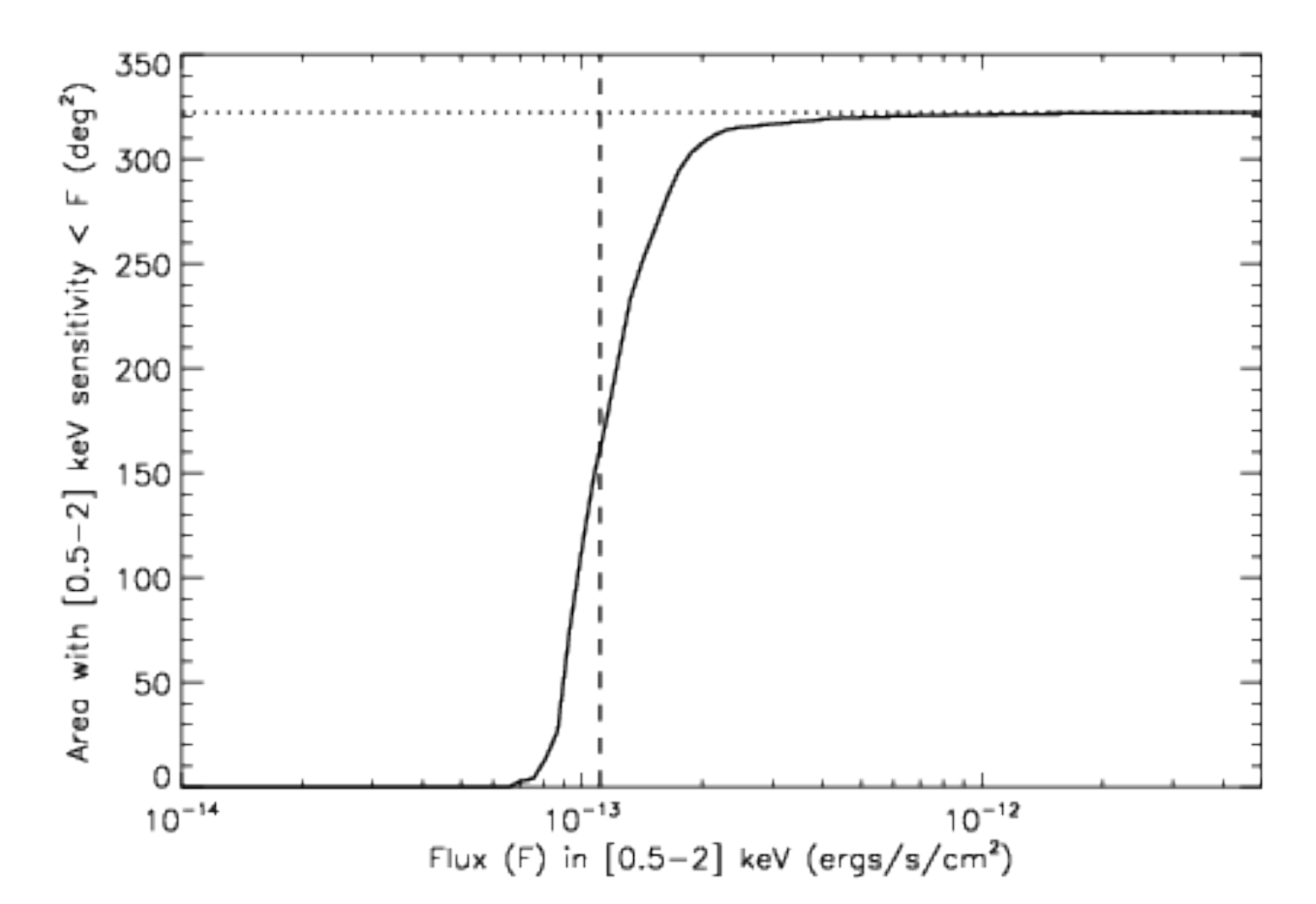}
		\caption{Effective area curve of the CODEX cluster survey, calculated over the footprint of the SPIDERS pilot area, expressed as a function of the X-ray flux sensitivity.}
	\label{fig:areacurve} 
\end{figure}

\begin{figure*}
	\includegraphics[width=\linewidth]{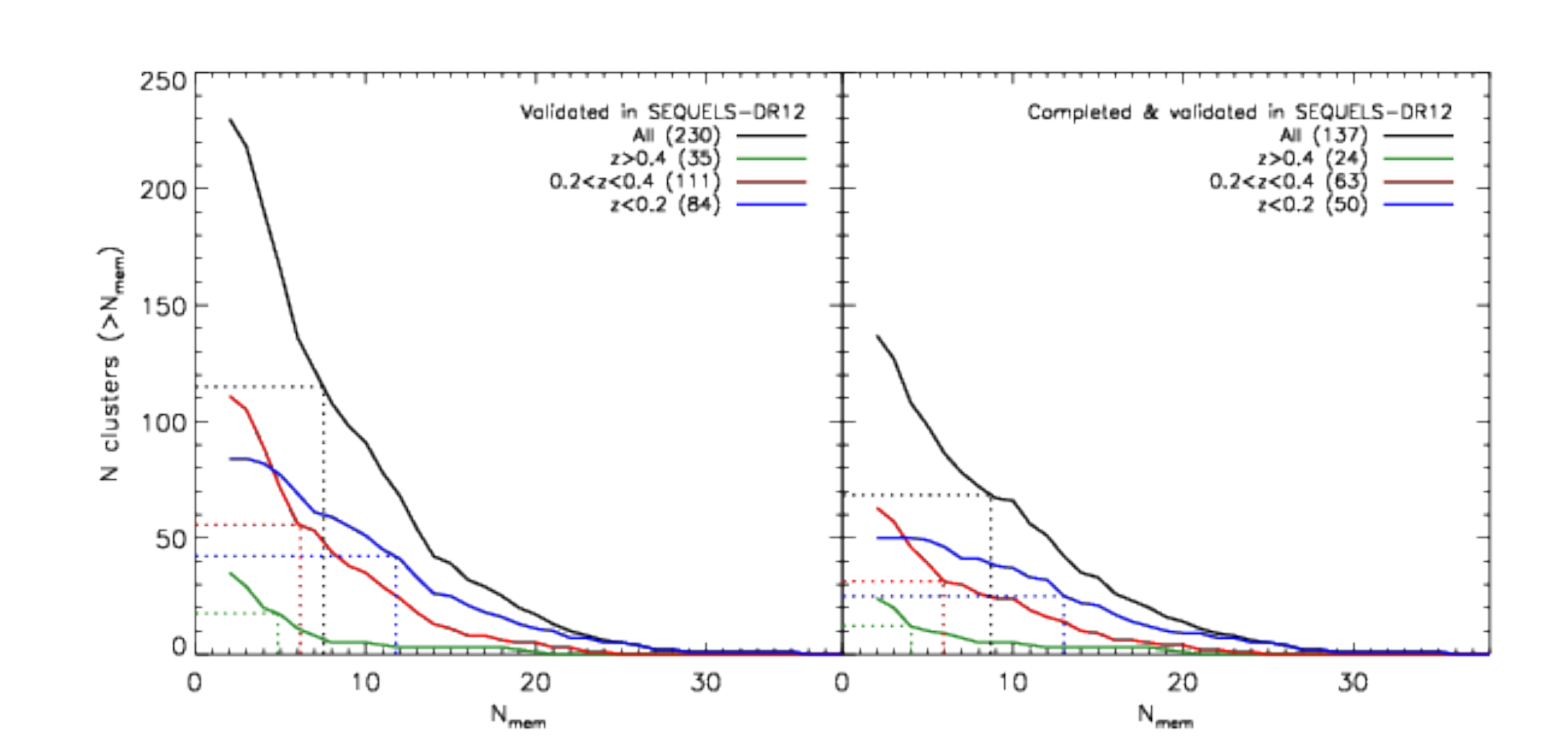}
		\caption{Cumulative distribution of CODEX clusters validated in SEQUELS-DR12, as a function of their number of spectroscopic members $N_{\rm mem}$. The right panel is for the case when all tiled targets within a cluster have been observed (completed observations) and is therefore more representative of the final outcome of SPIDERS. Colour lines represent cases when redshifts cuts are applied, as shown in legend together with the total number of objects. Dotted lines indicate the median number of spectroscopic members, and correspond to: 7.5/8.7 (all), 4.8/4.0 ($z>0.4$), 6.2/5.9 ($0.2<z<0.4$) and 11.7/13.0 ($z<0.2$) for all/completed clusters respectively.}
	\label{fig:nmem_dist} 
\end{figure*}

The SEQUELS-DR12 sample consists of 230 validated CODEX systems, out of an initial set of 351 CODEX candidate clusters within the SEQUELS footprint \citep[][]{alam2015}. Among those 230 clusters, 137 are fully observed within SEQUELS-DR12 (i.e.~all tiled targets have received a fiber). Let alone differences in target selection outlined in Sect.~\ref{sect:targeting}, this subsample offers a representative view of the expected, $\sim 20$ times larger, entire SPIDERS sample of clusters. Half of the validated clusters have more than 7 spectroscopic members (8 for completed clusters), this number increases with decreasing redshift, as shown in Fig.~\ref{fig:nmem_dist}. 
Fig.~\ref{fig:areacurve} illustrates the X-ray sensitivity of the CODEX survey integrated over the footprint of the area considered for the present catalogue. The median sensitivity in the [0.5-2]~keV band is $\sim 10^{-13}$\,ergs\,s$^{-1}$\,cm$^{-2}$.

The redshift distribution of clusters in bins of $\Delta_z = 0.04$ is shown in Fig.~\ref{fig:zrm_vs_zspec} and peaks at $z \sim 0.2$. A deficit is observed in one bin around $z=0.3$, which we attribute to a mixture of selection effects involving different redshift dependencies of the X-ray sensitivity and RedMapper efficiency, to the preliminary existence of numerous redshifts in SDSS pre-SPIDERS data peaking below and above $z \sim 0.3$, and to sample variance.

The X-ray properties of the 230 validated clusters were computed according to the updated redshift value, starting from the ROSAT counts \citep[e.g.][]{mirkazemi2015}. Their $L_X-z$ distribution is displayed in Figure~\ref{fig:lxzdist}.
A caveat in the computation of X-ray properties relates to the 10 clusters split into two components after visual inspection. Only one X-ray detection is associated to the original CODEX candidate, and current data do not allow to assign a flux to each of the components. In this work we considered only the primary component as the source of the X-ray emission and therefore discarded the 10 secondary components from the catalogue.

Two points in Fig.~\ref{fig:lxzdist} are labelled with '(C)', indicating likely contamination of the X-ray measurement by a (possibly unrelated) point-source in RASS data. These clusters are optically poor ($\lambda_{\rm OPT} < 10$), hence are among the less reliable sources in the CODEX sample. This class of sources is not targeted in the main SPIDERS program.

		\subsubsection[]{The SPIDERS-XCLASS clusters}

The SEQUELS-DR12 sample contains three XCLASS-RedMapper clusters, as listed in Table~\ref{tab:xclassrmdr12}, two being completely observed, ID-5117 still awaiting completion. One of the clusters (ID~157) is also found in the CODEX subsample. However, the higher quality of the X-ray data allows to measure its flux, luminosity and temperature with much greater accuracy than the RASS does. This system is in fact better known as Abell~851 (Table~\ref{tab:mcxc_matches}). The SPIDERS redshift, luminosity and $R_{500}$ values agree with those found in the literature \citep{piffaretti2011}. Our XMM-derived gas temperature is similar to the value of $T_{X,{\rm all}}= 5.7 \pm 0.5$~keV reported in \citet{mahdavi2013}. Our velocity dispersion estimate computed from 18 SPIDERS spectroscopic members is in agreement with the value of $\sigma_{v}=1067_{-96}^{+89}$~km/s derived by \citet{girardi2001} using 55 members; and with that of \citet{oemler2009}, $\sigma_{v}=1287$~km/s, using 101 members.

After masking point sources, XMM MOS1, MOS2 and PN spectra were extracted in the $[0.3-10]$~keV energy range and analysed with \textsc{XSpec} \citep{arnaudxspec}. An \textsc{APEC} model was fit to measure the cluster temperatures $T_X$, fixing the element abundance to $0.3\,Z_{\odot}$. The results shown in Table~\ref{tab:xclassrmdr12} involve a scaling relation linking $T_X$ and $R_{500c}$ \citep[][]{sun2009}, found by iteratively recomputing the temperature within the aperture. Fluxes were extracted on $[0.5-2]$~keV XMM images following a growth curve analysis \citep[][]{reiprichboehringer2002, suhada2012, clerc2014, pacaud2016} and converted into rest-frame $[0.1-2.4]$~keV luminosities assuming the best-fit APEC spectral model found earlier.

This analysis, summarized in Table~\ref{tab:xclassrmdr12}, illustrates the gain in information brought by the XCLASS subsample of SPIDERS clusters (originating from XMM data) in comparison with the CODEX subsample (originating from the shallower RASS data). For instance, the CODEX $[0.1-2.4]$~keV luminosity relative uncertainty on ID~157 is $\sim 25\%$, while the XCLASS one is only a few percent.
However, given the low number of XCLASS clusters within the SEQUELS-DR12 demonstration sample, we do not consider them further and defer the interpretation of the full SPIDERS/XCLASS sample to a future study.

\begin{table*}
	\centering
\caption{\label{tab:xclassrmdr12}The XCLASS-RedMapper clusters validated in SEQUELS-DR12. Their absorbed fluxes, luminosities (in the rest-frame $[0.1-2.4]$~keV band) and temperatures are derived from XMM data and computed within the $[0-R_{500c}]$ radial range. The radius $R_{500c}$ is derived via a $T_X-R_{500c}$ scaling relation \citep{sun2009}. Line-of-sight velocity dispersions and uncertainties are estimated as described in Sect.~\ref{sect:zveldisp}. The ID in the first two columns refer to the RedMapper (RM) and XCLASS (XC) catalogue IDs respectively. ($^{*}$: is present in the SPIDERS-CODEX subsample and also known as Abell~851.)}
		\begin{tabular}{@{}ccccccccccc@{}}
\hline
RM	&	XC	&	R.A.	&	Dec	&	$z_{\rm spec}$	&	$N_{\rm mem}$	&	$\sigma$	&	$f_X^{[0.5-2]}$	&	$L_X^{[0.1-2.4]}$	&	$T_X$	&	$R_{500c}$	\\
ID	&	ID	&	(J2000)	& (J2000)	&	&	&	km/s		&	($10^{-14}$~ergs/s/cm$^2$)	&	($10^{43}$~ergs/s)	&	(keV)	&	(Mpc)	\\
\hline
\hline
5117			&	1288	&	122.586	&	48.347	&	$0.534 \pm 0.002$	&	11	&	$1060 \pm 250 $	&	$28.7 \pm 0.9$	&	$41.7 \pm 1.3$	&	$4.8 ^{+0.5}_{-0.4}$	&	0.84	\\
157$^{*}$	&	1678	&	145.754	&	46.992	&	$0.408 \pm 0.002$	&	18	&	$1270 \pm 210$	&	$56.2 \pm 0.6$	&	$45.9 \pm 0.5$	&	$5.3 \pm 0.1$	&	0.93	\\
15756		&	1451	&	170.746	&	46.988	&	$0.478 \pm 0.004$	&	6	&	-	&	$3.3 \pm 0.4$		&	$3.9 \pm 0.5$	&	$3.0_{-0.8}^{+1.6}$	&	0.67	\\
\hline
		\end{tabular}
\end{table*}

		\subsubsection[]{The pilot sample catalogue}
We provide in Table~\ref{tab:recap_samples} a condensed summary of the samples and catalogues discussed in this paper.

The list of 230 SPIDERS/CODEX clusters is available online\footnote{https://data.sdss.org/sas/dr13/eboss/spiders/analysis/catCluster-SPIDERS\_RASS\_CLUS-v1.0.fits}. The content of the columns in the catalogue are summarized in Table~\ref{tab:cat_cols}. Column names starting with {\tt SCREEN} result from visual inspection of the system. The luminosity and cluster radius are computed according to the cluster redshift assigned after visual inspection. Note the presence of 10 additional entries in this catalogue, flagged with {\tt COMPONENT} set to 2, corresponding to putative groups along the line of sight of a given cluster.

\begin{table}
	\centering
\caption{\label{tab:recap_samples}Summary of the number of objects in the different survey areas mentioned in this work. The number of candidates ("cand.") and validated ("val.") clusters in each sample are shown. "MCXC" refers to the compilation of X-ray detected galaxy clusters by \citet{piffaretti2011}, as detailed in Sect.~\ref{sect:mcxc}.}
		\begin{tabular}{@{}lcccc@{}}
\hline
Area:			&	\multicolumn{2}{c}{BOSS imaging}	&	\multicolumn{2}{c}{SEQUELS-DR12}	\\
\hline
				&	cand.	&	val.	&	cand.	& val. \\
\hline
\hline
SPIDERS/CODEX	&	10\,415 		& 	-		&	351  		& 230	\\
SPIDERS/XCLASS	&	278			&  	-		&	7			&  3		\\
MCXC 			&	-			&	718		&	-			& 24		\\
\hline
		\end{tabular}
\end{table}

\begin{table*}
	\centering
	\caption{\label{tab:cat_cols}Description of the columns entering the catalogue of validated SPIDERS/CODEX clusters (230 entries) in the SEQUELS-DR12 pilot area. The full table is available online (see text).}
	\begin{tabular}{@{}lllr@{}}
\hline
Column	&	Unit	&	Description	&	Example	\\
\hline
{\tt CLUS\_ID}	&	&	SPIDERS/CODEX identification number &	1\_4601	\\
{\tt COMPONENT}	&	&	Component index of the system	&	1	\\
{\tt RA}			&	deg	&	CODEX X-ray detection right ascension (J2000)	&	185.497	\\
{\tt DEC	}		&	deg	&	CODEX X-ray detection declination (J2000)	&	45.310	\\
{\tt RA\_OPT}		&	deg	&	CODEX optical detection right ascension (J2000)	&	185.522	\\
{\tt DEC\_OPT}	&	deg	&	CODEX optical detection declination (J2000)	&	45.404	\\
{\tt LAMBDA\_CHISQ\_OPT}	&	&	Richness ($\lambda_{\rm OPT}$) of the CODEX optical detection	&	47.2	\\
{\tt Z\_LAMBDA}	&	&	Photometric redshift ($z_{\lambda}$) of the CODEX optical detection	&	0.266	\\
{\tt Z\_LAMBDA\_ERR}	&	&	Uncertainty on {\tt Z\_LAMBDA}	&	0.009	\\
{\tt NMEM}		&	&	Number of objects in the CODEX red sequence ($p_{\rm mem} > 5\%$)	&	64	\\
{\tt NOKZ}		&	&	Number of red-sequence members with a spectroscopic redshift	&	21	\\
{\tt SCREEN\_CLUZSPEC}		&	&	Galaxy cluster redshift	&	0.2630	\\
{\tt SCREEN\_CLUZSPECBOOT}	&	&	Bootstrap uncertainty on {\tt SCREEN\_CLUZSPEC}	&	0.0009	\\
{\tt SCREEN\_CLUVDISP\_GAP}	&	km/s	&	Gapper estimate of the cluster velocity dispersion	&	869.8	\\
{\tt SCREEN\_CLUVDISP\_BWT}	&	km/s	&		Square root of the biweight variance velocity dispersion	&	868.0	\\
{\tt SCREEN\_CLUVDISPTYPE}	&	&	Type of the "best" velocity dispersion (gapper or bi-weight)	&	SIG-BWT	\\
{\tt SCREEN\_CLUVDISPBEST}	&	km/s	&	Value of the "best velocity dispersion"	&	868.0	\\
{\tt SCREEN\_DAZSPEC}	&	Mpc	&	Angular diameter distance computed at $z=$ {\tt SCREEN\_CLUZSPEC}	&	836.9	\\
{\tt SCREEN\_NMEMBERS}	&	&	Number of red-sequence members retained as cluster members	&	20	\\
{\tt SCREEN\_STATUS}	&	&	Validation status of the cluster assigned by the visual inspector	&	validated	\\
{\tt LX0124}	&	ergs/s	&	Luminosity in the rest-frame 0.1-2.4 keV band in $R_{500c}$	&	$1.3\times10^{44}$	\\
{\tt ELX}	&	ergs/s	&	Uncertainty on {\tt LX0124}	&	$0.4\times10^{44}$	\\
{\tt R200C\_DEG}	&	deg	&	Apparent $R_{200c}$ radius of the galaxy cluster	&	0.093	\\
{\tt FLUX052}	&	ergs/s/cm2	&	Galaxy cluster X-ray flux in the 0.5-2.0 keV band	&	$4.1\times10^{-13}$\\
{\tt EFLUX052}	&	ergs/s/cm2	&	Uncertainty on {\tt FLUX052}	&	$1.3\times10^{-13}$	\\
{\tt MCXC}		&			&	Identifier in the MCXC catalogue \citep{piffaretti2011}, if present &	n/a	\\
{\tt ANAME}		&			&	Alternative name in \citet{piffaretti2011}, if present	&	n/a	\\
\hline
	\end{tabular}
\end{table*}

	\subsection[]{Cluster $L_X-\sigma$ relation from individual measurements}
	
\begin{figure}
	\includegraphics[width=84mm]{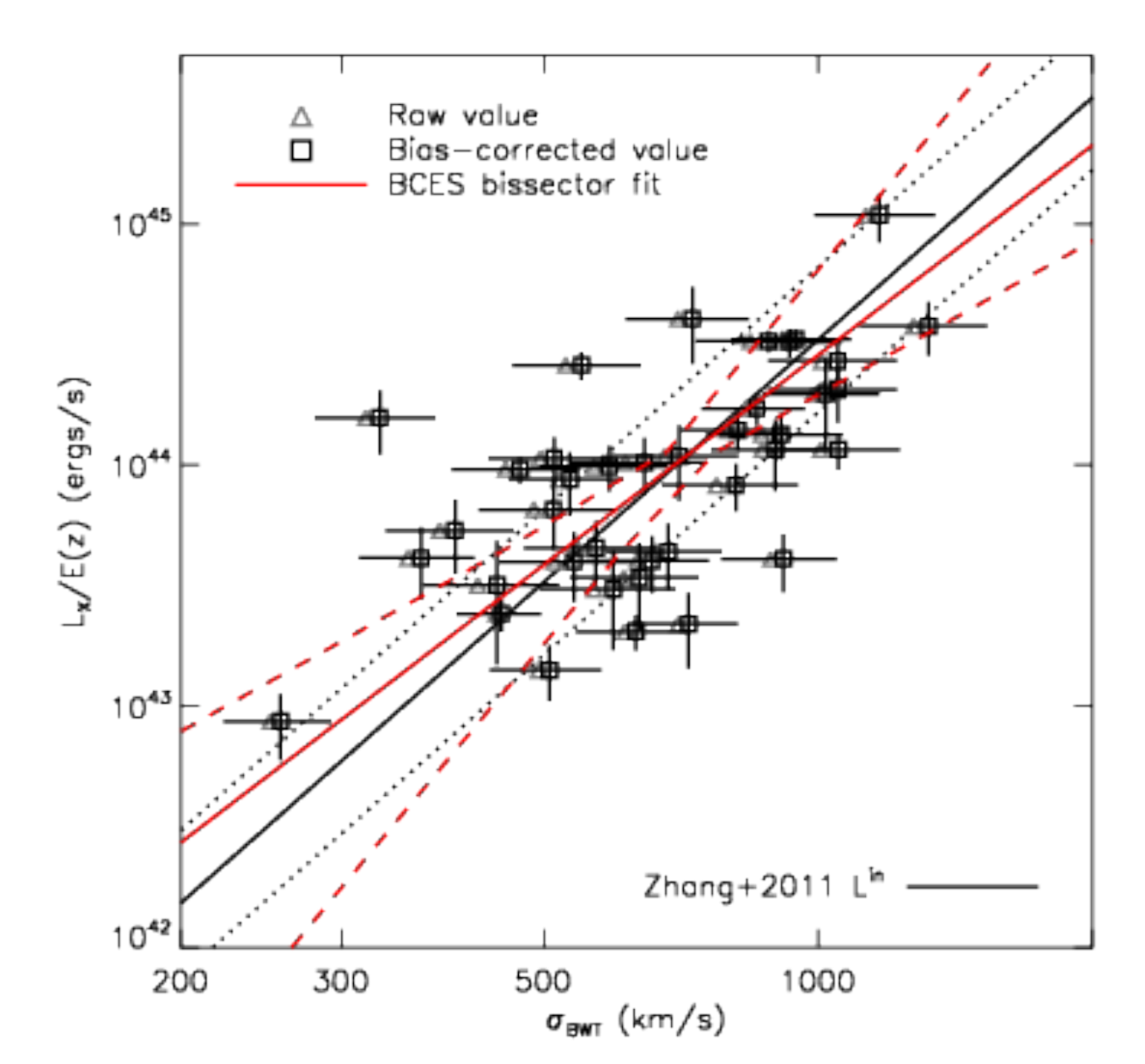}
		\caption{Individual SPIDERS-CODEX clusters in the $L_C-\sigma$ plane. Points represent CODEX clusters validated in the SEQUELS-DR12 demonstration sample with more than 15 spectroscopic members. The raw biweight variance calculations are indicated with light triangles, the bias-corrected values with squares, together with the uncertainty (see text). The plain and dotted red lines show the BCES fit to the bias-corrected values and 1-$\sigma$ uncertainty range. The solid line corresponds to the scaling relation from \citet[][]{zhang2011} and is not fit to the data. A typical 0.3\,dex intrinsic dispersion is materialized as dotted lines.}
	\label{fig:lx_vs_vdisp} 
\end{figure}

Fig.~\ref{fig:lx_vs_vdisp} shows the distribution of SPIDERS/CODEX clusters in the $L_C$-$\sigma_{\rm BWT}$ plane, where $\sigma_{\rm BWT}$ is computed using the biweight sample variance estimate. Only 39 clusters with more than 15 spectroscopic members are considered here.
The raw $\sigma_{\rm BWT}$ was corrected from its expected bias, according to the model described in Sect.~\ref{sect:zveldisp}. This model is also used to assign error bars to the velocity dispersion measurements, based on the number of members within each system.

Fig.~\ref{fig:lx_vs_vdisp} also shows the scaling relation derived from the HIFLUGCS cluster sample. This relation was derived from observations of 62 low-redshift clusters, with much denser spectroscopic coverage \citep[][]{zhang2011} than the current SPIDERS sample. For this comparison we considered the core-included luminosity-velocity dispersion relation. Considering the intrinsic scatter residing in the $L_X-\sigma$ relation (dotted lines in the figure), there is a satisfactory agreement between the position of these points and the HIFLUGCS scaling relations.

We computed the best-fit power-law using the BCES bissector method\footnote{We are thankful to C.~Sif\'on for making the Python implementation of the BCES algorithm available at http://home.strw.leidenuniv.nl/\~{}sifon/pycorner/bces/.} \citep{akritas1996}, as a rough indicator of the overall trend in our sample. For this exercise, we fitted constants $A$ and $B$, defined such as:
\begin{equation}
\log_{10} \left( \frac{\sigma_{\rm BWT}}{700 ~{\rm km\,s}^{-1}} \right) = A + B . \log_{10} \left(  \frac{L_X.E(z)^{-1}}{10^{44} ~{\rm ergs\,s}^{-1}} \right)
\end{equation}

The consistency between the best-fit power-law and our reference HIFLUGCS relation is encouraging. Proper derivation of scaling relations between X-ray quantities and velocity dispersions relying on a fully consistent statistical treatment and including covariances and selection effects, will constitute a major task once the SPIDERS sample of clusters grows up in size.

	\subsection[]{Cluster $L_X-\sigma$ relation from stacked velocity-distance diagrams}
	
In this section, we investigate how stacking together clusters of similar properties can enhance the statistical power in determining scaling relations between those properties and average velocity dispersion measurements. This method is used \citep[e.g.][]{carlberg1997,biviano2009,rines2013, munari2013} when the number of spectroscopic members per cluster is low and does not allow accurate, individual, velocity dispersion measurements. \citet{becker2007} in particular could measure with accuracy the relation between optical richness and velocity dispersion of optically selected galaxy clusters up to $z=0.3$ by means of stacking systems in richness and redshift. Our approach here is similar and uses X-ray luminosity instead of richness.
	
		\subsubsection[]{An adaptive $L_X-z$ space binning}
	
We first selected the 108 clusters with at least 8 members. This threshold ensures the uncertainty on the cluster rest velocity to be $\lesssim 200$~km/s for a typical 500~km/s velocity dispersion cluster (Eq.~\ref{eq:ruelzerr}).
We split the sample in 3 redshift slices, namely $[0.03-0.26]$, $[0.26-0.50]$ and $[0.50-0.73]$. Each of them is subdivided into a number of $L_C$ bins, according to an adaptive procedure. Starting from the highest luminosity, each bin is enlarged until the clusters it contains bring $N_{\rm bin} > 150$ galaxies within $\pm 4000$~km\,s$^{-1}$ of their own cluster rest velocity, and we ensure the size of a bin in luminosity exceeds $\Delta \ln[L_C] > 0.35$. This value is indeed comparable to the typical uncertainty in a CODEX cluster luminosity (see Sect.~\ref{sect:catalogue_prod}). An additional constraint was added to the adaptive binning algorithm, such that each bin contains at least 50~\% of the number of galaxies $\lambda_{\rm scal}(L_X^{\rm cen})$ expected\footnote{$\lambda_{\rm scal}(L_X)$ was estimated from \citet[][their Eq.~29]{rykoff2012}.} to pertain to a single cluster at the centre bin luminosity. This last requirement ensures that each "stacked" cluster contains a high enough number of galaxies, thus avoiding biases in the resulting velocity dispersions \citep[see e.g.][their Fig.~12]{zhang2011}.

\begin{figure}
	\includegraphics[width=84mm]{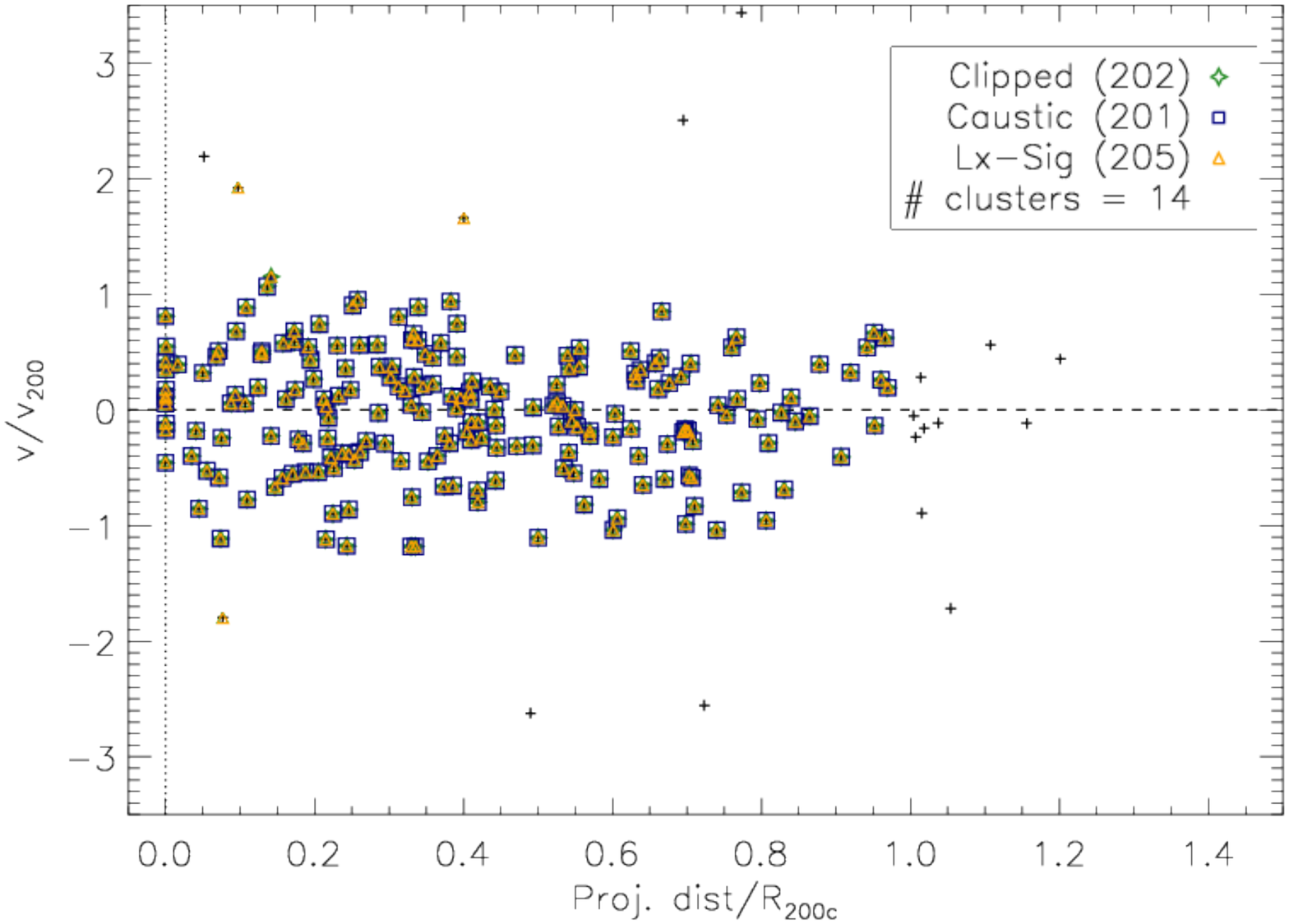} 
		\caption{Stacked phase-space diagram of 14 SPIDERS clusters with $0.03<z<0.26$ and $0.75 < L_C/ (10^{44} ~{\rm ergs\,s}^{-1}) < 1.1$. Black crosses represent all red-sequence members with a spectroscopic redshift, within $\pm 4000$~km\,s$^{-1}$ of their parent cluster redshift. Coloured symbols show galaxies selected by three cleaning techniques, as indicated in legend (iterative clipping, caustic method, clipping with $L_X-\sigma$ prior).}
	\label{fig:stacked-phase-space} 
\end{figure}

Considering all clusters within a bin, red-sequence members with a spectroscopic redshift were assembled into phase-space diagrams, as shown in Fig.~\ref{fig:stacked-phase-space} for the specific bin $0.03<z<0.26$ and $0.75 \times 10^{44} ~{\rm ergs\,s}^{-1} < L_C < 1.1 \times 10^{44} ~{\rm ergs\,s}^{-1}$. In this particular example, 14 clusters are stacked together and 220 spectroscopic galaxies contribute to the stack (corresponding to the black crosses).
To produce such stacked diagrams, the projected distance of each member is scaled by $R_{200c}$ of its host cluster, as estimated from the X-ray data\footnote{Following the scheme described in Sect.~\ref{sect:catalogue_prod}, this involves a scaling relation $L_X \rightarrow M_{200c} \rightarrow R_{200c}$.}. The cluster centre was chosen to be the optical centre, as derived by the RedMapper algorithm for each CODEX cluster. Individual galaxy velocities were rescaled by their parent cluster $v_{200} = \sqrt{G\,M_{200c}/R_{200c}}$ so to provide normalized velocities $v/v_{200}$.

Finally, each stack is assigned a typical X-ray luminosity and a representative $\langle v_{200} \rangle$ by taking the error-weighted averages of the luminosities and $v_{200}$ values of all clusters in the stack.

\begin{table*}
	\centering
\caption{\label{tab:stacked_bins}Velocity dispersion results from the stacked phase-space analysis. The 108 SPIDERS galaxy clusters are binned according to their redshift in $[z_{min}, z_{max}]$ and their luminosity $L_C$ in $[L_{min}, L_{max}]$, expressed in units $10^{44}$~ergs/s. The bins contain $N_{clu}$ clusters, and $N_{bin}$ galaxies contribute initially to each stack. The velocity dispersions obtained after selection from three different techniques are computed using the $\sigma_{\rm BWT}$ estimator.}
		\begin{tabular}{@{}ccccccc|cc|cc|cc@{}}
\hline
\multicolumn{7}{r}{Membership method:}	&	\multicolumn{2}{c}{Iterative clipping}	&	\multicolumn{2}{c}{Caustic}	&\multicolumn{2}{c}{$\sigma(L_X)$ clipping}		\\
\hline
	&&&&&&&&&&\\
ID	&	$z_{min}$	&	$z_{max}$	&	$L_{min}$	&	$L_{max}$	&	$N_{\rm clu}$	&	$N_{\rm bin}$	&	$N_{\rm sel}$	&	$\sigma_{\rm BWT}$	&	$N_{\rm sel}$	&	$\sigma_{\rm BWT}$	&	$N_{\rm sel}$	&	$\sigma_{\rm BWT}$	\\
&	&	&	&	&	&	&	&	(km/s)	&	&	(km/s)	&	&	(km/s)	\\
\hline
\hline
1	&	0.03	&	0.26	&	0.02	&	0.2	&	10	&	131	&	123	&	$330 \pm 27$	&	120	&	$313 \pm 24$	&	123	&	$330 \pm 27$		\\
2	&	0.03	&	0.26	&	0.2	&	0.3	&	12	&	188	&	171	&	$503 \pm 34$	&	162	&	$444 \pm 26$	&	171	&	$503 \pm 34$		\\
4	&	0.03	&	0.26	&	0.3	&	0.45	&	15	&	218	&	206	&	$557 \pm 30$	&	196	&	$499 \pm 24$	&	206	&	$557 \pm 30$		\\
5	&	0.03	&	0.26	&	0.45	&	0.75	&	12	&	156	&	145	&	$508 \pm 41$	&	143	&	$491 \pm 38$	&	145	&	$508 \pm 41$		\\
6	&	0.03	&	0.26	&	0.75	&	1.1	&	14	&	220	&	202	&	$546 \pm 25$	&	201	&	$541 \pm 24$	&	205	&	$568 \pm 30$		\\
0	&	0.03	&	0.26	&	1.1	&	2	&	12	&	182	&	171	&	$695 \pm 50$	&	159	&	$582 \pm 34$	&	171	&	$695 \pm 50$		\\
3	&	0.03	&	0.26	&	2	&	20	&	7	&	116	&	114	&	$871 \pm 65$	&	108	&	$779 \pm 52$	&	114	&	$871 \pm 65$		\\
	&&&&&&&&&&\\
8	&	0.26	&	0.50	&	0.09	&	0.92	&	2	&	16	&	15	&	$355 \pm 107$	&	-	&	-	&	16	&	$374 \pm 155$		\\
9	&	0.26	&	0.50	&	0.92	&	2	&	14	&	172	&	161	&	$561 \pm 39$	&	134	&	$392 \pm 24$	&	164	&	$585 \pm 44$		\\
7	&	0.26	&	0.50	&	2	&	20	&	8	&	97	&	96	&	$1011 \pm 80$	&	72	&	$686 \pm 50$	&	95	&	$985 \pm 72$	\\
	&&&&&&&&&&\\
10	&	0.50	&	0.73	&	2	&	20	&	2	&	39	&	23	&	$1061 \pm 184$	&	12	&	$773 \pm 179$	&	23	&	$1061 \pm 184$		\\

\hline
		\end{tabular}
\end{table*}

		\subsubsection[]{Member identification in stacked diagrams}
		
Although stacked diagrams are pre-filtered such as to contain only red-sequence members within $4000$~km/s of their cluster parent, they still contain a fraction of potential interlopers.
We investigate three methods to clean stacked diagrams and converge to more precise membership, within the limitations of our present catalogue:
\begin{enumerate}
\item The first method is very similar to the one used precedently for individual cluster velocity dispersions. It relies on an iterative 3-$\sigma$ clipping technique using the bi-weight average and bi-weight variance as estimates of the centre and velocity dispersions of the stacked clusters. Only members at $R_{\rm proj}/R_{200c} < 1$ are considered in this analysis.
\item The second method relies on the identification of the caustic \citep{diaferio1999} in each diagram, similarly implemented as in \citet{zhang2011}. The caustic is a characteristic shape in the phase-space diagrams, it isolates interlopers from virialized members in a cluster. It effectively makes full use of the two-dimensional structure of the diagrams.
\item The third method starts by estimating the expected velocity dispersion $\sigma_{\rm exp}$ of a galaxy cluster of luminosity $L_X^{\rm cen}$ using \citet{zhang2011} scaling relations. Galaxies with offset velocities larger than $3 \times \sigma_{\rm exp}$ are excluded.
\end{enumerate}

Results are illustrated in Fig.~\ref{fig:stacked-phase-space}, where 202/220, 201/220 and 205/220 members were selected by each of the respective methods, the most stringent selection originating from the caustic identification.

		\subsubsection[]{Velocity dispersions from stacked diagrams}

Considering only members identified by one of the 3 'cleaning' methods, two numerical estimators of the velocity dispersion and their respective uncertainties are derived. In both cases only members within a projected radius less than $R_{200c}$ enter the computation.

The first method computes the bi-weight variance $\sigma_{\rm BWT}$ of the selected members, and the uncertainty is based on 1000 bootstrap resamplings of the data.

The second method is similar to \citet{rozo2015}. It is based on maximizing the likelihood:
\begin{equation}
\mathcal{L} = \prod_{i} \left[ p G(v_i ; 0,\sigma)+(1-p) \frac{1}{2 v_m} \right]
\end{equation}
with $v_i$ the velocities of individual members, $G(x ; \mu, \sigma)$ the Gaussian function of mean $\mu$ and standard deviation $\sigma$. Here $v_m$ is the maximal velocity, i.e.~$3 \, \sigma_{\rm BWT}$, $4 \langle v_{200} \rangle$ and $3\, \sigma_{\rm exp}$ for each of the cleaning method (i), (ii) and (iii) respectively. The parameters $\sigma_{\rm gauss}$ and $p$ that maximize $\mathcal{L}$ are found using the {\scshape Amoeba} algorithm and 1000 bootstrap resamplings are performed to estimate the uncertainty on $\sigma_{\rm gauss}$.

Combining the three 'cleaning' methods to the two estimators leads to 6 estimates of the velocity dispersion for a given stacked phase-space diagram.

	\subsubsection[]{The $L_X - \sigma$ relation of stacked SPIDERS clusters}

\begin{figure*}
	\includegraphics[width=\linewidth]{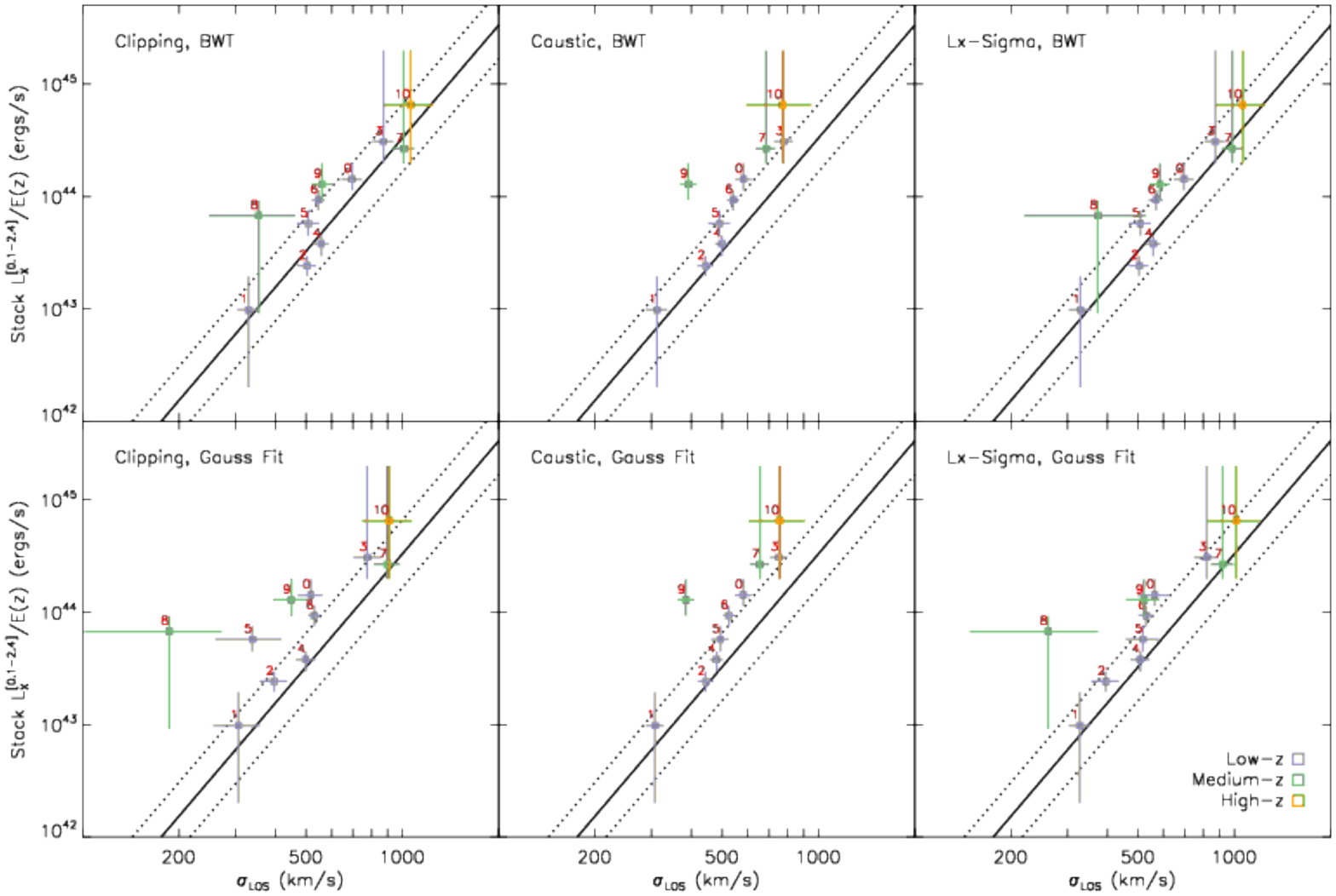}
		\caption{The radial velocity dispersion ($\sigma_{\rm LOS}$) versus X-ray luminosity ($L_X$ in [0.1-2.4]~keV band) as drawn from the stacked phase-space analysis. Six different methods are used to extract a velocity dispersion estimate from phase-space diagrams binned in the $(z, L_C)$ plane, as shown in Fig.~\ref{fig:stacked-phase-space}. Columns from left to right correspond to the three different cleaning techniques: iterative $\sigma$-clipping, caustic identification and $\sigma(L_X)$-clipping. The top row corresponds to the bi-weight variance estimate $\sigma_{\rm BWT}$, while the bottom row correspond to the gaussian fit estimate $\sigma_{\rm gauss}$ (see text). Red numbers refer to the bin ID as listed in Table~\ref{tab:stacked_bins}. Plain, dashed and dotted line are identical in each panel, and correspond to the scaling relations derived in \citep[][]{zhang2011}. Colours encode the redshift binning used for this analysis.}
	\label{fig:compa_stack_sigestimates} 
\end{figure*}

The values of the velocity dispersion in each $(z, L_C)$ bin are reported in Fig.~\ref{fig:compa_stack_sigestimates} (bin numbering listed in Table~\ref{tab:stacked_bins}). The externally derived scaling relation superimposed to guide the eye is taken from \citet[][]{zhang2011} who fit the $L_X-\sigma$ relation on individual, bright, X-ray clusters in the HIFLUGCS sample. This relation is the same as the black solid line in Fig.~\ref{fig:lx_vs_vdisp}. The dotted lines represent the typical intrinsic scatter ($\sim 0.3$ dex) in this relation.

We defer quantitative measurements and a thorough assessment of the stacked $L_X-\sigma$ relation to further studies, that will rely on the entire SPIDERS sample of galaxy clusters and detailed treatment of numerical simulations. We note at this stage a broad agreement between the location of the data points and our reference $L_X-\sigma$ relation. Our results differ according to the combination of cleaning and fitting method employed.
The clipping-based method may lead to more complete but less clean member sampling, the prior-based method is very similar and possibly introduces auto-correlation to some extent.

In the present work, the caustic method filters out more members than the clipping- and prior-based techniques do, and it provides lower velocity dispersion values, hence higher deviations from the fiducial scaling relation (central column in Fig.~\ref{fig:lx_vs_vdisp}).
Simulations \citep[e.g.][]{serra2013} indicate that the caustic method better distinguishes cluster members from interlopers than other methods~; however, small number statistics impact the precision of the determination of the amplitude of the caustic and thus the caustic mass distribution. A lower number of members tends to provide a slightly reduced amplitude, which causes underestimation of the total mass. 
Moreover, since the caustic-based filtering makes full use of the projected radius information enclosed in phase-space diagrams, one expects an increased sensitivity of this method to centering uncertainties, to uncertainties in the computation of the normalizing $R_{200c}$ and to sparsity in the 2-dimensional diagrams. Further studies based on numerical simulations tailored to SPIDERS stacks will assess the absolute and relative performances of the methods when combining higher-quality X-ray data (\emph{eROSITA} data) to the entire, larger, SPIDERS dataset.

Interestingly, data points corresponding to the 'medium-redshift' bin ($z \in [0.26,0.50]$) deviate from this relation at low-$\sigma$ values, and do so more strongly than 'low-redshift' data points. Part of this deviation can be attributed to sample selection effects and Eddington bias. In App.~\ref{app:selbias} we describe a modeling of X-ray selection biases and their impact on scaling relations, by comparing the measured cluster luminosity. As shown in Fig.~\ref{fig:massmatrix}, Eddington bias makes low-mass system (equivalently, low-velocity dispersion systems) appear more luminous on average, and the effect increases with redshift, in agreement with the trend seen in this analysis.
The results shown in App.~\ref{app:selbias} assume perfect association of the optical spectra to the X-ray emitting intra-cluster gas. Studying the reliability of such identification, as well as possible contaminants to the X-ray (due for instance to the increased presence of X-ray AGN in group-like halos) and optical signals, is beyond the scope of this paper. These additional sources of bias, likely to dominate in the low-count/low-richness regime, need to be addressed with further simulations.

\section[]{Comparison with previous X-ray cluster catalogues}
\label{sect:mcxc}

\begin{figure*}
	\includegraphics[width=\linewidth]{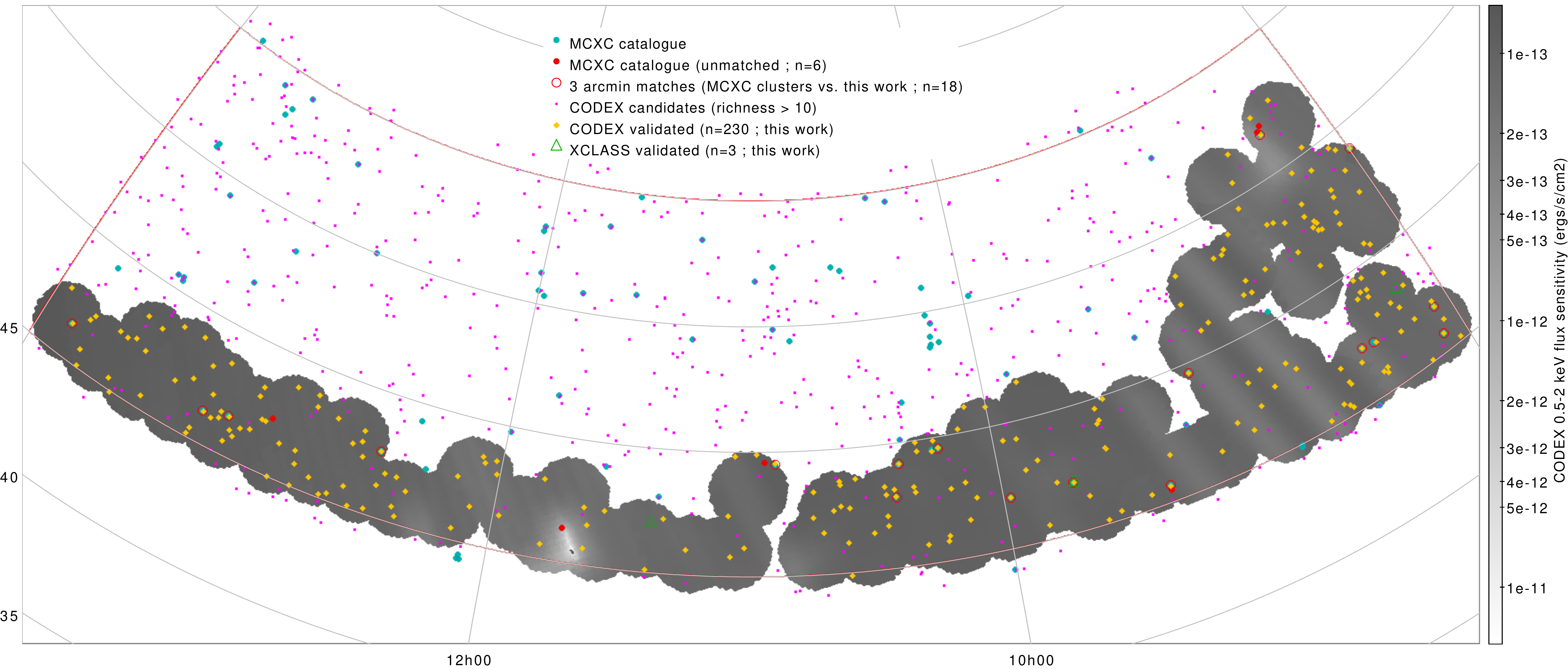}
		\caption{Distribution in equatorial coordinates of the main samples of objects discussed in this work. The grey shaded area corresponds to the X-ray flux sensitivity of the ROSAT All-sky survey in the footprint of the SPIDERS pilot area. Galaxy clusters from the MCXC catalogue \citep{piffaretti2011} are displayed and those not matching any validated SPIDERS clusters in this area are highlighted red (6 objects, see text).}
	\label{fig:pilotareasamples} 
\end{figure*}

\begin{figure}
	\includegraphics[width=84mm]{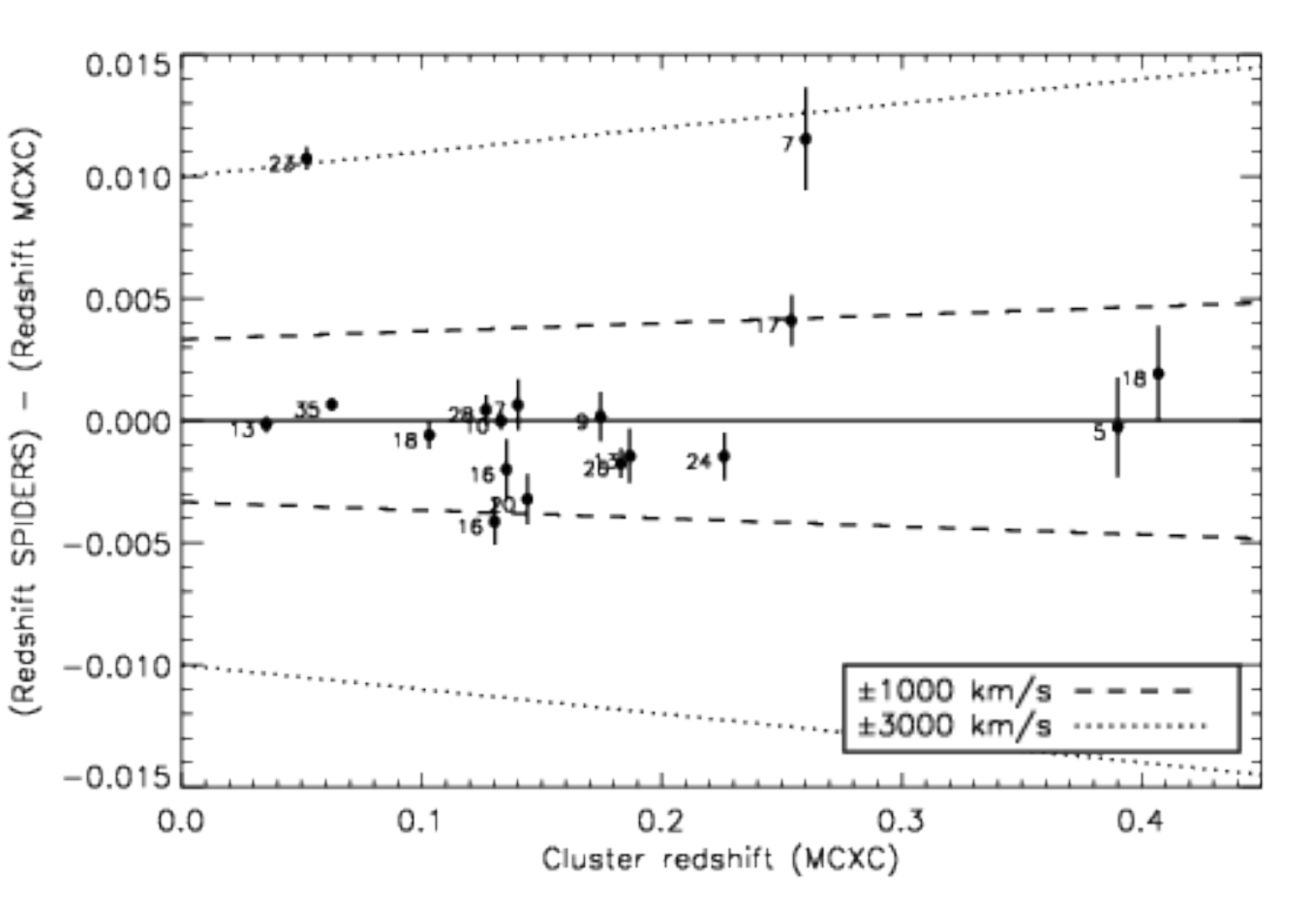}
		\caption{Comparison of redshift values for the 18 matches between the SPIDERS pilot validated sample presented in this work and the MCXC meta-catalogue \citep{piffaretti2011}. The figures in parenthesis indicate the number of spectroscopic members entering the computation of the cluster redshift, error bars indicate the SPIDERS redshift uncertainty. Lines indicate the velocity offset at the comparison redshift.}
	\label{fig:compazmcxc} 
\end{figure}

We compare our work to the objects extracted from the MCXC compilation of catalogues \citep{piffaretti2011}, which contains most of the ROSAT-based samples, including serendipitous detections from deep pointed observations. Fig.~\ref{fig:pilotareasamples} shows the distribution on sky of the samples discussed in this paper. We find 18 matches between the SPIDERS pilot sample and the MCXC database within 3\,arcmin of the CODEX X-ray position. Their properties are summarized in Table~\ref{tab:mcxc_matches} and compared with values extracted from the MCXC compilation. Five systems exhibit a richness lower than 20 in the overlap between the two catalogues. Fig.~\ref{fig:compazmcxc} compares the redshift values in both catalogues and demonstrates their good agreement. All but two agree within 1000~km\,s$^{-1}$ of the MCXC redshift, the other two agree within 3000~km\,s$^{-1}$.

Within the survey footprint, 6 MCXC clusters could not be matched to a SPIDERS validated cluster (Table.~\ref{tab:mcxc_nonmatches}). Five of them have values of luminosity and redshift (Fig.~\ref{fig:lxzmcxc}) consistent with sources below or at the edge of the CODEX X-ray detectability -- these are sources detected in deep ROSAT pointed observations \citep{vikhlinin1998,mullis2003}. In particular, MCXC~J0921.2+4528 at $z=0.315$ is the brightest of these five and shows a flux significant at a 1.3-$\sigma$ level only (in the ROSAT All-sky survey) at the value of $1.2\times10^{-13}$~ergs\,s$^{-1}$\,cm$^{-2}$, hence it lies in the largely incomplete part of the sensitivity range.
The sixth source is Abell~1361 and is located within a masked region of the ROSAT all-sky data used as input of the CODEX X-ray finding algorithm (a degree north of the sensitivity dip visible in Fig.~\ref{fig:pilotareasamples} at R.A. $\sim$11h40m and Dec. $\sim +45\deg$).

\begin{table*}
	\centering
\caption{\label{tab:mcxc_matches}Comparison of the 18 galaxy clusters in common between the MCXC compilation \citep{piffaretti2011} and the SPIDERS-DR12 pilot sample presented in this work. Luminosities $L_{500}$ are expressed within the $R_{500}$ radius taken from each catalogue, in the [0.1-2.4]~keV band. $^*$: see also Table~\ref{tab:xclassrmdr12}.}
		\begin{tabular}{@{}cccccccc@{}}
\hline
SPIDERS	&	MCXC	&	Alternative	&	$z_{\rm spec}$	&	$z_{\rm lit}$	&	$L_{500}$	&	$L_{500}$	&	$\lambda_{\rm OPT}$ \\
	ID		&	ID	&		name		&	(SPIDERS)	&	(MCXC)	&	(SPIDERS)	&	(MCXC)	&\\
			&		&				&							&						&	$10^{44}$\,ergs\,s$^{-1}$	&	$10^{44}$\,ergs\,s$^{-1}$	&	\\
\hline
1\_2952	&	J1053.7+4929	&		&	$0.141	\pm	0.001$		&	0.140	&	$0.5	\pm	0.1$	&	1.7	&	12.2 \\
1\_4189	&	J0921.1+4538	&	3C 219	&	$0.175	\pm	0.001$	&	0.175	&	$1.5	\pm	0.3$	&	1.4	&	13.4 \\
2\_2449	&	J0907.8+4936	&	VV 196	&	$0.0351	\pm	0.0003$	&	0.035	&	$0.14 \pm 0.03$	&	0.1	&	17.1 \\
1\_2848	&	J1013.6+4933	&	VMF98 87	&	$0.1330	\pm	0.0003$	&	0.133	&	$0.2	 \pm 0.1$	&	0.3	&	17.7 \\
1\_4021	&	J0822.1+4705	&	A0646	&	$0.1262	\pm	0.0009$	&	0.130	&	$3.5	\pm	0.3$	&	3.0	&	19.0 \\
1\_2788	&	J1025.0+4750	&	A1003	&	$0.0627	\pm	0.0005$	&	0.052	&	$0.2	1\pm	0.04$	&	0.1	&	24.4 \\
1\_4240	&	J0958.3+4702	&			&	$0.390	\pm	0.002$	&	0.390	&	$3.0	\pm	0.8$	&	1.9	&	28.6 \\
2\_4405	&	J1351.7+4622	&			&	$0.0632	\pm	0.0003$	&	0.062	&	$0.2	5\pm	0.04$	&	0.3	&	31.4 \\
1\_1172	&	J0759.7+5400	&	Zw 0755.8+5408	&	$0.1026	\pm	0.0006$		&	0.103	&	$1.0		\pm	0.1$	&	1.1	&	34.9 \\
1\_1198	&	J0819.9+5634	&	VMF98 50	&	$0.272	\pm	0.002$	&	0.260	&	$1.7	\pm	0.8$	&	0.9	&	37.1 \\
2\_3671	&	J0804.3+4646	&	A0616	&	$0.185	\pm	0.001$	&	0.187	&	$1.1	\pm	0.3$	&	1.4	&	53.1 \\
2\_3682	&	J0805.7+4541	&	A0620	&	$0.133	\pm	0.001$	&	0.135	&	$1.2	\pm	0.2$	&	0.9	&	64.3 \\
2\_2602	&	J1023.6+4907	&	A0990	&	$0.141	\pm	0.001$	&	0.144	&	$3.6	\pm	0.3$	&	3.9	&	72.5 \\
2\_4317	&	J1313.1+4616	&	A1697	&	$0.1813	\pm	0.0006$	&	0.183	&	$1.5	\pm	0.3$	&	2.6	&	82.5 \\
1\_3111	&	J1229.0+4737	&	A1550	&	$0.258	\pm	0.001$	&	0.254	&	$3.1	\pm	0.6$	&	3.3	&	108.4 \\
2\_4315	&	J1306.9+4633	&	A1682	&	$0.224	\pm	0.001$	&	0.226	&	$3.6	\pm	0.5$	&	5.1	&	123.9 \\
2\_3664	&	J0825.5+4707	&	A0655	&	$0.1271	\pm	0.0006$	&	0.127	&	$1.8	\pm	0.3$	&	2.8	&	131.8 \\
1\_4241$^{*}$	&	J0943.1+4659	&	A0851	&	$0.409	\pm	0.002$	&	0.407	&	$4.7	\pm	1.2$	&	4.9	&	148.8 \\
\hline
	\end{tabular}
\end{table*}

\begin{table}
	\centering
\caption{\label{tab:mcxc_nonmatches}The 6 galaxy clusters found in the MCXC compilation \citep{piffaretti2011} within the footprint of the SPIDERS-DR12 pilot area, but not present in the SPIDERS-DR12 pilot sample. The last 5 entries correspond to faint clusters detected in deep ROSAT pointed observations, therefore they are unseen in shallower RASS data (see text).}
		\begin{tabular}{@{}cccc@{}}
\hline
MCXC	&	Alternative	&	$z_{\rm lit}$	&	$L_{500}$ \\
	ID	&		name		&	(MCXC)	&	(MCXC)\\
		&				&			&	$10^{44}$\,ergs\,s$^{-1}$	\\
\hline
J1143.5+4623	&	A1361	&	0.117	&	2.8 \\
J1256.6+4715	&	VMF98 129	&	0.404	&	0.5 \\
J0818.9+5654	&	VMF98 48	&	0.260	&	0.3 \\
J0820.4+5645	&	VMF98 51	&	0.043	&	0.02 \\
J0921.2+4528	&	VMF98 70	&	0.315	&	1.0 \\
J1056.2+4933	&	VMF98 94	&	0.199	&	0.2 \\
\hline
	\end{tabular}
\end{table}


\section[]{Conclusions}
	\label{sect:conclusions}

This paper introduces the SPIDERS spectroscopic follow-up of X-ray galaxy clusters with particular emphasis on the selection of targets. The galaxy cluster component in SDSS-IV/SPIDERS will obtain optical spectra of 40\,000-50\,000 galaxies identified as potential members of 5\,000 to 6\,000 massive (from $10^{13.5}$ to $10^{15.5} M_{\odot}$, peaking at $\sim 10^{14.5} M_{\odot}$) X-ray galaxy clusters in the Northern hemisphere, up to redshift 0.6 and above. This massive observational effort will bring the average number of galaxy spectra within their respective red-sequences from 2 to 10, therefore allowing precise calculation of galaxy cluster redshifts, relying on a median number of 8 member galaxies per system.
Until the launch of the \emph{eROSITA} satellite (2017), the observed sample of X-ray galaxy clusters originates from the ROSAT all-sky survey and from XMM-Newton archival observations. The target selection heavily relies on the RedMapper algorithm, that is able to assign membership probabilities to galaxies photometrically identified as red-sequence members across the SDSS imaging data.
The \emph{eROSITA} all-sky survey will complement this preliminary observational tier by bringing denser samples than ROSAT and a more detailed X-ray information over the entire surveyed area.

The achieved spectral quality will allow secure redshift measurements of the targeted red galaxies (up to $i=21.2$ in $2\arcsec$ aperture), relying particularly on the extensive developments achieved for the BOSS and eBOSS surveys: observation planning and realization, processing pipelines, infrastructure, databases and analysis tools. A number of steps are envisaged to construct reliable catalogues of X-ray validated clusters with redshift, by assigning membership of galaxies within their parent clusters. These procedures will mix automatized algorithms -- to treat the bulk of the dataset in a most efficient way -- and visual screening -- to address peculiar cases, especially in the low-member regime.
Throughout this paper, the feasibility of this programme was demonstrated based on a pilot sample of 230 galaxy clusters. All were confirmed with spectroscopic data, providing accurate redshifts at the $\Delta_z/(1+z) \sim 0.001$ level. We highlighted the difficulties implied by reduced X-ray information for the poorer systems (projection effects, ambiguous associations, etc.) Better X-ray data are required for these low-mass, high-redshift systems, as already provided by XMM or by \emph{eROSITA} in the future.

The SPIDERS cluster follow-up programme is essential to achieve the cosmological analysis of the mass function and tridimensional distribution of X-ray galaxy clusters. Indeed, precise redshift information enables precise determination of cluster X-ray properties (luminosity, temperature, gas mass, etc.) related to the host halo mass, and accurate localization of these objects in the cosmic web. Moreover, a wealth of additional science cases will be addressed via the SPIDERS survey. Among them, we have shown that dynamical mass estimates are accessible for a subset of the clusters (in this paper through the radial velocity dispersion proxy), despite the low number of spectroscopic members per individual system. In particular, the stacking of X-ray clusters offers a promising avenue to the study of average properties in such a large sample. Specifically, the results of our pilot study could establish that the radial velocity dispersions correlate with the X-ray luminosity of (stacked) clusters in a similar way as local galaxy clusters do. The methods introduced in this paper are meant to evolve during the course of the survey, most likely including state-of-the art and most recent techniques.
The quality and number of galaxy spectra within or in the line-of-sight of galaxy clusters will be exploited to address several science topics, ranging from galaxy formation and evolution to properties of the intergalactic medium.

Finally, besides the exceptional dataset provided by the programme and its predicted science outcome, SPIDERS is already starting to pave the way for future, large-area, spectroscopic surveys. In particular, the 4MOST instrument on the ESO-VISTA telescope \citep{dejong2014} will follow-up \emph{eROSITA} clusters in the Southern hemisphere in the early 2020s and will largely benefit from the science and technical developments pursued within SPIDERS in SDSS-IV.

\section*{Acknowledgments}

We thank the referee for useful discussion that helped in improving the content of this paper.

Funding for the Sloan Digital Sky Survey IV has been provided by
the Alfred P. Sloan Foundation, the U.S. Department of Energy Office of
Science, and the Participating Institutions. SDSS-IV acknowledges
support and resources from the Center for High-Performance Computing at
the University of Utah. The SDSS web site is www.sdss.org.

SDSS-IV is managed by the Astrophysical Research Consortium for the 
Participating Institutions of the SDSS Collaboration including the 
Brazilian Participation Group, the Carnegie Institution for Science, 
Carnegie Mellon University, the Chilean Participation Group, the French Participation Group, Harvard-Smithsonian Center for Astrophysics, 
Instituto de Astrof\'isica de Canarias, The Johns Hopkins University, 
Kavli Institute for the Physics and Mathematics of the Universe (IPMU) / 
University of Tokyo, Lawrence Berkeley National Laboratory, 
Leibniz Institut f\"ur Astrophysik Potsdam (AIP),  
Max-Planck-Institut f\"ur Astronomie (MPIA Heidelberg), 
Max-Planck-Institut f\"ur Astrophysik (MPA Garching), 
Max-Planck-Institut f\"ur Extraterrestrische Physik (MPE), 
National Astronomical Observatory of China, New Mexico State University, 
New York University, University of Notre Dame, 
Observat\'ario Nacional / MCTI, The Ohio State University, 
Pennsylvania State University, Shanghai Astronomical Observatory, 
United Kingdom Participation Group,
Universidad Nacional Aut\'onoma de M\'exico, University of Arizona, 
University of Colorado Boulder, University of Oxford, University of Portsmouth, 
University of Utah, University of Virginia, University of Washington, University of Wisconsin, 
Vanderbilt University, and Yale University.

Y.Y.Z acknowledges support by the German BMWI through the Verbundforschung under grant 50\,OR\,1506.

Funding for SDSS-III has been provided by the Alfred P. Sloan Foundation, the Participating Institutions, the National Science Foundation, and the U.S. Department of Energy Office of Science. The SDSS-III web site is {http://www.sdss3.org/}.

\appendix

\section[]{Impact of a lower priority threshold on the targeting}
\label{app:eboss3}

\begin{figure}
	\includegraphics[width=84mm]{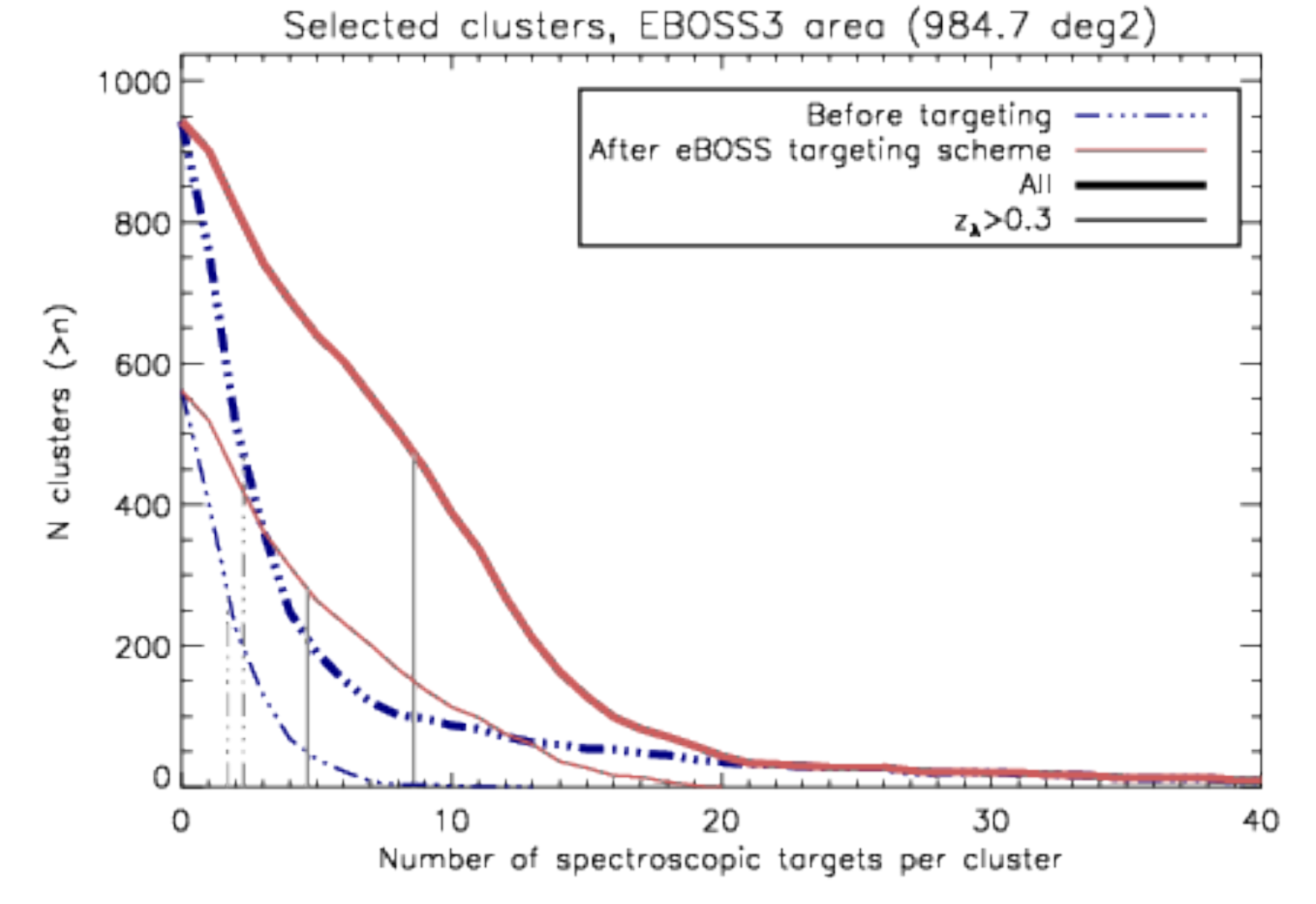}
		\caption{Same as Fig.~\ref{fig:eboss1_tilingforecasts} for the {\tt eboss3} chunk. The density of SPIDERS clusters in this chunk is higher than average ($\sim 1$~deg$^{-2}$), hence a lower cut in priority was imposed to the target list submitted to tiling ($priority < 33$ instead of $80$). The median number of spectroscopic redshifts in a cluster red-sequence is hence slightly lower than in other chunks (8 instead of 10). These median numbers change from 2 to 4 when considering only high-redshift candidates (thin lines).}
	\label{fig:eboss3_tilingforecasts} 
\end{figure}

\begin{figure*}
\begin{tabular}{c}
	\includegraphics[width=\linewidth]{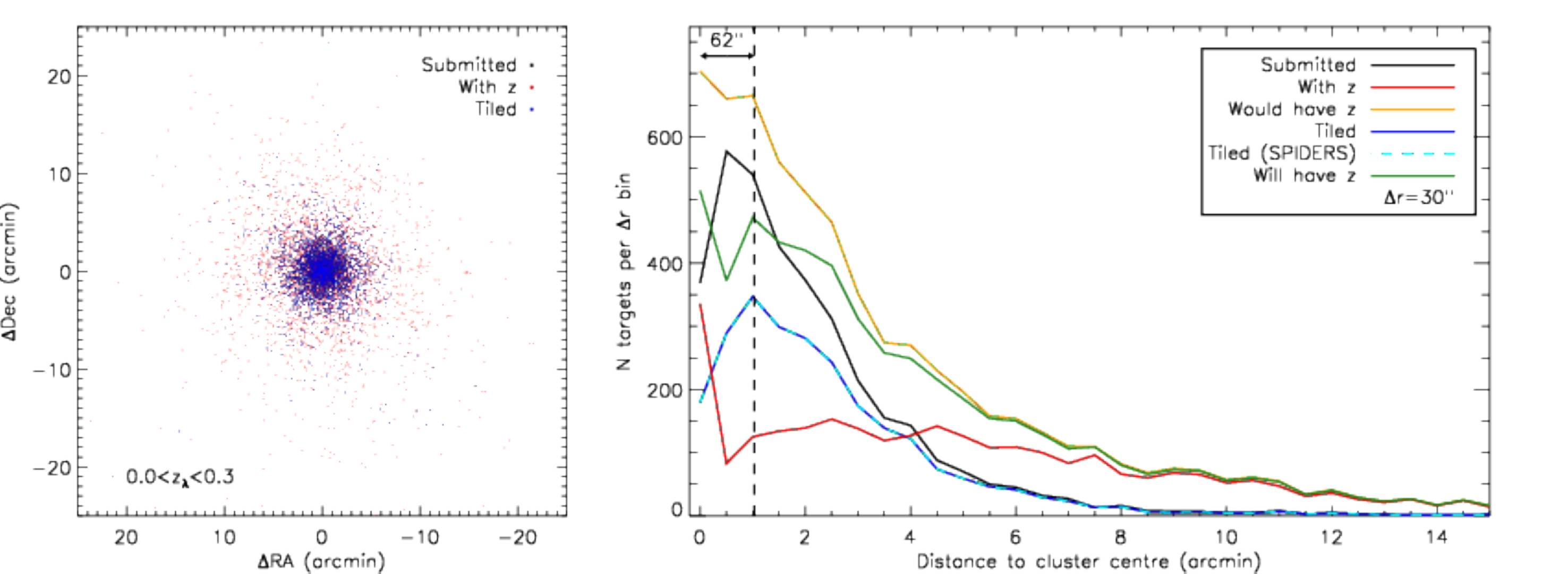} \\
	\includegraphics[width=\linewidth]{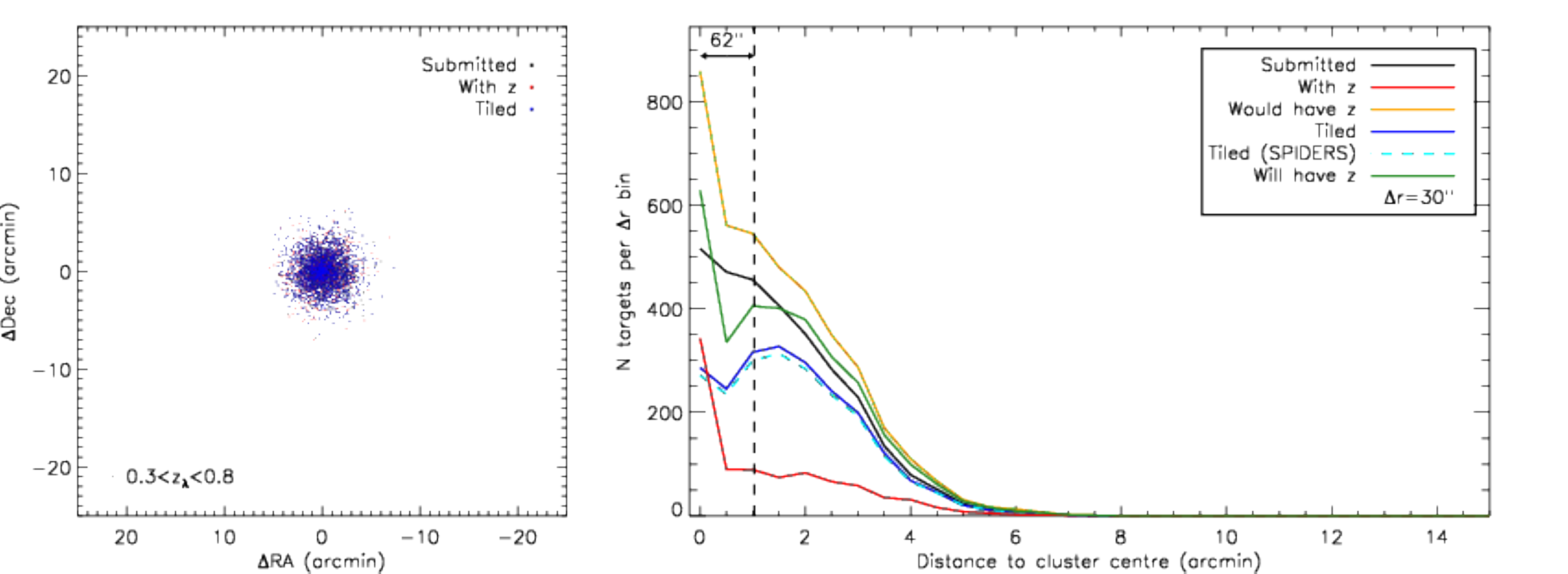}
\end{tabular}
		\caption{Equivalent of Fig.~\ref{fig:eboss1_targetsky} for the {\tt eboss3} chunk (943 systems). In this chunk the priority threshold for SPIDERS\_RASS\_CLUS galaxy cluster targets has been set to 33 instead of 80. This change in targeting strategy is motivated by the higher density of X-ray clusters in this area of sky, due to the corresponding local increase of exposure time of the ROSAT All-sky survey. The main consequence is a lower tiling rate at higher cluster-centric radii.}
	\label{fig:eboss3_targetsky} 
\end{figure*}

We presented in Sect.~\ref{sect:targetsel} our choice to lower the priority threshold in the submitted target list in order to accommodate for the higher density of unique galaxy clusters in certain area of sky.
In the particular case of the {\tt eboss3} chunk, a threshold of 33 (instead of 80) was set. Fig.~\ref{fig:eboss3_tilingforecasts} shows the impact of this change, resulting in a decrease in the net number of targets per system (from $\sim 10$ to $\sim 8$, see Fig.~\ref{fig:eboss1_tilingforecasts}).
Fig.~\ref{fig:eboss3_targetsky} shows that targets in the outer regions of clusters are more severely impacted by this incompleteness. This is easily explained by the link between priority and RedMapper probability that in particular accounts for the galaxy's cluster-centric distances through a radial density profile weighting \citep[][]{rykoff2014}. As the interloper rate is expected to decrease with decreasing cluster-centric distance \citep[e.g.][]{saro2013}, the impact of a lower number of spectroscopic redshifts per cluster should be mitigated by a lower interloper rate.

\section[]{Description of the target catalogue}

The target catalogue contains the list of targets submitted to the tiling algorithm. It is made available as part of SDSS Data Release 13\footnote{http://www.sdss.org}. Nominally, SPIDERS targets galaxies selected within the optical red-sequence of individual galaxy clusters. In the Tier-0 phase of SPIDERS, these clusters are drawn from the CODEX and XCLASS samples.

The targets of SPIDERS clusters are split into three categories:
\begin{itemize}
\item in the 117 SEQUELS plates (66 of them have already been released as part of SDSS DR12), the {\tt SPIDERS\_RASS\_CLUS} target flag corresponds to objects indifferently selected among CODEX or XCLASS red-sequences; 
\item in all eBOSS plates, the {\tt SPIDERS\_RASS\_CLUS} target flag corresponds to object selected in CODEX red-sequences; 
\item in all eBOSS plates, the {\tt SPIDERS\_XCLASS\_CLUS} target flag corresponds to object selected in X-CLASS red-sequences.
\end{itemize}

In BOSS and eBOSS, all SEQUELS targets are tracked by the {\tt EBOSS\_TARGET0} bitmask and the {\tt SPIDERS\_RASS\_CLUS} clusters targets correspond to bit 21.
In eBOSS, all SPIDERS targets (clusters and AGN) are tracked by the {\tt EBOSS\_TARGET1} with bit 31. The {\tt EBOSS\_TARGET\_2} bitmask distentangles between the various SPIDERS targets: the {\tt SPIDERS\_RASS\_CLUS} targets correspond to bit 1 and {\tt SPIDERS\_XCLASS\_CLUS} targets to bit 5.

The SPIDERS clusters target catalogues contain the minimal information necessary for the tiling algorithm, as well as supplementary information relative to their parent galaxy cluster. Table~\ref{tab:example_targetcat} shows a subset of the SPIDERS target catalogue. {\tt FIBER2MAG} is the $2\arcsec$ aperture magnitude of the source in each of the five SDSS filters. {\tt TARGETSELECTED} is the confirmation rank of the galaxy in the red-sequence. The cluster photometric redshift $z_{\lambda}$ (and uncertainty $\Delta z_{\lambda}$) and the richness $\lambda_{\rm OPT}$ are given together with the internal cluster identifier {\tt CLUS\_ID}. The priority column plays an important role in the tiling algorithm as it defines how fiber collisions are resolved. Note that SPIDERS AGN all have a priority set to 1.

\begin{table*}
	\centering
\caption{\label{tab:example_targetcat}Sample of the SPIDERS target catalogue of galaxy clusters from the CODEX sample ({\tt SPIDERS\_RASS\_CLUS} targets, see text). The full table is available electronically as part of SDSS Data Release 13 (http://www.sdss.org).}
		\begin{tabular}{@{}ccccccccc@{}}
\hline
RA	&	Dec	&	{\tt FIBER2MAG} ($u,g,r,i,z$)	&	{\tt TARGETSELECTED}	&	$z_{\lambda}$	&	$\Delta z_{\lambda}$	&	$\lambda_{\rm OPT}$	&	{\tt CLUS\_ID}	&	Priority		\\
deg	&	deg	&	mag.	&\\
\hline
117.9716&	27.3276	&	$(24.1, 22.4, 21.0, 20.4, 20.0)$	&	1	&	0.265	&	0.019	&	12.5	&	1\_10000	&	0	\\
112.1157&	26.6917	&	$(22.4, 20.5, 19.4, 18.9, 18.5)$	&	1	&	0.137	&	0.005	&	30.2	&	1\_10008	&	0	\\
...	&	...	&	...	&	...	&	...	&	...	&	...	&	...	&	... \\
134.8594	&	55.3670	&	(23.0, 22.0, 20.6, 20.1, 19.7)	&	2	&	0.269	&	0.023	&	11.8	&	1\_1284	&	3 \\
...	&	...	&	...	&	...	&	...	&	...	&	...	&	...	&	... \\
201.2000 & 28.4698	&	(24.3, 22.3, 20.9, 20.3, 20.1)	&	36	&	0.266	&	0.009	&	48.3	&	2\_9973	&	79 \\
\hline
		\end{tabular}
\end{table*}


\section[]{Modeling the selection and measurement biases}
\label{app:selbias}

\begin{figure}
	\includegraphics[width=84mm]{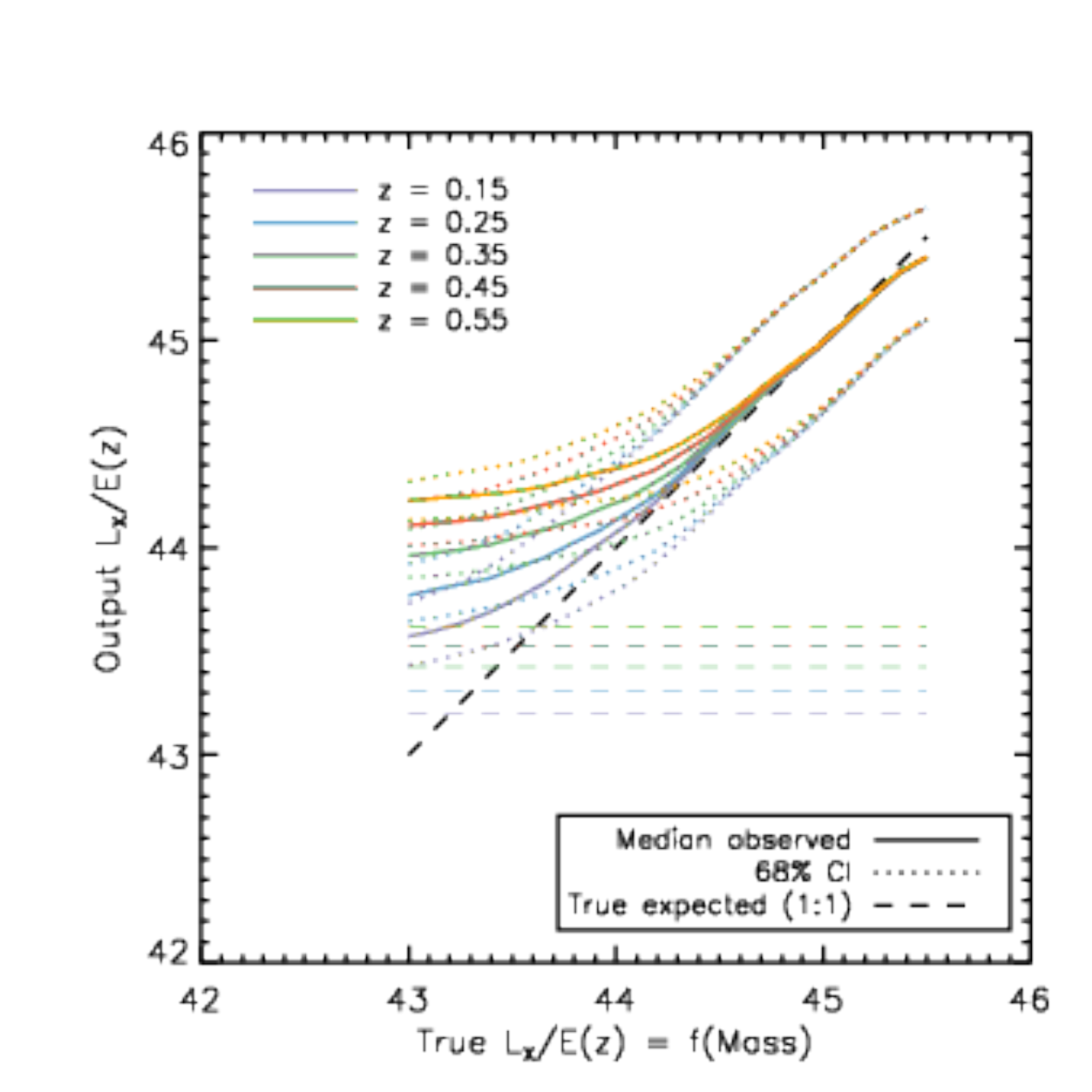}
		\caption{Median and 68\% confidence intervals of simulated cluster luminosity measurements in the SPIDERS pilot sample (CODEX clusters only), as a function of the X-ray luminosity of a cluster whose mass is given as input (x-axis). Horizontal dashed lines show for each redshift the level of flux contamination by X-ray AGN as calculated by our model (see text) and that is added to each curve.}
	\label{fig:massmatrix} 
\end{figure}	

The SPIDERS/CODEX sample is based on a sample of faint X-ray sources, down to 4 photon counts in RASS data. Measurement of scaling relations and population studies are therefore subject to selection and measurement biases, that can be addressed via simulations. As an illustration, we show in Fig.~\ref{fig:massmatrix} the outcome of a series of simulations of CODEX clusters. The sensitivity curves were calculated specifically according to the footprint of the pilot sample presented in this paper (see Fig.~\ref{fig:pilotareasamples} and Fig.~\ref{fig:areacurve}). The 'true' luminosity is that expected from a simple mass-luminosity relation and is not necessarily equal to the quantity measured in real data. Due to Eddington bias and in presence of intrinsic scatter in the mass-luminosity relation, faint objects are on average measured with higher luminosities, hence the upturn at low luminosities. The effect is redshift-dependent and more pronounced for high-redshift objects.

Given that the RASS luminosity might be affected by the AGN, we have included into the calculation of the observed luminosity the effect of unresolved AGN. Our model combines the evolution of AGN X-ray luminosity function of \citet{ebrero2009} with the halo occupation distribution (HOD) study of \citet{allevato2012}. The resulting mean contamination of AGN per cluster is $(1+z)^{3.3} \times 6.6\times 10^{42}$\,ergs\,s$^{-1}$ in the rest-frame 0.5--2\,keV band, which has to be rescaled up self-consistently using cluster K-correction factors of $\sim$ 1.5. This model is in a very good agreement with the results of \citet{martini2013}, who measured AGN activity in $1<z<1.5$ clusters at an average level of $10^{44}$\,ergs\,s$^{-1}$ in the 0.5--8\,keV band.
In our highest redshift bin, the model predicts an AGN contribution of $4.7 \times 10^{43}$ erg s$^{-1}$, which on average is still a subdominant contribution to the observed flux. This model predicts that the $z<0.1$ and low $L_X$ ($L_X<10^{43}$\,ergs\,s$^{-1}$) sample requires significant correction for AGN contamination.

The SPIDERS/XCLASS sample is based on a different selection technique and different X-ray datasets. The galaxy cluster X-ray selection function are presented in \citet{clerc2012b} along with the X-ray flux measurement calibrations.

\bsp

\label{lastpage}

\end{document}